\ifdefined\pldisubmission
  \documentclass[acmsmall,screen,review,anonymous,nonacm]{acmart}
\else
  \documentclass[acmsmall,screen,nonacm]{acmart}
\fi
\newif\ifanon
\ifdefined\pldisubmission\anontrue\else\anonfalse\fi

\setcopyright{none}
\renewcommand\footnotetextcopyrightpermission[1]{}
\usepackage{booktabs}
\usepackage{multirow}
\usepackage{makecell}
\usepackage{tabularx}
\usepackage{longtable}
\usepackage{placeins}
\usepackage{array}
\usepackage{siunitx}
\usepackage{xspace}
\usepackage{relsize}
\usepackage{listings}
\usepackage{subcaption}
\usepackage{pifont}
\usepackage{enumitem}
\usepackage{algpseudocode}
\usepackage[capitalise,noabbrev]{cleveref}

\makeatletter\def\@ACM@checkaffil{}\makeatother
\graphicspath{{../}{./}}
\newcommand{\crosstl}{CrossTL\xspace}
\newcommand{\crossgl}{CrossGL\xspace}

\newcommand{\code}[1]{\texttt{\relsize{-0.5}#1}}
\newcommand{\crosstlr}{CrossTL-R\xspace}
\newcommand{\repolink}[1]{\ifanon(link removed for double-blind review)\else\url{#1}\fi}
\newcommand{\rul}{\rho}
\newcommand{\undefrule}{\top}
\newcommand{\rtop}{\rul_{\top}}
\newcommand{\loader}{\mathcal{L}}
\newcommand{\refines}{\sqsubseteq}
\newcommand{\nrefines}{\not\sqsubseteq}
\newcommand{\sem}[1]{\llbracket #1\rrbracket}
\newcommand{\RQ}[1]{\textbf{RQ#1}\xspace}
\newcommand{\cmark}{\ding{51}}
\newcommand{\xmark}{\ding{55}}
\newcommand{\outcome}[1]{\textsf{#1}}

\InputIfFileExists{tables/numbers}{}{}

\definecolor{lstbg}{RGB}{248,248,250}
\definecolor{lstkw}{RGB}{36,62,170}
\definecolor{lstcm}{RGB}{96,104,116}
\definecolor{lsttype}{RGB}{122,40,120}
\lstdefinestyle{paper}{
  basicstyle=\ttfamily\footnotesize,
  backgroundcolor=\color{lstbg},
  keywordstyle=\color{lstkw}\bfseries,
  commentstyle=\color{lstcm},
  frame=single, framerule=0.3pt, rulecolor=\color{black!20},
  xleftmargin=4pt, xrightmargin=4pt, aboveskip=6pt, belowskip=4pt,
  columns=fullflexible, keepspaces=true, showstringspaces=false,
  breaklines=true, tabsize=4, captionpos=b, escapeinside={(*@}{@*)}
}
\lstdefinelanguage{GLSL}{
  morekeywords={void,int,uint,float,bool,vec2,vec3,vec4,ivec2,ivec3,ivec4,uvec3,mat3,mat4,layout,in,out,
    buffer,uniform,readonly,writeonly,shared,precise,if,else,for,while,return,break,struct,const,
    local_size_x,std430,std140,binding,location},
  sensitive=true, morecomment=[l]{//}, morecomment=[s]{/*}{*/}, morestring=[b]",
  moredelim=[s][\color{lstcm}]{\#}{\ }
}
\lstdefinelanguage{HLSL}{
  morekeywords={void,int,uint,uint3,float,float2,float3,float4,bool,StructuredBuffer,RWStructuredBuffer,
    ByteAddressBuffer,RWByteAddressBuffer,cbuffer,register,numthreads,precise,groupshared,if,else,for,while,
    return,struct,static,const,SV_DispatchThreadID,SV_Target,SV_Position},
  sensitive=true, morecomment=[l]{//}, morecomment=[s]{/*}{*/}, morestring=[b]"
}
\lstdefinelanguage{Metal}{
  morekeywords={kernel,device,constant,threadgroup,thread,void,int,uint,uint3,float,float2,float3,float4,half,
    bfloat,bool,using,namespace,template,typename,struct,if,else,for,while,return,const,static,constexpr,
    as_type,simd_sum,simd_shuffle,threadgroup_barrier,mem_flags,thread_position_in_grid,buffer,
    instantiate_kernel},
  sensitive=true, morecomment=[l]{//}, morecomment=[s]{/*}{*/}, morestring=[b]",
  moredelim=[s][\color{lstcm}]{\#}{\ }
}
\lstdefinelanguage{CrossGL}{
  morekeywords={shader,compute,vertex,fragment,struct,void,int,uint,float,half,bool,vec2,vec3,vec4,uvec3,
    if,else,for,while,return,let,mut,fn,match,const,StructuredBuffer,RWStructuredBuffer},
  sensitive=true, morecomment=[l]{//}, morecomment=[s]{/*}{*/}, morestring=[b]",
  moredelim=[s][\color{lsttype}]{@}{\ }
}
\lstdefinelanguage{WGSL}{
  morekeywords={fn,let,var,const,struct,return,if,else,for,loop,break,storage,read,read_write,uniform,
    workgroup,compute,workgroup_size,builtin,group,binding,select,array,vec3,u32,i32,f32,bool},
  sensitive=true, morecomment=[l]{//}, morecomment=[s]{/*}{*/}
}
\definecolor{hzgood}{HTML}{0B8A0B}
\definecolor{hzwarn}{HTML}{A87400}
\definecolor{hzbad}{HTML}{C62F2F}
\definecolor{hzserious}{HTML}{C4561F}
\definecolor{hzgrey}{HTML}{6B6B6B}

\crefname{definition}{Definition}{Definitions}
\Crefname{definition}{Definition}{Definitions}
\crefname{theorem}{Theorem}{Theorems}
\crefname{proposition}{Proposition}{Propositions}
\crefname{lemma}{Lemma}{Lemmas}
\crefname{corollary}{Corollary}{Corollaries}

\begin{document}

\title[Hub-Based Translation of GPU Programs]{CrossTL: Hub-Based Translation of GPU Programs and the Preservation of Source-Language Rules}

\author{Nripesh Niketan}
\orcid{0009-0008-2066-1937}
\email{nripesh@crossgl.net}
\affiliation{\institution{CrossGL}}
\author{Vaatsalya Shrivastva}
\affiliation{\institution{CrossGL}}
\author{Bhushan Srivastava}
\affiliation{\institution{CrossGL}}
\author{Sushant Mishra}
\affiliation{\institution{CrossGL}}

\ifanon\else\renewcommand{\shortauthors}{Niketan et al.}\fi

\begin{abstract}
GPU programs are written in vendor languages that share a syntax but disagree on what operations mean: on integer overflow, division by zero, NaN handling, floating-point contraction, data layout and dispatch. A translated program can therefore compile on its target and still compute something else.
We study what a hub language that connects many GPU languages must record to preserve meaning, using \crosstl, a source-to-source translator that routes ten languages to eleven targets.
We define faithful translation as refinement up to rejection under a compile configuration and a recorded state encoding. In a model of operation rules, recording each operation's rule and lowering it through one table per target preserves meaning when annotation is sound and complete and the table is sound; given an exact frame, every target construct must refine the source's rule, and a hub that does not refuse must record an operation's source exactly where defined rules conflict.
We ran 66 hazard kernels in six languages, 44 of them registered by hash before they ran, on one Apple GPU and on Mesa's CPU drivers: SPIRV-Cross, naga, Tint, Slang and \crosstl each lose some source rule, and outcomes change with the compile configuration and the Vulkan implementation.
Carrying the rules removes 32 of 37 divergences on the carried rules, and none of 462 held-out executions diverges; the five that remain are MoltenVK wrap-around failures, three of which a conformant Vulkan driver for Metal reproduces. An SMT check of the rule table finds 13 of 131 entries unsound as recorded.
On 2,585 third-party programs, established translators produce valid graphics code more often than \crosstl in 31 of 39 comparisons, and carrying the rules changes validity by at most 0.2 points per target.
\end{abstract}



\maketitle

\section{Introduction}
\label{sec:intro}

GPU programs are written in vendor languages: game engines keep their shaders in GLSL or HLSL, machine-learning frameworks for Apple silicon write their kernels in the Metal Shading Language (MSL), and scientific codes for NVIDIA hardware use CUDA.
Moving such code to another platform means translating it, and the established tools each cover part of the space: compilers and decompilers around SPIR-V~\citep{khronos2026glslang,microsoft2026dxcspirv,khronos2026spirvcross}, translators with their own intermediate representations, such as naga, Tint and Slang~\citep{gfxrs2026naga,google2026tint,he2018slang}, and one-directional translators for compute code~\citep{amd2026hipify,intel2026syclomatic}.

Getting the syntax right is the easy part.
\Cref{fig:motivation} shows a one-line GLSL kernel and what ten translation routes make of it on seven runtime configurations.
GLSL defines signed addition to wrap around, so \code{x + 1 > x} is false for the largest integer; MSL inherits C++'s rule that signed overflow is undefined, and Apple's compiler folds the comparison to true.
Every route that copies the operator into MSL, including SPIRV-Cross and \crosstl, the hub-based translator that this paper studies, yields a program that compiles and computes the wrong answer, and so does HLSL compiled by DXC, which marks signed arithmetic as non-wrapping.
naga and Tint succeed, because their intermediate representations give signed arithmetic wrap-around semantics and their back ends polyfill it~\citep{nicol2025nagaoverflowfix,tint2021overflow,dawn2025tintsignedpolyfill}, while MoltenVK, a Vulkan layer over Metal, miscompiles even correct SPIR-V, and so, for addition, does KosmicKrisp, a conformant Vulkan driver for the same GPU.
The languages also disagree on unsigned division by zero, NaN operands of \code{min}, contraction of $a\cdot b+c$, buffer layouts and partial work-groups (\cref{sec:background}), and none of these rules is visible in a program's text.

\begin{figure}[t]
\centering
\begin{minipage}[c]{0.33\linewidth}
\begin{lstlisting}[style=paper,language=GLSL,numbers=none,basicstyle=\ttfamily\scriptsize]
layout(local_size_x = 64) in;
layout(std430, binding = 0)
  buffer Inp { int a[]; };
layout(std430, binding = 1)
  buffer Outp { int o[]; };
void main() {
  uint i = gl_GlobalInvocationID.x;
  int x = a[i];
  o[i] = (x + 1 > x) ? 1 : 0;
}
\end{lstlisting}
\end{minipage}\hfill
\begin{minipage}[c]{0.65\linewidth}
\small\setlength{\tabcolsep}{3.5pt}
\begin{tabular}{@{}lccccccc@{}}
\toprule
Route & \rotatebox{90}{Metal, safe} & \rotatebox{90}{Metal, default} & \rotatebox{90}{WGSL, wgpu} & \rotatebox{90}{lavapipe} & \rotatebox{90}{MoltenVK} & \rotatebox{90}{D3D12} & \rotatebox{90}{llvmpipe} \\
\midrule
CrossTL & \xmark & \xmark & \cmark & \cmark & \xmark & \xmark & \cmark \\
CrossTL-R & \cmark & \cmark & \cmark & \cmark & \xmark & \cmark & \cmark \\
glslang$\to$SPIRV-Cross & \xmark & \xmark &  &  &  & \xmark & \cmark \\
glslang (SPIR-V) &  &  &  & \cmark & \xmark &  &  \\
naga (GLSL) & \cmark & \cmark & \cmark &  &  & \cmark &  \\
glslang$\to$naga & \cmark & \cmark & \cmark &  &  & \cmark &  \\
glslang$\to$Tint & \cmark & \cmark & \cmark &  &  & \cmark &  \\
HLSL: DXC$\to$DXIL &  &  &  &  &  & \xmark &  \\
HLSL: DXC$\to$MSC & \xmark & \xmark &  &  &  &  &  \\
HLSL: Slang & \xmark & \xmark & \cmark &  &  &  & \cmark \\
\bottomrule
\end{tabular}

\end{minipage}
\caption{A GLSL kernel (left) through ten translation routes and seven runtime configurations on two code generators, Apple's GPU compiler and Mesa's CPU drivers (right). GLSL defines the addition to wrap, so \code{o[i]} must be 0 for $x=2^{31}-1$ and 1 otherwise. \cmark: all 64 outputs match. \xmark: the output is 1 for $x=2^{31}-1$, because the target compiler treated the overflow as undefined and folded the comparison. Every output shown was accepted by its target compiler; blank: no output for that runtime. D3D12 is emulated by converting DXIL to SPIR-V for lavapipe. The HLSL rows translate the same kernel written in HLSL under the legacy Direct3D rule that signed arithmetic wraps; DXC does not keep it (DXIL row), and Apple's Metal Shader Converter (MSC) inherits its folding. \crosstlr: \crosstl with the passes of \cref{sec:rulehub}. On KosmicKrisp, every SPIR-V route also fails (\cref{sec:eval:hazards}, \cref{app:probes}).}
\label{fig:motivation}
\Description{A GLSL compute kernel computing (x + 1 > x) and a grid showing, for ten translation routes and seven runtime configurations, whether the translated program preserves the wrap-around result. Routes that copy the operator fail on Metal and Direct3D 12. naga and Tint preserve it on every runtime they target, and CrossTL-R on every runtime except MoltenVK, where every SPIR-V route fails.}
\end{figure}

The difficulty is sharpest for translators built around a hub language, which need one importer per source and one emitter per target instead of one translator per pair~\citep{strong1958uncol}.
A hub that records only an operator cannot tell whether the operation follows GLSL's rule or MSL's.
Compiler intermediate representations avoid the problem by fixing one semantics per operation and recording deviations from it, as LLVM does with \code{nsw} flags, poison values and fast-math flags and SPIR-V with decorations and float-control modes~\citep{llvm2026langref,lee2017tamingub,khronos_spirv16}; shader translators inherit the semantics of one language, WGSL for naga and Tint, SPIR-V for SPIRV-Cross.
A hub that accepts many source languages has no single semantics to inherit. We ask what such a hub must record to preserve the meaning of the programs it translates.

\paragraph{This paper.}
We study \crosstl, an open-source translator that imports ten GPU languages (GLSL, HLSL, Metal with its C++ templates, SPIR-V assembly, CUDA, HIP, OpenCL~C, Rust-GPU, Mojo and Slang) into \crossgl, a textual hub language, and emits eleven, WGSL and GLSL~ES among them. We define faithful translation, show that carrying each operation's source-language rule through the hub suffices for it and when it is necessary, implement rule carrying in \crosstl as \crosstlr, and evaluate both, with a suite of hazard kernels, an analysis plan and three hypotheses registered by hash before \crosstlr ran. Our contributions are:
\begin{itemize}[leftmargin=1.2em, itemsep=1pt]
\item a semantic account of GPU program translation (\cref{sec:background,sec:fidelity}): a specification-derived catalogue of 18 classes of operations on which eight GPU languages disagree, faithfulness as refinement up to rejection, obligations and their treatments, and the bases on which a passing test can rest;
\item rule-carrying translation (\cref{sec:rulehub}): a model of operation rules ordered by refinement; a theorem that reduces faithfulness to sound and complete annotation and a sound table per target, with a partial converse; a characterisation of when one rule per operator suffices and when the source must be recorded; and an implementation whose rule table we check mechanically with an SMT solver;
\item a description of \crosstl as a whole (\cref{sec:system}), a hub-based translator for ten source and eleven target GPU languages, with a monomorphiser for Metal's C++ templates and artifacts whose contracts loaders check;
\item an evaluation (\cref{sec:eval}) with 66 registered hazard kernels, eight third-party translators and seven runtime configurations, 102 held-out random kernels, and 2,585 translation units written by third parties.
\end{itemize}

\paragraph{Results.}
SPIRV-Cross, naga, Tint, Slang and \crosstl each lose some source rule, and outcomes change with Metal's math mode and the Vulkan implementation. Carrying the rules removes 32 of the 37 divergences on the carried rules, 24 of them failures of \crosstl's own SPIR-V output, against 14 for local fixes to that emitter, and on the held-out kernels \crosstlr diverged in none of 462 executions; the five that remain are MoltenVK wrap-around failures, three of which KosmicKrisp, a conformant driver for the same GPU, reproduces in an exploratory run. A mechanical check finds 13 of the table's 131 operator entries unsound as recorded; the suite's matches on them hold only by implementation. A single strongest semantics would keep the same rules with about nine times as many rewrites (exploratory). On third-party code, 31\,\% of \crosstl's translations are usable, established translators produce valid output more often in 31 of 39 comparisons, and carrying the rules changes no target's accepted share by more than 0.2 points.

\section{Where GPU Languages Disagree}
\label{sec:background}

GPU languages share a vocabulary of vectors, buffers, barriers and stage entry points, but they give the same operations different meanings.
From the specifications of GLSL, HLSL, MSL, WGSL, SPIR-V, CUDA and OpenCL~C we derived a catalogue of 18 classes of such differences (H1--H18)~\citep{khronos_glsl460,khronos_gl46,ms_d3d113_functional_spec,ms_dxil,ms_hlsl_spec_draft,ms_learn_hlsl_operators,ms_dxc_wiki_buffer_packing,apple_msl41,w3c_wgsl,khronos_spirv16,khronos_glslstd450,khronos_vulkan14,nvidia_cuda_programming_guide,nvidia_cuda_math_api,amd_hip_language_extensions,iso_cpp_n4950,khronos_opencl_c,iso_c99_n1256}.
\Cref{tab:hazard-matrix} gives the rules on values, and \cref{tab:hazard-matrix-env} in \cref{app:hazards} those on the floating-point environment, memory and execution. Each cell rests on a clause of the language's specification, quoted in the artifact, with one exception.
HLSL's specification is a draft that is silent on most rows~\citep{ms_hlsl_spec_draft}, so we state HLSL's rules twice: the legacy rules of the Direct3D bytecode (DXBC) that the FXC compiler emits~\citep{ms_d3d113_functional_spec}, against which many shaders were written, and the rules under which DXC, the current compiler, translates them to DXIL~\citep{ms_dxil}, which we established by inspecting and executing the code it emits. The evaluation uses the current rules; \cref{app:readingA} repeats it under the legacy ones.

\begin{table}[t]
\caption{How GPU languages define operations on values. HLSL is given twice: under the legacy bytecode (FXC/DXBC) and under the current reference compiler (DXC/DXIL). The DXC column records the code that DXC emits, which we inspected and executed; every other cell rests on a clause of the language's specification. Codes: \textsf{wrap} two's-complement wrap-around; \textbf{\textsf{UB}} undefined behaviour; \textsf{poison} a poison value (SPIR-V; LLVM for DXC); \textsf{uval} unspecified value; \textsf{mask} shift count reduced modulo the width; \textsf{def}($v$) defined result $v$; \textsf{sat} saturating conversion; \textsf{num} the non-NaN operand is returned (as IEEE~754-2019 minimumNumber, but C's \code{fmin} and MSL's need not order signed zeros); \textsf{undef} an LLVM undefined value; \textsf{impl} implementation-defined; \textsf{n/s} no rule stated. MSL's ``undefined'' results are read as undefined behaviour, as in C++; GLSL's as unspecified values.}
\label{tab:hazard-matrix}
\footnotesize
\setlength{\tabcolsep}{2.2pt}
\renewcommand{\arraystretch}{1.08}
\begin{tabularx}{\linewidth}{@{}>{\raggedright\arraybackslash}p{1.85cm}>{\raggedright\arraybackslash}X>{\raggedright\arraybackslash}p{1.6cm}*{5}{>{\raggedright\arraybackslash}X}@{}}
\toprule
 & GLSL & HLSL (DXBC / DXC) & MSL & WGSL & SPIR-V (Vulkan) & CUDA / HIP & OpenCL~C \\
\midrule
Signed overflow ($+,-,\times$) & \textsf{wrap} & \textsf{wrap} / \textsf{poison}$^{a}$ & \textbf{\textsf{UB}} & \textsf{wrap} & \textsf{wrap}$^{b}$ & \textbf{\textsf{UB}} & \textbf{\textsf{UB}} \\
Shift count $\geq$ width & \textsf{uval} & \textsf{mask} / \textsf{mask} & \textsf{mask} & \textsf{mask} & \textsf{poison} & \textbf{\textsf{UB}} & \textsf{mask} \\
Unsigned $x/0$ & \textsf{n/s} & \textsf{def}(\texttt{0xFFFF\allowbreak FFFF}) / \textbf{\textsf{UB}}$^{c}$ & \textsf{uval} & \textsf{def}($x$) & \textbf{\textsf{UB}} & \textbf{\textsf{UB}} & \textsf{uval} \\
Signed \% with negative operand & \textsf{uval} & same-sign only / trunc. & \textbf{\textsf{UB}} & trunc. & \textsf{poison}$^{d}$ & trunc. & trunc. \\
\code{int} vs \code{uint} compare & unsigned & unsigned & unsigned & error & by opcode & unsigned & unsigned \\
Float$\to$int: NaN; out of range & \textsf{n/s}; \textsf{n/s} & 0, \textsf{sat} / \textsf{undef}$^{e}$ & 0; \textbf{\textsf{UB}} & \textsf{uval}; \textsf{sat} to $[-2^{31},$ $2^{31}{-}128]$ & \textbf{\textsf{UB}} & \textbf{\textsf{UB}} & \textsf{impl} (\code{\_sat}: 0, \textsf{sat}) \\
\code{min}/\code{fmin} with a NaN & \textsf{uval} & \textsf{num} / \textsf{num} & \textsf{num}$^{f}$ & \textsf{uval} & \textsf{uval}$^{g}$ & \textsf{num} (\code{fminf}) & \textsf{num} (\code{fmin}) \\
\bottomrule
\end{tabularx}
\par\smallskip
{\raggedright\scriptsize
$^{a}$\,DXC emits \code{add nsw}/\code{mul nsw}, whose overflow yields an LLVM poison value, and folds \code{x+1>x} to \code{true} even at \code{-Od}. $^{b}$\,\textsf{UB} only with the \code{NoSignedWrap} decoration. $^{c}$\,DXC emits LLVM \code{udiv}, whose division by zero is undefined in the LLVM~3.7 semantics that DXIL inherits; DXIL's \code{UDiv} operation, which defines \texttt{0xFFFFFFFF}, is not used. $^{d}$\,unless the \code{maintenance8} feature is enabled. $^{e}$\,DXC emits \code{fptosi}, whose result is undefined (\code{undef}) in the LLVM~3.7 semantics that DXIL inherits; the HLSL draft specification makes the conversion undefined. $^{f}$\,undefined under fast math. $^{g}$\,The SPIR-V instruction set leaves \code{FMin} with a NaN operand implementation-defined and defines \code{NMin} to return the other operand, but in Vulkan an implementation may assume that the operands and results of both are not NaN. \code{SignedZero\allowbreak InfNan\allowbreak Preserve} lifts this only for bit-preserving operations and the core instructions that the specification lists, such as \code{OpIsNan} and the comparisons, not for GLSL.std.450 instructions; under Vulkan~1.4 a module can instead state its floating-point environment explicitly.\par}
\end{table}

\paragraph{Values.}
On signed overflow the languages fall into three groups. GLSL, WGSL and SPIR-V define signed arithmetic to wrap around. CUDA and OpenCL~C inherit from C and C++ the rule that signed overflow is undefined, and MSL restates it; Apple's Metal compiler uses the rule (\cref{fig:motivation}). MSL departs from C++ elsewhere: it masks shift counts and leaves \code{\%} with a negative operand undefined. HLSL falls between: DXBC's \code{iadd} wraps, but DXC marks signed additions and multiplications \code{nsw} and folds \code{x + 1 > x} to true even without optimisation.
Shift counts, unsigned division by zero, float-to-integer conversion and \code{min} with a NaN operand each have three or more answers.

\paragraph{The floating-point environment.}
By default every language lets a compiler contract $a\cdot b+c$ into a fused multiply-add, and most defaults let it assume that NaNs and infinities do not occur.
The opt-outs differ in kind: per operation (\code{precise} in GLSL and HLSL, \code{NoContraction} in SPIR-V), per scope (\code{\#pragma METAL fp} in MSL, \code{FP\_CONTRACT} in OpenCL~C), per entry point, instruction or type (SPIR-V's float-control execution modes and its per-instruction fast-math flags, core since Vulkan~1.4~\citep{khronos2026floatcontrols,khronos2024floatcontrolstwo}), per compilation (Metal's compile options), or not at all (WGSL).
Some opt-outs are optional device features, and some are partial: MSL's safe mode still allows contraction within a statement~\citep[\S1.6.3]{apple_msl41}. A program's meaning therefore depends on a configuration that its text does not contain, and the effect can be large: compiled for an Apple M2~Max with the API default of fast math and fast functions, MSL's \code{sin(a)} for $|a|\le 10^4$ has a median error of 40 units in the last place (ULP) and a 99th percentile of 80,263~ULP, against 0 and 2~ULP with precise functions.

\paragraph{Memory and execution.}
A Metal \code{float3} occupies 16 bytes in a buffer, whereas an HLSL structured buffer of \code{float3} has a 12-byte stride, so a translation that copies the type changes the layout that the host must upload.
Metal can launch a partial work-group, which most other APIs cannot, so a Metal kernel that omits its bounds check is correct only on Metal; and MSL leaves the SIMD-group width implementation-defined, whereas Direct3D wave sizes range from 4 to 128.

\section{\crosstl}
\label{sec:system}

\crosstl has four layers (\cref{fig:architecture} in \cref{app:example}). \emph{Importers} print native programs as \crossgl text, one \emph{front end} parses \crossgl into a syntax tree, eleven \emph{emitters} walk the tree, and a \emph{project layer} turns a repository into a planned translation whose artifacts carry contracts that loaders check. Ten importers and eleven emitters cover 110 directed language pairs. \crosstl is written in Python and is open source; statements about it refer to the implementation archived in our artifact.

\subsection{The Hub}
\label{sec:system:hub}

\crossgl is a C-family language with extensions from Rust and WGSL: \code{fn} declarations, \code{let} bindings, \code{match} expressions, generics with trait bounds, WGSL-style address spaces, and \emph{stage blocks} for sixteen canonical stages, from vertex to ray-tracing callable; \cref{fig:cgl} in \cref{app:example} shows a kernel in \crossgl. Two design decisions shape the system.
First, \emph{the hub is text}. Each importer builds its own native syntax tree and prints \crossgl, which the front end parses again. The hub is therefore inspectable, but in practice it is a family of dialects: the front end accepts variant spellings of the same metadata and treats unknown type names as named types, so that importers can stay close to their sources, and every emitter must accept every dialect.
Second, \emph{the syntax tree is the intermediate representation}, and languages are registered by binding them to an importer or an emitter. There is no typed or lowered representation; the front end validates metadata, stage layouts and stage interfaces but does not check types, and each emitter infers types itself, recursively over tables of declarations, struct members and swizzles, resolving overloads by argument types. The hub therefore cannot tell which language's rule an operation follows (\cref{sec:rulehub}).

\subsection{Importers and Emitters}
\label{sec:system:importers}
\label{sec:system:emitters}

An importer maps its language's resource model onto \crossgl resources, keeping the source's binding attributes (descriptor sets and bindings, registers and spaces, buffer indices, specialisation constants) for the emitters to map into their targets' index spaces, explicit bindings first.
MSL is a C++ dialect, and production Metal libraries use templates heavily: the 49 \code{.metal} files of Apple's MLX framework~\citep{hannun2023mlx} instantiate templates into 17,832 host-visible kernel entries. Since no target has the same templates, the Metal importer \emph{monomorphises} the program before parsing it, in a sequence of source-to-source steps (\cref{app:mono}); an argument that cannot be deduced is reported, not guessed.
The SPIR-V importer reads assembly and decompiles it: it rebuilds one structured construct per function from the merge instructions, turns two-way $\phi$-nodes into conditional expressions, and flattens other control flow, a source of silent defects (\cref{app:defects}).
Some rules are discharged in the hub itself. For \code{precise::sin}, \code{cos}, \code{exp}, \code{log} and seven other Metal functions, the importer writes an explicit \crossgl implementation into the program, adapted from Arm's optimized-routines, so that an emitter has to translate only integer arithmetic, bit reinterpretation and binary32 operations faithfully, without reassociating or contracting them.

The eleven emitters are hand-written tree walkers. The SPIR-V emitter prints assembly, producing structured control flow by construction and keeping locals in function-scope variables rather than $\phi$-nodes. The GLSL and HLSL emitters carry most of the semantic machinery: a C-family conversion model for integer promotion and narrowing, binary16 emulated in binary32 with re-rounding after every operation~\citep{figueroa1995doublerounding}, lowering of device pointers to (resource, offset) pairs, emulation of 32-lane subgroups through shared memory, which they refuse outside provably uniform control flow, and narrowing of indices that a constant, an interval analysis or a declared range bounds.
When an emitter meets a construct it cannot lower, it \emph{refuses} with a typed diagnostic, \emph{falls back} to a default value, type or padding while leaving a comment, or \emph{passes through} an identifier it does not know and leaves the failure to the target compiler. A fallback is a silent loss that only a comment marks, and formatting can strip comments, so our evaluation reads the raw emitter output.

\subsection{Projects, Contracts and Loaders}
\label{sec:system:projects}

A project file supplies the facts that single-file translation cannot know, such as work-group sizes, subgroup widths, specialisation constants and ranges of scalar inputs, and the run records a hash of every source and artifact.
Obligations (\cref{def:obligation}) that concern the host travel in the artifact as JSON contract lines: the exact grid that a kernel written for Metal's partial thread-groups assumes, \emph{checked} ranges of scalar inputs, folded specialisation constants, and the encoding of a buffer whose element type the target lacks, such as binary16 values stored as 16-bit integers for Direct3D without native 16-bit types. That record is an $R$ in the sense of \cref{def:faithful}; for other buffers the host reads the layout from the artifact by reflection, and the encoding is the identity. Caller-trusted index ranges are assumptions, and the run report lists them instead.
Before each dispatch, the loader checks the artifact's hash and contracts against its package record, the dispatch geometry, the sizes and aliasing of the bound views, and the scalar preconditions on the bytes actually uploaded; a violation rejects the dispatch rather than converting or clamping a value. This implements the check treatment of \cref{sec:fidelity}, which \cref{sec:eval:check} tests.

\section{Faithful Translation}
\label{sec:fidelity}

Our definitions follow the behavioural refinement of verified compilers~\citep{leroy2009compcert}, with data refinement for state encodings~\citep{hoare1972datarefinement,he1986datarefinement}, and add two things that GPU programs need: the compile configuration is part of a program's meaning, and a loader can refuse a dispatch.

\paragraph{Programs, configurations and states.}
Let $p$ be a program in a source language $S$, built under a configuration $\kappa$ that comprises compiler options, execution modes and device features.
An input state $\sigma\in\Sigma_S$ comprises the buffer contents, the launch geometry and any scalar inputs that the host passes.
We write $\sem{p}_{S,\kappa}(\sigma)$ for the set of outcomes that the specification of $S$ allows, each outcome being final buffer contents or non-termination, to which a deadlock also counts.
The set is never empty: it is a singleton for a fully specified computation, larger for unspecified or implementation-defined results, and contains every outcome when $p$ has undefined behaviour on $\sigma$.
An \emph{unspecified value} is any value of its type, and the program stays defined; a \emph{poison} value, as in SPIR-V and LLVM, becomes undefined behaviour only when an operation needs a valid value, such as a branch condition, an address or a divisor~\citep{lee2017tamingub,llvm2026langref,khronos_spirv16}.

\paragraph{Artifacts, encodings and loaders.}
A translator produces an artifact $a$ in a target language $T$, the configuration $\kappa_T$ that the artifact requires, and an encoding $R$ that relates source input states to target input states and target outcomes to source outcomes; under $R$, for example, a launch may be rounded up to whole work-groups. $R$ must relate every input state that satisfies the assumption $\Phi$ below to some target state; on outcomes it must be the converse of a function, the host's decoding, which the buffers' declared types fix independently of $p$ and which maps non-termination to itself; and, like $\kappa_T$, it must be recorded with the artifact in a form that the host or the loader applies. Otherwise an encoding that relates every outcome to every other would make any artifact faithful.
A layout that the module declares in the target's own terms, such as a SPIR-V \code{ArrayStride} or a std430 block, fixes the target's encoding but does not record $R$, because it does not say how the source's data map onto it. An artifact that records no $R$ is read with the identity on host-visible data.
We write $\sem{a}_{T,\kappa_T,\Psi}$ for the outcomes that $T$'s specification allows under $\kappa_T$ to the implementations that satisfy an assumption $\Psi$, such as support for NaN. The artifact's loader $\loader$ may check a predicate on the target state before it dispatches the artifact: $\sem{\loader\circ a}_{T,\kappa_T,\Psi}(\sigma')$ is $\{\bot_{\mathit{rej}}\}$ if the check fails on $\sigma'$, and $\sem{a}_{T,\kappa_T,\Psi}(\sigma')$ otherwise. An artifact that $T$'s compiler rejects counts as a refusal.

\begin{definition}[Faithful translation]
\label[definition]{def:faithful}
The artifact $(a,\kappa_T,R)$ is \emph{faithful} to $(p,\kappa)$ under an assumption $\Phi$ on input states and an assumption $\Psi$ on implementations if, for every $\sigma\models\Phi$ and every $\sigma'$ with $R(\sigma,\sigma')$, every outcome of $\sem{\loader\circ a}_{T,\kappa_T,\Psi}(\sigma')$ is $\bot_{\mathit{rej}}$, or $p$ has undefined behaviour on $\sigma$, or the outcome decodes under $R$ to an outcome in $\sem{p}_{S,\kappa}(\sigma)$.
\end{definition}

Faithfulness is refinement up to rejection and encoding; execution can only refute it, because the definition quantifies over every implementation of $T$ that conforms under $\kappa_T$ and satisfies $\Psi$. A checked predicate is enforced by $\loader$, so the states that violate it end in $\bot_{\mathit{rej}}$, whereas an assumed predicate enters $\Phi$ or $\Psi$ and removes those states or implementations from the quantifier.

\begin{definition}[Obligation]
\label[definition]{def:obligation}
Let $\mu$ be a translator's mapping from source constructs to target constructs, under the configuration $\kappa_T$ that the artifact records. A construct $c$, an operator occurrence whose operands its context supplies, carries an \emph{obligation} under $\mu$ if, for some program context $C$, some translation of $C[c]$ that maps $c$ to $\mu(c)$ and is otherwise faithful (\cref{app:proofs}) is unfaithful to $C[c]$ under $(\mathit{true},\mathit{true})$.
\end{definition}

For constructs other than operations, such as layouts and contraction, we use the definition informally. Obligations depend on the mapping: for MSL's \code{fmod}, emitting SPIR-V's \code{OpFRem} instead of \code{OpFMod} removes the obligation for non-zero remainders but leaves one for zero remainders, whose sign SPIR-V does not specify. The catalogue of \cref{sec:background} lists the rule pairs for which mapping an operator to its namesake is unfaithful.
A translator can treat an obligation in five ways. It can \emph{discharge} it, by emitting a construct or a configuration under which the target's rule refines the source's; \emph{check} it, by having the loader reject the states for which it cannot; \emph{assume} it, by restricting $\Phi$ or $\Psi$ and recording the restriction; \emph{refuse}, by stopping with a diagnostic; or \emph{drop} it, by doing none of these.

\begin{definition}[Drop-free translation]
\label[definition]{def:dropfree}
A translator is \emph{drop-free} for a class of obligations if it never drops an obligation of that class.
It is \emph{fail-closed} for the class if, in addition, it never relies on an assumption in $\Phi$ or $\Psi$: every obligation of the class is discharged, checked or refused.
\end{definition}

Both properties are relative to a class of obligations: every translator is drop-free for the empty class, and our verdicts are only as complete as the catalogue that defines the classes.

\paragraph{Why a passing test passed.}
A run whose result equals the oracle has one of four \emph{bases}. The match is \emph{defined by the target} when the target's rule refines the source's under $\kappa_T$; it holds \emph{by construction} when the artifact discharges the obligation, possibly under a configuration or an assumption that it records; it holds \emph{by configuration} when the run matched only under a configuration that the artifact does not record, such as Metal's safe math mode; and it holds \emph{by implementation} when neither the target's rule nor the artifact guarantees it.
Only the first two bases support faithfulness; they are the per-occurrence condition of \cref{thm:sufficiency}, read off the artifact. A match on either of the other two, for an obligation that the artifact neither discharges nor records, is a \emph{latent drop}: the obligation was dropped, but the runtime that executed the artifact did not expose it.

\section{Carrying Source-Language Rules Through the Hub}
\label{sec:rulehub}

\crossgl records the syntax of an operation, not which language's rule it follows: an \code{x + 1} printed by the GLSL importer wraps on overflow, the same text printed by the Metal importer does not, and an emitter that sees only the hub cannot tell the two apart. This section models the rules, states when carrying them through the hub preserves them, and describes and checks \crosstlr, which implements rule carrying in \crosstl.

\subsection{Operation Rules}
\label{sec:rulehub:rules}

Fix a set of operators, such as signed 32-bit addition, left shift, unsigned division, remainder, narrowing conversion, float-to-integer conversion and \code{min}, each with an arity and operand types, and let $V$ be the set of values.
A \emph{rule} $\rul$ for an $n$-ary operator maps each argument tuple $v\in V^n$ either to a non-empty set $\rul(v)\subseteq V$ of permitted results or to $\undefrule$, which stands for undefined behaviour; $\rtop$ is the rule that is undefined everywhere.
A language $L$ gives each operator $\mathit{op}$ a rule $\rul_{L,\kappa}(\mathit{op})$ under each configuration $\kappa$; we write $\rul_L(\mathit{op})$ for its reference configuration (\cref{tab:hazard-matrix}). For signed addition, $\rul_{\mathrm{GLSL}}(+)(x,y)=\{\mathrm{wrap}_{32}(x+y)\}$, whereas $\rul_{\mathrm{MSL}}(+)(x,y)=\undefrule$ whenever $x+y$ overflows.
We read the catalogue as follows: an unspecified result is the set of all values of its type; a result that may be poison or LLVM \code{undef} is that set for a source and $\undefrule$ for a target; \textsf{impl} is unspecified and \textsf{n/s} is $\undefrule$. This over-approximates what a target may do and under-approximates what a source permits, except that \textsf{impl} and \textsf{n/s} are read permissively in a source; the guarantees below are relative to this reading.

\begin{definition}[Refinement of rules]
\label[definition]{def:refine}
A rule $\rul'$ \emph{refines} a rule $\rul$, written $\rul'\refines\rul$, if for every argument tuple $v$, either $\rul(v)=\undefrule$, or $\rul'(v)\neq\undefrule$ and $\rul'(v)\subseteq\rul(v)$.
\end{definition}

A refinement may resolve nondeterminism and may define what was undefined, but it may not add a result or make undefined what was defined. The relation $\refines$ is a partial order with greatest element $\rtop$.
Under the identity encoding, mapping $\mathit{op}$ from $S$ to its namesake in $T$ creates an obligation (\cref{def:obligation}) only if $\rul_{T,\kappa_T}(\mathit{op})\nrefines\rul_{S,\kappa}(\mathit{op})$, and exactly then when $S$'s rule is exact at a conflicting tuple that a kernel can load (\cref{prop:obligation} in \cref{app:proofs}). GLSL's wrapping addition is thus an obligation for MSL, whereas MSL's undefined overflow is refined by every rule and is an obligation for no target.

\subsection{Rule-Carrying Translation}
\label{sec:rulehub:carrying}

A syntax-only hub maps an operator from every source to one hub operator, which each emitter maps to its target's namesake by a mapping $\mu_T$, so whether a rule survives depends on the pair $(S,T)$.
A \emph{rule-carrying} hub instead has an operator $\mathit{op}_{\rul}$ for each operator $\mathit{op}$ and each rule $\rul$ that it carries.
An \emph{annotation} $\alpha$ of a program $p$ from $S$ maps some operation occurrences $o$ of $p$ to carried rules $\alpha(o)$; lowering rewrites each such $o$ into $\mathit{op}_{\alpha(o)}$, and the emitter maps every other occurrence by $\mu_T$. The exact annotation sets $\alpha(o)$ to $\rul_{S,\kappa}(o)$, the rule that $S$ gives the operator of $o$.
For each target $T$, a \emph{rule table} $\Lambda_T$ maps each carried operator $\mathit{op}_{\rul}$ to an entry of one of three kinds:
\begin{itemize}[leftmargin=1.2em, itemsep=1pt]
\item \emph{native} (N): $T$'s namesake;
\item \emph{polyfill} (P): a fragment $f_{\rul}$, written once in the hub, that uses only the operators of a \emph{trusted base} $B$;
\item \emph{refusal} (Ref): the translator stops with a diagnostic.
\end{itemize}
An N or P entry may also be valid only under a condition that the artifact then records: a configuration that it requests, such as a math mode (+C), or an assumption that it cannot enforce (+A). The recorded configuration $\kappa_T$ is $T$'s default together with the +C conditions of the entries used, and obligations and rules are taken under it; a +A condition, such as support for NaN, joins $\Psi$. An entry is \emph{sound} if, under $\kappa_T$ and $\Psi$, the rule of what $T$ executes for it refines $\rul$: $T$'s namesake for an N entry, and for a P entry $f_{\rul}$ with each operator of $B$ replaced by the construct to which $T$'s emitter maps it. By \cref{lem:monotone}, a P entry is sound if $f_{\rul}$ computes a rule that refines $\rul$ when the operators of $B$ follow their hub rules, and the emitter maps each operator of $B$ that $f_{\rul}$ uses to a construct whose rule refines its hub rule.

\begin{theorem}[Carrying rules]
\label{thm:sufficiency}
Let a translator map a program $p$ of $S$, built under $\kappa$, to an artifact $(a,\kappa_T,R)$ for $T$ by annotating $p$ with $\alpha$ and lowering through $\Lambda_T$, and let $\rul_T(o)$ be the rule, under $\kappa_T$ and $\Psi$, of the construct to which it maps the occurrence $o$: its namesake under $\mu_T$, or what $T$ executes for the entry of $\Lambda_T$ that $o$'s annotation selects. Suppose that
\begin{description}[leftmargin=2.4em, labelwidth=2em, itemsep=1pt, font=\normalfont]
\item[(P0)] the artifact is faithful under $(\Phi,\Psi)$ to $p$ read with the rules $\rul_T$, that is, with $\sem{p}_{\rul_T}$ (the program model of \cref{app:proofs}) in place of $\sem{p}_{S,\kappa}$ in \cref{def:faithful};
\item[(P1)] the annotation is sound and complete: $\alpha(o)\refines\rul_{S,\kappa}(o)$ for every annotated occurrence $o$, and $\rul_T(o)\refines\rul_{S,\kappa}(o)$ for every unannotated $o$; and
\item[(P2)] every entry of $\Lambda_T$ that the translation uses is sound.
\end{description}
Then the translator refuses $p$, or the artifact is faithful to $(p,\kappa)$ under $(\Phi,\Psi)$.
\end{theorem}

(P0) is the frame: the translation is faithful except for which rule each operation follows, so control flow, memory, layouts, contraction and the execution geometry live there. The proof (\cref{app:proofs}) derives $\rul_T(o)\refines\rul_{S,\kappa}(o)$ for every occurrence from (P1) and (P2), and lifts it to programs by monotonicity (\cref{lem:monotone}).

\begin{corollary}
\label[corollary]{cor:execution}
Under (P0)--(P2), on an implementation that conforms to $T$ under $\kappa_T$ and satisfies $\Psi$, every execution from a state that encodes some $\sigma\models\Phi$ ends in $\bot_{\mathit{rej}}$ or in an outcome that decodes to one of $\sem{p}_{S,\kappa}(\sigma)$. Against a correct oracle, a divergence under these conditions therefore refutes (P0), (P1) or (P2); a divergence on a non-conforming implementation need not refute any of them.
\end{corollary}

If (P1) and (P2) hold for every program, with $\Psi$ the assumptions that the artifacts record, the translator is drop-free (\cref{def:dropfree}) for the classes that its table covers, and fail-closed if, in addition, $\Psi$ is empty.
Conversely (\cref{prop:converse}), given (P0) with equality and exact rules, a translator that neither refuses nor rejects is faithful on every program of $S$ only if each operator's construct refines the source rule on every tuple that a kernel can load and pass to it; (P1) and (P2) ask for this on every tuple, and split the $M\times N$ conditions into $M$ on annotations and $N$ on tables.

\paragraph{When must the source be recorded?}
\Cref{thm:sufficiency} requires each annotation to refine the operation's source rule, not to name its source.

\begin{proposition}[Recording the source]
\label[proposition]{prop:provenance}
Let $D(\mathit{op})$ be the set of rules $\rul_{S,\kappa}(\mathit{op})$, other than $\rtop$, that the sources give $\mathit{op}$ under the configurations of the programs translated.
(a) The rules in $D(\mathit{op})$ have a common refinement iff, for every tuple $v$, the sets $\rul(v)$ of the rules $\rul\in D(\mathit{op})$ that are defined at $v$ share an element; their weakest common refinement is then the pointwise meet $\sqcap D(\mathit{op})$, which is the least element of $D(\mathit{op})$ when there is one. Annotating every occurrence of $\mathit{op}$, from every source, with one rule $\rul^{\circ}$ makes (P1)'s first half hold iff $\rul^{\circ}\refines\sqcap D(\mathit{op})$.
(b) If there is no common refinement, witnessed by a tuple $v$ at which the conflicting sources' rules are exact, then for any hub whose importers map the witness kernels of the conflicting sources (load $v$, apply $\mathit{op}$, store the result) to one hub program, the artifact that any emitter produces for it is refused, rejects the witness state, or is unfaithful under $(\mathit{true},\mathit{true})$ to one of the kernels.
(c) A construct of $T$ that refines $\sqcap D(\mathit{op})$ also refines each $\rul_{S,\kappa}(\mathit{op})$, but not conversely; so a hub that carries only $\sqcap D(\mathit{op})$ must refuse wherever one that carries each source's rule must, and possibly elsewhere.
\end{proposition}

For every rule that \crosstlr carries, $D(\mathit{op})$ has a least element over \crosstl's sources: wrapping arithmetic, masked shifts, Direct3D's division by zero, the truncated remainder, truncating narrowing, saturating conversion with NaN to zero, the NaN-ignoring \code{min}, an exact \code{isnan}, and unsigned mixed comparison. A hub that gives every operation this rule, as naga and Tint do for wrap-around, would keep the carried rules too (a), but must refuse wherever a target cannot implement the strongest rule, even for sources that do not ask for it (c); \cref{sec:eval:rulehub} measures that price.
Elsewhere, defined rules conflict: with a WGSL importer, WGSL's $x/0=x$ would meet legacy Direct3D's \texttt{0xFFFFFFFF}, an instance of (b); Direct3D~11's flushing of binary32 subnormals against CUDA's preservation, and the 12- against the 16-byte \code{float3} stride, are analogues of (b) for the floating-point environment and for encodings. There a hub that does not refuse must record the source's rule, or the artifact an encoding; \crosstlr does neither. Records must be per operation when one hub program mixes provenance, as \crosstl's do: importers insert \crossgl implementations of Metal's precise functions, and the lowering inserts polyfills written against the hub's rules.

\subsection{Rule Carrying in \crosstl}
\label{sec:rulehub:impl}

\crosstlr implements rule-carrying translation in \crosstl as two passes between the front end and the emitters, enabled by a switch; with the switch off, the output is unchanged. \Cref{fig:cgl} in \cref{app:example} shows both passes on the kernel of \cref{fig:motivation}.

\paragraph{Annotation.}
The annotation pass knows the source language and infers operand types from a table of declarations, struct members and function results. It rewrites each operation whose source rule it carries into a call to a reserved intrinsic: signed $+$, $-$ and $\times$ from GLSL, SPIR-V and HLSL become \code{\_\_cgl\_wrap\_add} and its siblings, shifts from MSL, OpenCL~C and HLSL \code{\_\_cgl\_shl\_u} and its siblings, unsigned division from HLSL \code{\_\_cgl\_udiv\_d3d}, and a NaN-sensitive \code{fmin} \code{\_\_cgl\_fmin\_num}; mixed signed-unsigned comparisons and stores into 8- and 16-bit variables become explicit conversions. For HLSL it carries the legacy rules, which refine the current ones for every operator it annotates.
The annotation is incomplete in known ways, each a failure of (P1): it covers scalar operations only, so vector and 64-bit integer arithmetic keeps its namesake; Rust-GPU, Mojo and Slang programs pass through unannotated; an operation whose operand types the pass cannot infer is left unchanged (\cref{sec:eval:cost}); and neither GLSL's and SPIR-V's oversized shifts, which give some value where CUDA has undefined behaviour, nor HLSL's truncated \code{\%} under the current rules is carried.

\paragraph{Lowering.}
The rule table is data: a set of (rule, target) pairs lowered natively, refusals of the NaN-dependent rules for WGSL, and a polyfill for every other pair (\cref{tab:rule-table}); of its $9\times11=99$ entries, 15 are native, 80 are polyfill entries and 4 refusals, and 4 also record a configuration and 8 an assumption. The 80 entries use 17 polyfills, 35 statements of \crossgl in all, under reserved names; the lowering injects those that a program uses, and the ordinary emitters translate them, so each polyfill is only as faithful as each emitter's lowering of the trusted base: unsigned 32-bit arithmetic and \code{max}, bit reinterpretation, select, signed and unsigned comparisons, binary32 comparisons and \code{clamp}, and conversions of in-range values (\cref{app:proofs}). Wrapping addition, for example, becomes \code{asint(asuint(x) + asuint(y))}.
A +C record is self-enforcing, because the Metal compiler applies the pragma; a +A record is a comment line that no loader checks. The passes do not run for units that the project layer specialises or equips with contracts.

\begin{table}[t]
\caption{\crosstlr's rule table for seven of its eleven targets. Rows are the carried rules; \emph{Defined by} lists the sources whose rule a row carries. N: native operation; P: polyfill written once in \crossgl; +C: the artifact also records a configuration (\code{\#pragma METAL fp math\_mode(safe)}); +A: it records an assumption (that the GLSL implementation supports NaN); Ref: refusal. $\ddagger$: the mechanical check finds an entry of the cell unsound as recorded.}
\label{tab:rule-table}
\footnotesize
\begin{tabularx}{\linewidth}{@{}>{\raggedright\arraybackslash}p{3.3cm}>{\raggedright\arraybackslash}Xccccccc@{}}
\toprule
Rule & Defined by & Metal & SPIR-V & HLSL & GLSL & WGSL & CUDA & HIP \\
\midrule
wrapping $+,-,\times$ & GLSL, HLSL, SPIR-V & P & N & P & N & N & P & P \\
shift, count masked & HLSL, MSL, OpenCL & N & P & N & P & N & P$^{\ddagger}$ & P$^{\ddagger}$ \\
$x/0$, $x\,\%\,0$ $\to$ \texttt{0xFFFFFFFF} & HLSL & P & P & P & P & P & P & P \\
truncated \code{\%} & CUDA, HIP, OpenCL & N$^{\ddagger}$ & P & P & P & N & N$^{\ddagger}$ & N$^{\ddagger}$ \\
8/16-bit store truncates & MSL, CUDA, HIP, OpenCL & N & P & P & P & P & N & N \\
float$\to$int saturates, NaN$\to$0 & HLSL, OpenCL (\code{convert\_int\_sat}) & P+C & P$^{\ddagger}$ & P$^{\ddagger}$ & P+A & Ref & P & P \\
float$\to$int, NaN$\to$0 & MSL & N+C & P$^{\ddagger}$ & P & P+A & Ref & P & P \\
\code{min}/\code{max} return the number & HLSL, MSL, CUDA, HIP, OpenCL & P+C & P$^{\ddagger}$ & P$^{\ddagger}$ & P+A & Ref & P & P \\
\code{isnan} exact & MSL & P+C & P$^{\ddagger}$ & P & P+A & Ref & P & P \\
\midrule
\code{int}--\code{uint} compare & GLSL, HLSL, MSL, CUDA, HIP, OpenCL & \multicolumn{7}{c}{explicit conversion in the hub (no table entry)} \\
\bottomrule
\end{tabularx}

\end{table}

\paragraph{Floating-point rules.}
The NaN-aware rules test for NaN on the bit pattern rather than with \code{isnan}, which a target that may assume NaNs absent can fold; the test is exact wherever bit reinterpretation is. The Metal artifact requests safe math with the scoped pragma that MSL defines (+C); WGSL lets an implementation assume that NaNs are absent and offers no opt-out, so the table refuses these rules for WGSL; and for GLSL, whose implementations need not support NaN, the artifact records the assumption that they do (+A).
Two targets defeat these entries through their trusted base, so (P2) fails there. Unless a SPIR-V entry point declares \code{SignedZeroInfNanPreserve} (or, under Vulkan~1.4, states its floating-point environment), Vulkan lets an implementation assume that no floating-point operation sees a NaN or an infinity; the mode lifts this only for bit-preserving operations and a listed set, such as \code{OpIsNan} and the comparisons, and \code{OpBitcast}, through which the polyfill reads the bits, is in neither~\citep{khronos_vulkan14}. A sound entry would test with \code{OpIsNan} under the mode, or clear NotNaN with per-instruction flags under Vulkan~1.4; both need the device property \code{shaderSignedZeroInfNanPreserveFloat32}, and without it the rule must be refused. DXC, in turn, marks every non-precise floating-point comparison with LLVM's fast-math flags~\citep{ms_dxil}, so the HLSL polyfills that compare a possibly infinite value are sound only with \code{precise} on the compared values or under \code{/Gis}, and the artifacts have neither.
The implementation also departs from its registered design (\cref{app:prereg}, A11).

\paragraph{Mechanical check.}
We checked every N and P entry for the eight targets whose rules the catalogue states (GLSL~ES read with GLSL's rules), 131 entries at the granularity of operators, as refinement queries in the style of Alive2~\citep{lopes2021alive2}: Z3~\citep{demoura2008z3} searches for arguments, and for choices of the target's unspecified results, on which the entry's construct, under the catalogue's rules for the target and its base, does not refine the carried rule (\cref{app:tablecheck}). In this model, which leaves subnormal flushing aside as the rule model does, 118 entries are sound and 13 are not, as recorded: the 5 SPIR-V NaN entries; the 3 HLSL entries above; Metal's native \code{\%}, which MSL leaves undefined for a negative operand~\citep{apple_msl41}; CUDA's and HIP's native \code{\%}, undefined for the zero divisor to which OpenCL~C gives some value; and CUDA's and HIP's signed right-shift polyfill, which needs an arithmetic shift that C++17 leaves implementation-defined, an unrecorded assumption ($\Psi$). Without its record, every +C and +A entry is unsound in the model. If comparisons may flush subnormals, as GLSL and MSL permit, \code{min} and \code{max} also fail on GLSL, GLSL~ES and Metal (19 of 131). The check does not cover which construct each emitter chooses for the base.

\section{Evaluation}
\label{sec:eval}

We ask five questions. \RQ{1}: On programs written by third parties, how often does \crosstl produce an output that the target accepts and that has visibly lost nothing, and why does it fail otherwise? \RQ{2}: How do established translators fare on the same programs? \RQ{3}: Which translation routes preserve the source's rules, on which runtimes, and on what basis? \RQ{4}: Does carrying rules preserve them as \cref{thm:sufficiency} predicts, and what does recording each operation's source save over a single strongest semantics? \RQ{5}: Do the loader's checks and the translator's refusals behave as \cref{def:faithful,def:dropfree} require?

\subsection{Methodology}
\label{sec:eval:setup}

\paragraph{Platform.}
All experiments ran on one Apple M2~Max. Of our seven runtime configurations, four compile through Apple's GPU compiler: Metal under safe math and under the API default of fast math (for the hazard suite also relaxed math), MoltenVK, which translates SPIR-V to MSL with SPIRV-Cross and compiles it with fast math, and wgpu, which translates WGSL to MSL. Three execute on Mesa's LLVM-based CPU drivers: lavapipe for Vulkan, llvmpipe for OpenGL, and lavapipe for DXIL, which dxil-spirv converts to SPIR-V as vkd3d-proton does, in place of Direct3D~12~\citep{hk2026dxilspirv,vkd3dproton2026}. Apple's Metal Shader Converter turned the same DXIL into Metal libraries~\citep{apple2026metalshaderconverter}. After the registered runs, we also executed every SPIR-V artifact on KosmicKrisp, a Vulkan driver for Apple GPUs certified conformant to Vulkan~1.4 (exploratory)~\citep{khronos2026conformant,lunarg2026kosmickrisp}. CUDA and HIP outputs were compiled by nvcc but not executed. The validators were each target's reference tools, with naga for WGSL; Rust and Mojo outputs had none (\cref{app:repro}).

\paragraph{Corpus.}
The corpus has 22 slices from 12 permissively licensed projects (\cref{app:corpora}). A \emph{unit} is a translation unit that declares an entry point, with its include closure resolved as the project's build resolves it. Of 2,713 units in ten languages, 2,404 compiled, 128 did not, and 181 Rust-GPU and Mojo units have no native check, so the experiments use 2,585 units in 719 \emph{program families} of units that share code or authorship. Two slices, the MLX kernels and the Mojo puzzles, were used during \crosstl's development (\emph{seen}).
Each unit was translated to every target with a 300-second limit, and a stage-support matrix fixed in advance marks each (unit, target) cell feasible, partial or infeasible. A feasible cell is \outcome{usable} if the target's validator accepts the output, the raw output carries no fallback marker, and it declares an entry point; \outcome{degraded} if it is accepted but carries a fallback or declares no entry point; \outcome{invalid} if the validator rejects it; and \outcome{refused} if \crosstl stopped with a diagnostic, crashed or timed out.

\paragraph{Hazard suite.}
The suite has one compute kernel for every catalogue class and every source language that \emph{defines} the class's behaviour under its reference configuration (safe math for MSL, as MLX, PyTorch and ggml build it); where a language leaves the behaviour undefined or unspecified, there is no case, and weaker no-UB obligations are not tested. This gives 66 cases in six source languages, of which 56 carry an obligation. Each kernel processes 64 inputs with critical values ($2^{31}-1$, zero divisors, NaN, $\pm\infty$, $\pm0$, subnormals, oversized shifts) at fixed positions, and each oracle is a Python model of the cited rule. Every case went through \crosstl and \crosstlr to seven targets, and through every third-party chain that accepts its source language: glslang or DXC to SPIR-V followed by SPIRV-Cross, naga or Tint; naga's GLSL front end; Slang; DXC followed by the Metal Shader Converter; and HIPIFY.

\paragraph{Registration, amendments and codings.}
Before \crosstlr or any of the 44 new cases ran, we fixed the suite's selection rule, routes, oracles, analysis plan and three hypotheses in a document whose SHA-256 we recorded; the other 22 cases come from an earlier suite whose outcomes we knew. \emph{H-R1}: every \crosstlr output that its validator accepts preserves the source rule on every executed runtime, on the target's rule or by construction except where the table records an assumption. \emph{H-R2}: unmodified \crosstl's outcomes depend on the source language for a fixed rule and target. \emph{H-R3}: \crosstlr changes no target's accepted share on the corpus by more than one point. Thirteen dated amendments record later changes; two correct the catalogue (A12, from an internal review: MSL's \code{\%} and SPIR-V's \code{NMin}; A13, from the table check: DXC's comparisons), and \cref{app:prereg,app:ledger} report each registered item as registered and as corrected.
We coded the basis of every preserved (case, route) artifact (\cref{sec:fidelity}) and, as registered, the mechanism of every failure: a \emph{fallback} to a default, a \emph{wrong construct}, a \emph{dropped obligation}, or a \emph{runtime-stack} divergence of an artifact that is correct by its target's specification. Each artifact was coded twice under written codebooks, once with the translator's identity visible and once blind, by separate instances of a large language model (Claude Opus~5.5), so agreement measures consistency between instances, not human agreement; the blind packets still revealed \crosstlr through its polyfill names in 75 of its 225 packets. The codebooks' rules settled disagreements.

\paragraph{Statistics.}
RQ1, RQ2 and the sign-flip test were not registered. Corpus results are proportions with cluster-bootstrap intervals over program families (2,000 resamples). Paired comparisons with \crosstlr use exact McNemar tests, and comparisons with other translators a sign-flip test over families, both with Holm's correction. As registered, the purposive hazard suite is reported as counts.

\subsection{RQ1: Reach and Validity on Third-Party Code}
\label{sec:eval:reach}

The 2,585 units give 27,671 attempted (unit, target) cells, of which 6,986 ask a target for a stage it cannot express; they serve RQ5. Of the 20,583 feasible cells, 17,567 are for the nine targets that have a validator (\cref{fig:reach} in \cref{app:corpora} breaks them down by source, target and slice).
Of these, 31.1\,\% of the translations are usable, 3.9\,\% degraded and 35.0\,\% rejected by the target's validator, and in 30.0\,\% \crosstl stops.
The share depends on the source language and on the corpus's composition. From the four graphics languages, 35 to 57\,\% of the translations are usable; from Metal, CUDA, HIP, OpenCL~C and Mojo, below 5\,\%, mostly because their imports fail or hub type names and source built-ins pass through unchanged (\cref{app:corpora}). Two collections of Vulkan samples, one with its Rust-GPU ports, supply 71\,\% of the feasible validated cells and 98\,\% of the usable translations: averaged over projects instead of cells, the usable share is 11.6\,\% (19.8\,\% over source languages), and leaving out one project at a time gives 21.8 to 33.7\,\%. Removing the two seen slices changes it from 31.1 to 32.0\,\%.
Even a round trip through the hub loses programs: translated back into its own language, a program yields usable output in 75\,\% of the feasible GLSL cells but only 8\,\% of the Metal ones. Some losses are silent: 603 of the 1,489 SPIR-V outputs that \code{spirv-val} accepts (40\,\%) contain a fallback marker.

Across all 20,583 feasible cells, \crosstl stopped on 6,113, 171 of them with a crash or time-out and the rest with a diagnostic. Two blind model codings of a stratified sample of 706 diagnostics agreed on 95\,\% of it ($\kappa=0.92$); reweighted, they attribute 53 to 57\,\% of the diagnostics to unsupported constructs, 24 to 26\,\% to inconsistencies in \crosstl's own state, 17\,\% to parse failures, and 1.4 to 3.6\,\% to a meaning-preservation reason, which a rule-based classification of the messages, with which they agree only moderately ($\kappa\le0.57$), puts at 21\,\%.

\smallskip\noindent\textit{Answer to RQ1.} About a third of \crosstl's translations of third-party programs are usable, a share that two collections of Vulkan samples dominate. Invalid outputs outnumber stops, most stops report unsupported constructs, and 40\,\% of the SPIR-V outputs that validate have silently lost a construct.

\subsection{RQ2: Established Translators on the Same Programs}
\label{sec:eval:baselines}

We ran every third-party chain on every graphics unit of each source language it accepts, under the same validators: SPIRV-Cross, naga and Tint on SPIR-V that glslang or DXC produced, or that the corpus supplies; naga's GLSL front end; and Slang. A chain succeeds when the validator accepts its output and the output declares an entry point that compiles with its stage's profile; \crosstl's output must also carry no fallback marker, a requirement that changes no comparison's direction.

The established chains produce valid output significantly more often than \crosstl in 31 of the 39 comparisons (\cref{tab:baselines} in \cref{app:corpora}). SPIRV-Cross succeeds on 81 to 89\,\% of the units for every pair it offers, against 16 to 69\,\% for \crosstl; its largest lead, from SPIR-V to HLSL, is 66 percentage points (interval 59--73). \crosstl is significantly ahead in six comparisons: against naga, through either front end, from GLSL to Metal and HLSL and from SPIR-V to WGSL, and against Slang from Slang to HLSL (66 against 12\,\%), where Slang's entry points lack the semantics that DXC requires of a stage's interface; compiled as libraries, 88\,\% of Slang's HLSL outputs are accepted (\cref{app:corpora}).

\smallskip\noindent\textit{Answer to RQ2.} On the graphics pairs they support, the established chains produce valid output more often than \crosstl, in 31 of 39 comparisons. Validity is not fidelity, however: RQ3 shows that the chains with the highest validity, SPIRV-Cross and Slang, also lose source rules, though on the cases they accept no more often than \crosstl.

\subsection{RQ3: Which Routes Preserve the Source's Rules?}
\label{sec:eval:hazards}

\Cref{tab:hazards-by-route} counts the outcomes of every route on the 56 cases with an obligation, out of 3,896 outcome records across all arms; \cref{app:hazard-matrix} gives every cell and the findings by class.

\paragraph{Overflow.}
Every route whose intermediate representation carries the wrap rule keeps it on every registered runtime except MoltenVK, where every SPIR-V route fails (\cref{fig:motivation}): naga, through either front end, Tint, and \crosstlr. Routes that copy the operator, SPIRV-Cross and \crosstl, lose the rule on Metal, because the Metal compiler folds the comparison.
In the exploratory run on KosmicKrisp, every SPIR-V route of the three addition cases with an obligation diverges, glslang's output and the suite's reference SPIR-V included (\cref{app:probes}): the driver's MSL printer emits the addition as a signed \code{int} addition, which Apple's compiler folds, and multiplication by two survives only because Mesa's NIR first rewrites it into a shift.
Compiling the Metal overflow artifacts with Apple's \code{-fwrapv} (exploratory) kept the rule in all 10 overflow artifacts that diverged on Metal and changed none of the 18 that matched; an MSL file cannot request the flag, and \crosstl's artifacts do not record it.
DXC emits \code{add nsw} for HLSL's signed arithmetic and folds \code{x+1>x} even at \code{-Od}, so a translation into HLSL that copies a GLSL or SPIR-V addition, such as \crosstl's or SPIRV-Cross's, can compute a different result without a diagnostic, whereas naga's and Tint's polyfills survive.

\begin{table}[t]
\caption{Outcomes of every route on the 56 cases with an obligation, per (case, route, runtime), with Metal under the source configuration. \emph{Cases}: the cases a route accepts; the third-party routes take only GLSL, HLSL or SPIR-V (HIPIFY: CUDA). Pres., Div., Hang: executions that preserved the rule, diverged, or did not terminate. Inv.: rejected by the target compiler. Ref.: refused by the translator. Comp.: compiled by nvcc, not executed. N/r: not run (missing OpenGL subgroup extensions).}
\label{tab:hazards-by-route}
\footnotesize
\begin{tabular}{@{}lrrrrrrrr@{}}
\toprule
Route & Cases & Pres. & Div. & Hang & Inv. & Ref. & Comp. & N/r \\
\midrule
\crosstl & 56 & 184 & 87 & 1 & 105 & 9 & 57 & 4 \\
CrossTL-R & 56 & 220 & 55 & 1 & 94 & 16 & 57 & 4 \\
SPIRV-Cross & 15 & 26 & 12 & 0 & 0 & 0 & 0 & 1 \\
naga (GLSL) & 6 & 15 & 0 & 0 & 0 & 3 & 0 & 0 \\
naga (SPIR-V) & 15 & 31 & 8 & 0 & 0 & 0 & 0 & 0 \\
Tint & 15 & 30 & 8 & 0 & 1 & 0 & 0 & 0 \\
Slang & 6 & 12 & 4 & 0 & 1 & 0 & 0 & 1 \\
DXC (SPIR-V) & 6 & 10 & 2 & 0 & 0 & 0 & 0 & 0 \\
Metal Shader Converter & 6 & 6 & 0 & 0 & 0 & 0 & 0 & 0 \\
glslang (SPIR-V) & 6 & 9 & 3 & 0 & 0 & 0 & 0 & 0 \\
HIPIFY (CUDA) & 15 & 0 & 0 & 0 & 0 & 0 & 15 & 0 \\
\bottomrule
\end{tabular}

\end{table}

\paragraph{Other rules, configuration and the runtime stack.}
Every translator whose outputs ran loses some rule except two: glslang, whose only divergences arise on the Vulkan drivers for Metal, and the Metal Shader Converter. naga and Tint drop SPIR-V's \code{NoContraction} on the way to MSL and WGSL, and Slang drops HLSL's \code{precise}. DXC's SPIR-V back end gives HLSL structured buffers of \code{float3} a 16-byte stride unless \code{-fvk-use-dx-layout} is given; the module declares the stride but records no encoding (\cref{def:faithful}), so a host that keeps Direct3D's 12-byte elements reads the wrong data. \crosstl also loses contraction control, signed zeros and the sign of C's \code{fmod} on several targets, and subgroup reductions on SPIR-V and Direct3D (\cref{app:hazard-matrix}).
On Metal, 12 of the 191 (case, route) cells with an obligation change outcome between the three configurations, every change separating safe mode from the other two; \crosstl's Metal output for CUDA's \code{sinf}, for example, meets the source's error bound only in safe mode, which it does not request. Between lavapipe and MoltenVK, 33 of the 130 cells with an obligation differ (15 in the wrap rule, 10 in contraction), so that 19 of 95 distinct SPIR-V binaries behave differently on the two implementations.
On the suite's NaN cases, no runtime exploited the latitude for NaN: every one matched on Metal under fast math, \crosstl's unannotated outputs included, and \crosstl's SPIR-V failures on NaN are translator defects. Where fast math does change a result, as for \code{x != x}, \code{x * 0.0f} or \code{x - x} in a probe, the scoped pragma that \crosstlr's +C entries emit restored the IEEE result under both the API default and explicit fast math (\cref{app:probes}).

\paragraph{Why the matches matched, and why the failures failed.}
We coded the basis of each of the 477 preserved (case, route) artifacts; the informed and blind codings agreed on 472 ($\kappa=0.98$, 95\,\% interval 0.97--1.00). As registered, a match that changes between Metal's configurations, or between lavapipe and MoltenVK, holds by configuration or by implementation (A8), and a construct counts only where the target's rules guarantee it, which makes \crosstlr's NaN tests on SPIR-V (A9) and comparisons of infinite values on HLSL (A13) hold by implementation.
\Cref{tab:bases} gives the result. For \crosstl, 62\,\% of the matches rest on a basis that supports faithfulness, the target's own rule or an explicit construct; the other 38\,\% hold only under an unrecorded configuration (11 matches) or on the particular implementation (52). The share is 73\,\% for \crosstlr, 80 to 91\,\% for SPIRV-Cross, naga, Tint and DXC's SPIR-V, 67\,\% for Slang and 50\,\% for glslang. The third-party routes accept only the 15 GLSL, HLSL and SPIR-V cases; on these, \crosstl's share is 67\,\% (32 of 48 matches) and \crosstlr's 80\,\% (47 of 59). Across all routes, 115 matches rest on the implementation, most often contraction that the runtime happened not to perform (35) and NaN handling in \code{min} and conversions (29).
The mechanism coding of the 150 artifacts that diverged or hung agreed on 149 ($\kappa=0.99$). On the rules that \crosstlr carries, \crosstl's 25 failures comprise 9 wrong constructs, 8 dropped obligations, 3 fallbacks and 5 runtime-stack divergences, whereas all 5 of \crosstlr's are runtime-stack divergences. The third-party routes drop the wrap and contraction obligations (24 artifacts) and inherit or produce the 16-byte \code{float3} stride (9).

\begin{table}[t]
\caption{Preserved artifacts of the cases with an obligation by the basis of the match, with the share on the two bases that support faithfulness, and non-preserved artifacts by mechanism. Each artifact was coded twice, informed and blind, and counts once however many runtimes it ran on; artifacts that matched on one runtime and failed on another appear in both halves.}
\label{tab:bases}\label{tab:mechanisms}
\footnotesize
\setlength{\tabcolsep}{3.5pt}
\begin{tabular}{@{}lrrrrrrrrr@{}}
\toprule
 & \multicolumn{5}{c}{Preserved: basis of the match} & \multicolumn{4}{c}{Not preserved: mechanism} \\
\cmidrule(lr){2-6}\cmidrule(l){7-10}
Route & Target & Constr. & Config. & Impl. & Supported & Fallb. & Wrong & Dropped & Stack \\
\midrule
\crosstl & 54 & 47 & 11 & 52 & 62\,\% & 3 & 31 & 27 & 6 \\
CrossTL-R & 39 & 98 & 5 & 46 & 73\,\% & 0 & 22 & 19 & 6 \\
SPIRV-Cross & 7 & 15 & 0 & 4 & 85\,\% & 0 & 2 & 10 & 0 \\
naga (GLSL) & 6 & 15 & 1 & 1 & 91\,\% & 0 & 0 & 0 & 0 \\
naga (SPIR-V) & 6 & 15 & 0 & 2 & 91\,\% & 0 & 2 & 6 & 0 \\
Tint & 7 & 19 & 1 & 3 & 87\,\% & 0 & 2 & 6 & 0 \\
Slang & 3 & 5 & 1 & 3 & 8 of 12 & 0 & 2 & 2 & 0 \\
DXC (SPIR-V) & 1 & 3 & 0 & 1 & 4 of 5 & 0 & 1 & 0 & 0 \\
glslang (SPIR-V) & 0 & 3 & 0 & 3 & 3 of 6 & 0 & 0 & 0 & 3 \\
\midrule
All & 123 & 220 & 19 & 115 & 72\,\% & 3 & 62 & 70 & 15 \\
\bottomrule
\end{tabular}

\end{table}

\smallskip\noindent\textit{Answer to RQ3.} Of the translators whose outputs ran, all but glslang and the Metal Shader Converter lose some source rule, and outcomes depend on the compile configuration and on the Vulkan implementation. A passing test is weak evidence of faithfulness: on the 15 cases that the third-party routes take, 33\,\% of \crosstl's matches hold only under an unrecorded configuration or on the particular implementation, against 9 to 50\,\% for those routes.

\subsection{RQ4: Carrying the Rules Through the Hub}
\label{sec:eval:rulehub}

We compared four arms on the full suite: \crosstl; a registered control that fixes in place the SPIR-V emitter defects behind five failures of an earlier suite (\code{OpFRem}, unsigned promotion at equal rank, \code{NMin}/\code{NMax} and \code{OpIsNan}); \crosstlr, which with its switch off reproduced all 659 of \crosstl's records; and both together. \Cref{tab:rulehub} counts divergences on the 24 cases whose rule \crosstlr carries and that carry an obligation.

\begin{table}[t]
\caption{The rule-carrying experiment on the 24 carried cases with an obligation (Metal under safe math; both Vulkan implementations). Divergences, refusals and invalid outputs count (case, runtime) records; source-dependent cells are (rule, target) pairs with executed outcomes for at least two source languages that define the rule, which differ. \Cref{app:readingA} gives the legacy HLSL rules.}
\label{tab:rulehub}
\small
\begin{tabular}{@{}lrrrr@{}}
\toprule
Arm & Divergences & Source-dependent & Refusals & Invalid \\
\midrule
\crosstl & 37 & 2 of 26 & 0 & 53 \\
SPIR-V emitter fixes & 23 & 0 of 26 & 0 & 53 \\
CrossTL-R & 5 & 0 of 26 & 7 & 42 \\
CrossTL-R and emitter fixes & 7 & 0 of 26 & 7 & 42 \\
\bottomrule
\end{tabular}

\end{table}

\paragraph{Divergences.}
Of the 37 divergences, 29 are on SPIR-V, 24 of them failures of \crosstl's own SPIR-V output and 5 MoltenVK wrap failures, and 8 are wrap losses on Metal and Direct3D~12. Fixing the SPIR-V emitter locally removes 14 of the 37, all on SPIR-V. Carrying the rules removes 32, the 8 cross-language losses included; on Direct3D~12 the wrap polyfill survives DXC's folding. Source-agnostic emitter changes, as in naga and Tint, could remove the control's other 18, so the contrast measures the scope of the fixes, not where an obligation must be discharged. Where the target cannot express a rule, \crosstlr refuses: it refuses the seven NaN-dependent cases on WGSL, for which \crosstl had produced five invalid outputs and two that matched only because wgpu's Metal back end happens to honour NaN.
Combining the two arms adds two divergences: the SPIR-V emitter types the remainder polyfill's literal \code{0} as unsigned, and the control's unsigned-promotion fix then lowers its signed test \code{x < 0} as \code{OpULessThan}, which is always false. This is a failure of (P2) in a base operator, which no check of the table alone can reveal.

\paragraph{The premises of \cref{thm:sufficiency}.}
All five remaining divergences are wrap-around on MoltenVK: the emitted \code{OpIAdd} is correct by the SPIR-V specification, but MoltenVK's translation to MSL turns it into signed arithmetic that Apple's compiler may assume does not overflow~\citep{moltenvk2019issue549}. MoltenVK implements Vulkan's portability subset and is not a conformant implementation~\citep{khronos2026moltenvkreadme,khronos2026conformant}; but KosmicKrisp, which is certified conformant, fails the three addition cases by the same mechanism, so conformance testing does not exercise this fold, and on this point KosmicKrisp does not conform to the specification (\cref{app:probes}). In the theorem's terms, both drivers implement a weaker target whose signed addition does not wrap. An exploratory table keyed by the runtime stack, which polyfills SPIR-V bound for MoltenVK, preserved all 10 overflow runs on MoltenVK and lavapipe; on KosmicKrisp the same polyfill fails the three additions, because the driver prints its unsigned addition as a signed one. The divergences falsify the registered H-R1, which quantifies over the runtimes we executed; \crosstlr's first run, before amendment A1, also diverged on \code{m\_uchar\_wrap}.
H-R1's basis clause fails for 18 of \crosstlr's 102 matches on the carried rules. Thirteen trace to (P2), at entries that the mechanical check finds unsound: the 7 NaN tests on SPIR-V and 5 comparisons of infinite values on HLSL hold only by implementation, and so does CUDA's remainder through Metal's native \code{\%}. The other 5 are wrap matches on lavapipe that rest on the target's rule, which the registered basis rule counts as implementation-dependent because MoltenVK did not keep them.

\paragraph{Held-out kernels.}
Because \crosstlr was debugged on the suite, we froze it and tested it on 102 random kernels in five source languages, generated from a fixed seed, each a random expression over carried operations; under the corrected catalogue, 84 carry an obligation. \crosstl diverged in 120 of 444 executions, all on SPIR-V, in 10 of the 14 (rule, source) cells; \crosstlr diverged in none of 462. The kernels test less severely than the suite, since none compares an overflowing expression, and the registered prediction that \crosstl would diverge on every target where the suite shows a loss failed.

\paragraph{Source dependence and validity (H-R2, H-R3).}
For a fixed carried rule and target, \crosstl's outcome differs between source languages in only 2 of 26 cells, both caused by one SPIR-V emitter defect, and in none for \crosstlr: H-R2 holds, narrowly, and most of \crosstl's losses depend on the target alone.
On the corpus, H-R3 holds as registered: \crosstlr changed the accepted shares by between $-0.17$ and $0$ points against a one-point margin, no change is significant after Holm's correction, and no target's exact one-sided 95\,\% lower bound lies below $-0.5$ points (\cref{tab:hr3}). Pooled over targets, 13 cells lost acceptance and 1 gained it (exact $p=0.002$), through two defects of \crosstlr (9 cells) and intended refusals for WGSL (4; \cref{app:corpora}). The test is weak: the pass rewrote at least one operation in only 337 of the 1,840 units it processed.

\paragraph{What recording the source saves.}
We ran the annotation pass over 2,326 corpus units with each unit's own rules and with the strongest rule for every operation, and lowered both for eight targets (exploratory; \cref{tab:provenance}). The pass rewrites 5,390 operations in 355 units with the sources' own rules and 50,027 in 612 with the strongest, 99\,\% of the extra in CUDA, HIP, MSL and OpenCL~C programs, whose sources leave signed overflow undefined; and, as \cref{prop:provenance}(c) predicts, 92 units instead of 34 need Metal's safe mode, a NaN assumption on GLSL and a refusal on WGSL. We counted rewrites, not their cost, which for wrap-around would be lost \code{nsw}-based optimisation of index arithmetic.

\smallskip\noindent\textit{Answer to RQ4.} Carrying the rules removes 32 of the translator's 37 divergences on the carried rules; the five that remain are wrap-around failures that MoltenVK introduces, three of which KosmicKrisp, a conformant driver, reproduces in an exploratory run, and H-R1 is falsified. Each match on a weaker basis traces to an entry that the mechanical check finds unsound or to the registered basis rule. Validity changes by at most 0.17 points per target, and in an exploratory count a single strongest semantics would need about nine times as many rewrites.

\subsection{RQ5: Do Checks and Refusals Work?}
\label{sec:eval:check}

\paragraph{Checks.}
We tested the loader by fault injection, independently of \crosstl's own tests (\cref{app:faults}). In runtime packages for Metal, OpenGL and Direct3D around a kernel with a checked scalar precondition, package inspection rejected all 315 mutations of the contract lines, including 52 well-formed contracts with another meaning, and the host-side dispatch checks rejected every request outside the precondition and accepted every request inside it. The checks guard only what a contract records: none of the value-level obligations of RQ3 is a contract.

\paragraph{Refusals.}
On the suite, \crosstl refused 9 outputs and \crosstlr 16, adding the seven NaN-dependent cases on WGSL. Combining outcomes and codings per (class, target) cell (\cref{tab:dropfree}), and leaving out the 5 cells for partial work-groups, which single-file translation never exercises, \crosstl is drop-free in 30 of the remaining 100 cells, refuses in 5 and drops in 51, 22 of them latently; \crosstlr is drop-free in 42, refuses in 8 and drops in 39, 17 latently. Neither is fail-closed: both rely on assumptions, such as caller-trusted index ranges, and both drop obligations, \crosstl in 16 of the 21 classes and \crosstlr in 14.
On the corpus, \crosstl refused only 2,147 (31\,\%) of the 6,986 cells that ask a target for a stage it cannot express, unevenly across targets (\cref{app:corpora}).

\smallskip\noindent\textit{Answer to RQ5.} On one kernel with a checked scalar precondition, packaged for three targets, the checks rejected every mutated contract and every out-of-range request, but contracts record only host-side obligations. \crosstl's refusals are uneven, and carrying the rules makes 12 more (class, target) cells drop-free.

\subsection{Case Study: The MLX Kernel Library}
\label{sec:eval:mlx}

\crosstl's largest workload, Apple's MLX framework~\citep{hannun2023mlx}, was used during its development, so we only reproduced the project's integration tests: translated kernels of 42 of MLX's 46 families pass 1,493 tests on Metal, and of 38 families 1,495 on OpenGL, under recorded assumptions such as caller-trusted index ranges and MLX's 32-lane SIMD groups, a width that MSL leaves implementation-defined (\cref{app:mlx}).

\subsection{Costs}
\label{sec:eval:cost}

Importing and translating a graphics program takes milliseconds, a Metal program 0.7~s (\cref{app:costs}). On the GPU, carrying a rule cost nothing measurable on Metal for seven of eight micro-kernels (median time ratio 1.00); the NaN-aware \code{fmin} costs 1.39.
The annotation pass's coverage is the limit. It rewrote 0.16\,\% of the arithmetic, shift, division and comparison operations of the 1,840 corpus units in the seven languages it annotates, and it could not infer the operand types of 47, 46 and 52\,\% of these operations in GLSL, HLSL and SPIR-V programs, mostly vector and member expressions, against at most 7.4\,\% in the compute languages. Wherever such an operation carries an obligation, (P1) fails.

\section{Discussion}
\label{sec:discussion}

\paragraph{Implications for hub design.}
A record of the rule helps only if it is complete, and type inference over \crosstl's untyped tree leaves about half of the relevant operations in graphics programs unannotated, so a rule-carrying hub needs a typed intermediate representation. The configuration belongs to the artifact: \crosstl's Metal and HLSL outputs depend on a math mode or a \code{precise} that they do not request, and a sound SPIR-V NaN rule on a device property. A rule table is small enough to check mechanically, but the defect that broke the combined arm lies in an emitter's lowering of a base operator, which only a per-emitter test would catch. And a translator should lower for the stack that will execute the artifact: correct SPIR-V diverged on two Vulkan drivers for Metal, one certified conformant.

\paragraph{Scope of the claims.}
\Cref{thm:sufficiency} holds in a model of operation rules, relative to the reading of the catalogue in \cref{sec:rulehub:rules}; the implementation is not verified, divergence counts are lower bounds for the inputs and stacks tested, and the experiment varies one hub. Recording operation rules is not new (\cref{sec:related}); what is new is a source-to-source hub with many source languages, in which provenance is per source and some rules can be honoured only by configuring or refusing the target, and the account of when it must be recorded.

\paragraph{Threats to validity.}
All executions ran through two code generators, Apple's and Mesa's: Direct3D~12 was emulated, no NVIDIA, AMD, Intel or mobile driver ran anything, and CUDA and HIP outputs were only compiled; the NaN entries rest on the specifications, not on observed failures. Validity is measured by proxies, two collections of Vulkan samples dominate the corpus, no strongest-semantics arm ran on the suite, and \code{INT\_MIN/-1}, matrix and \code{bool} layouts and barrier scope are untested.
The catalogue, cases, oracles, rule table and codings were all prepared with one model under the authors' direction, so an error in the catalogue propagates to all of them; an internal review found four, two of which removed obligations (A12), and encoding the table for the mechanical check a fifth (A13). We corrected seven errors in our harness and one in the analysis of Slang's HLSL, none changing a conclusion (\cref{app:harness,app:corpora}); the held-out kernels, none of which compares an overflowing value, only partly offset the debugging of \crosstlr on the suite.

\section{Related Work}
\label{sec:related}

\paragraph{Intermediate languages and their semantics.}
Routing $M$ source languages and $N$ targets through one intermediate language is as old as UNCOL~\citep{strong1958uncol,conway1958uncol,steel1961uncol} and underlies the Amsterdam Compiler Kit, GCC, LLVM, MLIR and WebAssembly~\citep{tanenbaum1983ack,merrill2003gimple,lattner2004llvm,lattner2021mlir,haas2017wasm}. These IRs record the rules that source operators leave implicit: LLVM marks additions \code{nsw} or \code{nuw}, models their violation as poison and attaches fast-math flags~\citep{llvm2026langref,lee2017tamingub}, some of which (\code{nnan}, \code{ninf}, \code{nsz}) Alive-FP formalises~\citep{menendez2016alivefp}; MLIR's \code{arith} dialect has overflow attributes and distinct \code{minnumf} and \code{minimumf}~\citep{mlir2026arith}; SPIR-V has \code{NoSignedWrap} and \code{NoContraction}, separate \code{FMin} and \code{NMin}, float-control execution modes and, with the float-controls extension that Vulkan~1.4 makes core, per-instruction fast-math flags~\citep{khronos_spirv16,khronos_glslstd450,khronos2026floatcontrols,khronos2024floatcontrolstwo}; TDF and the CLI give arithmetic operations explicit error treatments~\citep{tdf1998spec,ecma2012cli}; WebAssembly instead defines integer arithmetic completely, and C2Rust emits \code{wrapping\_add} for C's unsigned arithmetic~\citep{c2rust2026}.
Per-target lowering of a recorded rule is established too: MLIR's conversion from \code{arith} to SPIR-V emulates the signed remainder for Vulkan and guards the NaN-propagating \code{minimumf} with \code{IsNan} unless fast-math flags allow otherwise~\citep{llvm2026arithtospirv}. In these IRs each operator variant has one fixed rule, and each front end picks the variant; a source-to-source hub must pick it per source for targets that are themselves languages with undefined behaviour and scoped configurations, and \cref{prop:provenance} states when the pick must name the source. Undefined behaviour, studied for C~\citep{wang2012undefined,wang2013stack,dietz2012integeroverflow,hathhorn2015undefinedness,memarian2016depthsofc}, reaches GPUs through MSL, CUDA, OpenCL~C and DXC.

\paragraph{Shader and compute translators.}
glslang and DXC compile GLSL and HLSL to SPIR-V~\citep{khronos2026glslang,microsoft2026dxc,microsoft2026dxcspirv}; SPIRV-Cross decompiles SPIR-V to GLSL, HLSL and MSL~\citep{khronos2026spirvcross}, and MoltenVK and IREE's Metal back end use it to run SPIR-V on Metal~\citep{khronos2026moltenvk,iree2026metal}; vkd3d-proton and the Metal Shader Converter run DXIL on Vulkan and Metal~\citep{vkd3dproton2026,hk2026dxilspirv,apple2026metalshaderconverter}. naga, a bidirectional hub around a validated module with configurable bounds-check policies, and Tint, which compiles WGSL through its own IR with configurable polyfills for division, remainder and shifts~\citep{gfxrs2026naga,gfxrs2026nagabounds,google2026tint,dawn2026tintpolyfill}, give their IRs WGSL's wrap-around semantics and polyfill it where the back end's language leaves overflow undefined~\citep{blandy2024nagaoverflow,nicol2025nagaoverflowfix,tint2021overflow,dawn2025tintsignedpolyfill,nicol2025nagaedgecases,tanasoaia2024loops,price2024tintloophardening}: a single fixed semantics, which keeps a source rule only where it refines that rule. Slang compiles an HLSL-derived language through its own IR, with target legalisation and direct SPIR-V emission~\citep{he2018slang,slang2026overview}, and has an open proposal to define integer semantics across targets~\citep{kiminki2025slangub}. clspv and clvk bring OpenCL~C to Vulkan~\citep{google2026clspv,kpet2026clvk}, and so face the OpenCL-to-Vulkan rows of \cref{tab:hazard-matrix}.
Compute translation has mostly started from CUDA~\citep{martinez2011cu2cl,harvey2011swan,stratton2008mcuda,amd2026hipify,intel2026syclomatic,castano2022dpct}; binary and IR-level approaches~\citep{diamos2010ocelot,velesko2026chipstar,han2022cupbop,moses2021polygeist} and portability layers and domain-specific languages~\citep{edwards2014kokkos,reyes2016sycl,raganKelley2013halide,chen2018tvm,tillet2019triton} avoid source translation, and learned translators can produce code that compiles but behaves differently~\citep{roziere2020transcoder,wen2022babeltower,pan2024lost}.
Earlier work described \crosstl's hub-and-spoke design and evaluated it on small sets of programs~\ifanon\citep{niketan2025crosstl}\else\citep{niketan2025crosstl,niketan2026crossglspringer}\fi; \cref{sec:system:hub,sec:rulehub} qualify its statement that \crossgl captures the essential semantics of all supported languages.

\paragraph{Correctness of translators and GPU programs.}
Translation validation checks each translation instead of the translator~\citep{pnueli1998tv,necula2000tv}, CompCert proves a compiler correct~\citep{leroy2009compcert}, and Alive2 checks LLVM optimisations as refinement queries to an SMT solver~\citep{lopes2021alive2}, as we check the rule table; our definitions borrow their refinement view.
Differential testing, random program generation and equivalence modulo inputs~\citep{mckeeman1998differential,yang2011csmith,le2014emi} have GPU descendants~\citep{lidbury2015manycore,donaldson2017graphicsfuzz,donaldson2021spirvfuzz,bernhard2024darthshader}; conformance suites check an implementation against one specification~\citep{khronos2026vkglcts}, yet a certified driver can still fold a wrapping addition (\cref{sec:eval:rulehub}); our suite checks translations against the source's rules. Other work verifies GPU kernels and their memory and progress models~\citep{betts2012gpuverify,li2012gklee,alglave2015gpuweak,sorensen2021progress}.

\section{Conclusion}
\label{sec:conclusion}

A hub that records only syntax loses the rules that GPU languages give the same operations. Recording each operation's rule and lowering it through one table per target keeps them when annotation is sound and complete and the table is sound; given an exact frame each target construct must refine the source's rule, and a hub that does not refuse must record the source where defined rules conflict. In \crosstl, every remaining failure traced to an unsound table entry, an emitter's lowering of the trusted base, or a runtime that miscompiles correct code, a certified-conformant driver included; a faithful translator must record its configuration and be tested on its users' stacks.

\section*{Data-Availability Statement}
\label{sec:dataavail}

\ifanon
An anonymised artifact accompanies this submission as supplementary material.
\else
The artifact is released with this preprint as arXiv ancillary files, and \crosstl is available at \repolink{https://github.com/CrossGL/crosstl}.
\fi
The artifact contains the hazard suite with its registration and amendments, the held-out kernel generator, the experiment harness and container files, the \crosstlr passes as a patch against the evaluated implementation of \crosstl, the mechanical check of the rule table, manifests and scripts that rebuild the corpus from pinned upstream revisions, the raw results of every experiment, the coding packets, codebooks and codes, and the scripts that regenerate every table, figure and number in this paper from those results.
The raw translator outputs and run records (about 3.8\,GB) exceed the size limit for supplementary material; they will be deposited in an archive with a persistent identifier for artifact evaluation, and the scripts regenerate them.
Regenerating the analyses needs only Python. Re-running the experiments needs an Apple-silicon Mac for the Metal, MoltenVK, KosmicKrisp, wgpu and OpenCL runs and a container runtime for Mesa, DXC, dxil-spirv and nvcc; the artifact cannot execute CUDA or HIP outputs, which this study only compiled.
The corpus files are not redistributed and remain under their original licences.
We will submit the artifact to Artifact Evaluation.

\ifanon\else
\section*{Acknowledgement}
The authors thank the open-source contributors to \crosstl, and the maintainers of the corpora, compilers and runtimes used in this study.

\section*{Declarations}
\phantomsection\label{sec:declarations}

\paragraph{Competing interests.}
All authors are affiliated with CrossGL, the company that develops \crosstl, and have a financial interest in it.
\crosstl is free software under the Apache~2.0 licence.
The evaluation reports the system's failures, including the defects listed in \cref{app:defects}.

\paragraph{Prior publication.}
Earlier descriptions of \crosstl appeared as an arXiv preprint~\citep{niketan2025crosstl} and in a Springer proceedings paper~\citep{niketan2026crossglspringer}. Both present the hub-and-spoke architecture and its language coverage and evaluate it on small sets of shaders.
This paper shares no text with them. It adds the semantic catalogue, the definitions, the model of operation rules and its theorem, rule carrying, the hazard suite and the corpus evaluation, and it qualifies the earlier claim that \crossgl is a semantic superset of its source languages.
\fi

\bibliographystyle{ACM-Reference-Format}
\bibliography{refs}

\appendix
\crefalias{section}{appendix}
\crefalias{subsection}{subappendix}

\section{Glossary}
\label{app:glossary}

\begin{description}[leftmargin=1.2em, itemsep=1pt, font=\normalfont\itshape]
\item[Artifact] A translator's output program, together with the configuration it requires and the encoding of the source's state (\cref{def:faithful}).
\item[Hub, \crossgl, dialect] \crosstl's textual intermediate language. Each importer prints its own variant of it, a \emph{dialect} (\cref{sec:system:hub}).
\item[Importer, emitter] A front end that turns a source language into \crossgl, and a back end that turns \crossgl into a target language.
\item[Route] One translation chain from a source language to a target language, such as glslang and SPIRV-Cross from GLSL to MSL.
\item[Runtime, runtime stack] What executes an artifact: a compiler, a driver and a device. MoltenVK, for example, is a stack that translates SPIR-V to MSL and compiles it with Apple's compiler.
\item[Configuration ($\kappa$)] Compiler options, execution modes and device features that fix a program's meaning, such as Metal's math mode.
\item[Current and legacy HLSL rules] HLSL's rules as DXC compiles it to DXIL (current; the registered primary reading, called B in the artifact), and under the Direct3D bytecode that FXC emits (legacy; reading A).
\item[Obligation] A construct whose mapping to the target can make a translation unfaithful (\cref{def:obligation}). A translator \emph{discharges}, \emph{checks}, \emph{assumes}, \emph{refuses} or \emph{drops} it.
\item[Drop-free, fail-closed] Never dropping an obligation of a class; in addition, never relying on an assumption (\cref{def:dropfree}).
\item[Rule, refinement] The set of results that a language permits for an operation on given arguments, or undefined behaviour; a rule refines another if it permits no more (\cref{def:refine}).
\item[Carried rule, rule table, \crosstlr] A source rule that the annotation pass records as a distinct hub operation; the per-target table that lowers it; \crosstl with both passes enabled (\cref{sec:rulehub}).
\item[Polyfill, trusted base] A replacement for a target operation, written once in \crossgl; the operations that every polyfill assumes each emitter translates faithfully.
\item[Preserved, diverged] An execution whose output equals the oracle's, or differs from it.
\item[Basis of a match] Why a preserved outcome matched: the target's own rule, an explicit construct, an unrecorded configuration, or the particular implementation.
\item[Latent drop] A match on an unrecorded configuration or on the particular implementation, for an obligation that the artifact neither discharges nor records.
\item[Mechanism of a failure] Why a non-preserved outcome failed: a fallback, a wrong construct, a dropped obligation, or a divergence of the runtime stack.
\item[Arm] One variant in the rule-carrying experiment: \crosstl, the SPIR-V emitter fixes, \crosstlr, or both together.
\item[Source-dependent cell] A (rule, target) pair whose executed outcomes differ between source languages that define the same rule.
\item[Unit, slice, family] A translation unit (a file with its include closure that declares an entry point); one project's units in one language; units that share code or authorship, the cluster of the bootstrap intervals.
\item[Feasible, partial cell] A (unit, target) pair whose target can express every stage of the unit, or only some of them.
\item[Usable, degraded, invalid, refused] Corpus outcomes: accepted without and with a fallback marker or missing entry point, rejected by the validator, or stopped by \crosstl.
\item[Fallback marker] A comment that an emitter leaves where it substituted a default value, type or padding for a construct it could not translate.
\item[Seen] A corpus slice used during \crosstl's development (MLX, Mojo puzzles).
\item[Contract, loader, runtime package] A JSON record in an artifact that must be honoured before dispatch; the code that checks contracts and dispatches; a bundle of hash-checked artifacts, their resource reflection and a loader interface.
\item[Held-out kernels] Random kernels generated from a fixed seed after \crosstlr was frozen (\cref{app:prereg}, A4).
\item[lavapipe, llvmpipe] Mesa's CPU implementations of Vulkan and of OpenGL.
\item[DXBC, DXIL] Direct3D's legacy bytecode and its LLVM-based successor.
\item[SIMD-group, subgroup, wave] The threads that execute in lockstep, so called in MSL, Vulkan and Direct3D.
\item[Poison, nsw] A value that becomes undefined behaviour when an operation needs a valid one (LLVM, SPIR-V); LLVM's flag under which signed overflow yields poison.
\end{description}

\FloatBarrier
\section{Results Under the Legacy HLSL Rules}
\label{app:readingA}

Under the legacy rules (FXC and DXBC), HLSL defines signed wrap-around, unsigned division by zero, float-to-integer conversion, subnormal flushing and a zero for out-of-bounds reads, so seven HLSL cases carry an obligation that the current rules do not give them: three signed-overflow kernels, unsigned division by zero, float-to-integer conversion, subnormal flushing and out-of-bounds reads. Under the legacy rules, 63 cases carry an obligation, against 56. This appendix gives the numbers of \cref{sec:eval:hazards,sec:eval:rulehub,sec:eval:check} under these rules. No conclusion changes; the differences are listed below.

\paragraph{RQ3.}
Slang, given the overflow kernels in HLSL, also copies the operator into MSL and loses the wrap rule on Metal, and MoltenVK also miscompiles DXC's SPIR-V for them. The \code{-fwrapv} probe kept the rule in all 19 overflow artifacts that diverged on Metal and changed none of the 27 that matched.
Three more losses appear. lavapipe breaks Direct3D~11's subnormal flush; every route to WGSL loses Direct3D's out-of-bounds zero, because naga and Tint clamp the index, which WGSL permits as an alternative to returning zero; and through the Metal Shader Converter, DXC's folding of signed overflow carries over, and the M2 returns 0 for $0/0$, where Direct3D requires \code{0xFFFFFFFF}. The converter's out-of-bounds reads matched only because the memory past the buffer happened to be zero: in a post-hoc probe in which the memory past the view held 999.0, the default conversion returned that value for all 60 out-of-bounds reads, and a conversion with the converter's bounds-check option returned zero for all 60.
On Metal, 12 of 233 (case, route) cells with an obligation change outcome between the three configurations. Between lavapipe and MoltenVK, 45 of 151 cells differ; \crosstlr's SPIR-V is byte-identical to \crosstl's in 17 of them, so 28 of 111 distinct binaries behave differently.
The basis coding covers 575 artifacts; the two codings agreed on 570 ($\kappa=0.99$). The registered rule A8 moves 20 matches, A9 moves 8 SPIR-V matches of \crosstlr and marks 8 GLSL matches as holding under a recorded assumption, and A12 moves the 2 remainder matches through Metal's native \code{\%}; A13 does not apply. The supported share is 58\,\% for \crosstl and 75\,\% for \crosstlr (62 and 82\,\% on the coders' codes alone), and 159 matches rest on the implementation. The mechanism coding covers 203 artifacts, of which 201 were agreed and 2 adjudicated.

\paragraph{RQ4.}
\Cref{tab:rulehub-A} repeats \cref{tab:rulehub} on the 29 carried cases with an obligation under the legacy rules. The local fixes remove 14 of 51 divergences, and carrying the rules removes 42 of 51. Two of the nine remaining divergences have a second cause, the SPIR-V emitter's loop defect S3, which breaks the faithfulness of the translation outside the carried operations. The stack-keyed variant preserved all 14 overflow runs. H-R1's basis clause fails for 16 of 125 matches (seven wrap matches, eight NaN tests on SPIR-V, and CUDA's remainder through Metal's native \code{\%}), or for 9 without the registered basis rule. For H-R2, \crosstl's outcome depends on the source in 5 of 30 cells, the local fixes leave 3 of 30, and \crosstlr 1 of 33, caused by S3.
On the held-out kernels, the 12 HLSL kernels for wrap-around and division by zero carry an obligation only under the legacy rules; in their 70 executions per variant, \crosstl diverged 14 times and \crosstlr never, so the totals are 134 of 514 executions for \crosstl and none of 532 for \crosstlr. \crosstl also lost Direct3D's division by zero on Metal, MoltenVK and WGSL, and the routes through DXC and Slang diverged in 42 executions, all on these kernels.

\paragraph{RQ5.}
Of 110 (class, target) cells, \crosstl is drop-free in 34, refuses in 5 and drops in 60; \crosstlr is drop-free in 50, refuses in 8 and drops in 43. \crosstlr is drop-free everywhere only for shift ranges (H2) and mixed comparisons (H4): signed wrap is not, because its HLSL loop case hangs on SPIR-V (S3). The two variants drop obligations in 17 and 16 of 22 classes.

\begin{table}[h]
\caption{The rule-carrying experiment under the legacy HLSL rules (\cref{tab:rulehub} gives the current rules).}
\label{tab:rulehub-A}
\small
\begin{tabular}{@{}lrrrr@{}}
\toprule
Arm & Divergences & Source-dependent & Refusals & Invalid \\
\midrule
\crosstl & 51 & 5 of 30 & 0 & 63 \\
SPIR-V emitter fixes & 37 & 3 of 30 & 0 & 63 \\
CrossTL-R & 9 & 1 of 33 & 8 & 52 \\
CrossTL-R and emitter fixes & 11 & 1 of 33 & 8 & 52 \\
\bottomrule
\end{tabular}

\end{table}

\FloatBarrier
\section{Proofs}
\label{app:proofs}

\paragraph{Program model.}
We model the semantics of \cref{sec:fidelity} by a small-step semantics parameterised by a rule assignment $\varrho$, which gives each operation occurrence $o$ of a program a rule $\varrho(o)$. Executions interleave the steps of the program's threads under sequential consistency. A step either evaluates an occurrence $o$ on argument values $v$, choosing a result $r\in\varrho(o)(v)$, or performs another action (control flow, memory access, synchronisation), whose effect, including going wrong (an out-of-bounds access, or a data race where the language makes it undefined), does not depend on $\varrho$. If $\varrho(o)(v)=\undefrule$, the execution goes wrong. If some execution from $\sigma$ goes wrong, $\sem{p}_{\varrho}(\sigma)$ is the set of all outcomes; otherwise it contains the final states of the terminating executions, and non-termination if some execution is infinite or ends in a state that is neither final nor able to step (a deadlock), so every state has an outcome. We assume that a source's specification contains its model, $\sem{p}_{\rul_{S,\kappa}}\subseteq\sem{p}_{S,\kappa}$. The reading of \cref{sec:rulehub:rules} ensures this unless $p$ reaches an \textsf{impl} or \textsf{n/s} case of $S$: an implementation-defined result is the same at every evaluation, which a choice per evaluation does not express, and an unstated rule need not be undefined. Weak memory, contraction, subnormal flushing, rounding, layouts and forward-progress guarantees are outside the model and inside (P0).
A rule is \emph{exact} for $L$ at a tuple $v$ if $L$'s specification allows exactly its results there, and undefined behaviour exactly where it is $\undefrule$; (P0) holds \emph{with equality} if the decodings of the artifact's outcomes on every accepted encoding of $\sigma\models\Phi$ are exactly $\sem{p}_{\rul_T}(\sigma)$.
A translation of $p$ is \emph{faithful apart from} a set $O$ of occurrences if it satisfies (P0) for the rule assignment that gives each occurrence in $O$ the rule of the target construct to which it is mapped and every other occurrence its source rule. This is the sense of ``otherwise faithful'' in \cref{def:obligation}.

\paragraph{Polyfills and the trusted base.}
A polyfill $f$ is a terminating function that reads only its arguments, each once, and touches no shared state. Its rule $\rul_f(v)$ is $\undefrule$ if some execution of $f$ on $v$ goes wrong, and the set of its results otherwise; because its steps commute with those of other threads, a call can be replaced by one step with rule $\rul_f$ without changing the outcomes. (The defect that amendment A1 fixed, a polyfill parameter that captured a program buffer, broke this condition.)
The hub rules of the trusted base $B$ are: unsigned 32-bit addition, subtraction, multiplication, bitwise operations, comparisons and \code{max}, exact modulo $2^{32}$; signed comparisons; shifts by counts below 32; unsigned division and remainder by a non-zero divisor; bit reinterpretation between \code{int}, \code{uint} and \code{float}, exact; conversion of \code{uint} values below $2^{31}$ to \code{int}; \code{select}, which evaluates both operands and returns one; binary32 comparisons and \code{clamp} as in IEEE~754, with a comparison involving NaN false and the \code{clamp} of a NaN unspecified; conversion of an in-range, non-NaN binary32 value to an integer, by truncation; signed subtraction, which the remainder polyfill uses only on values that cannot overflow; and the arithmetic right shift of a negative signed integer. MSL and the targets with wrapping semantics define the last one; for CUDA and HIP, under C++17, it is implementation-defined, an assumption in $\Psi$ that \crosstlr's artifacts do not record. Rust's and Mojo's emitters are outside the check: Rust's integer arithmetic panics on overflow when overflow checks are on, so its unsigned base is exact only through wrapping operations, which we did not examine.

\begin{lemma}[Monotonicity]
\label[lemma]{lem:monotone}
If $\varrho'(o)\refines\varrho(o)$ for every occurrence $o$, then $\sem{p}_{\varrho'}(\sigma)\subseteq\sem{p}_{\varrho}(\sigma)$ for every $\sigma$.
\end{lemma}

\begin{proof}
If some $\varrho$-execution from $\sigma$ goes wrong, the right-hand side contains every outcome. Otherwise, let $e'$ be a $\varrho'$-execution from $\sigma$. We show by induction on its length that every finite prefix of $e'$ is a $\varrho$-execution. Steps that evaluate no operation do not depend on the rules. At a step that evaluates $o$ on $v$, the prefix before it is a $\varrho$-execution by the induction hypothesis, so $\varrho(o)(v)\neq\undefrule$, since otherwise a $\varrho$-execution would go wrong. By \cref{def:refine}, $\varrho'(o)(v)\neq\undefrule$ and $\varrho'(o)(v)\subseteq\varrho(o)(v)$, so the result chosen in $e'$ is permitted by $\varrho$, and $e'$ does not go wrong at this step. Hence $e'$ never goes wrong; a terminating $e'$ is a terminating $\varrho$-execution with the same final state, and an infinite or stuck $e'$ is an infinite or stuck $\varrho$-execution.
\end{proof}

In particular, if $p$ has undefined behaviour on $\sigma$ under $\varrho'$, it has undefined behaviour under $\varrho$.

\begin{proposition}
\label[proposition]{prop:obligation}
Let a translation map the occurrences of an operator $\mathit{op}$ from $S$ to a target construct with rule $\rul_T$ and be faithful apart from them. If $\rul_T\refines\rul_{S,\kappa}(\mathit{op})$, the occurrences carry no obligation (\cref{def:obligation}). If $\rul_T$ is the rule that $T$'s specification gives the construct, and a tuple $v$ that can be loaded from an input buffer has $\rul_{S,\kappa}(\mathit{op})(v)\neq\undefrule$, $S$'s rule exact at $v$, and either $\rul_T(v)=\undefrule$ or $\rul_T(v)\not\subseteq\rul_{S,\kappa}(\mathit{op})(v)$, they carry one.
\end{proposition}

\begin{proof}
If $\rul_T\refines\rul_{S,\kappa}(\mathit{op})$, then in every context the artifact's outcomes are, up to rejection and decoding, among those of the source program with the rule of $\mathit{op}$'s occurrences replaced by $\rul_T$, which by \cref{lem:monotone} are outcomes of the source program; so there is no obligation.
Otherwise take the translation of the kernel $K_v$ that loads $v$, applies $\mu(\mathit{op})$ and stores the result, with the identity encoding and no loader check; it is faithful apart from $\mathit{op}$, and on the state that holds $v$, where $K_v$ is defined and, by exactness, stores only elements of $\rul_{S,\kappa}(\mathit{op})(v)$, $T$'s specification lets it go wrong or store an element of $\rul_T(v)\setminus\rul_{S,\kappa}(\mathit{op})(v)$, neither of which is an outcome of $K_v$. So this translation is unfaithful under $(\mathit{true},\mathit{true})$.
\end{proof}

\begin{proof}[Proof of \cref{thm:sufficiency}]
If the translation uses a refusal entry, the translator refuses. Otherwise let $o$ be an occurrence of $p$. If $o$ is not annotated, $\rul_T(o)\refines\rul_{S,\kappa}(o)$ by (P1). If it is, $\alpha(o)\refines\rul_{S,\kappa}(o)$ by (P1), and $\Lambda_T$ maps $\mathit{op}_{\alpha(o)}$ to an entry that is sound by (P2), under $\kappa_T$ and $\Psi$, so $\rul_T(o)\refines\alpha(o)$. (For a polyfill $f$, the sufficient condition of \cref{sec:rulehub:carrying} gives this by \cref{lem:monotone}, applied to $f$ as a program, whose call can be replaced by one step with $f$'s rule.) By transitivity, $\rul_T(o)\refines\rul_{S,\kappa}(o)$ for every $o$.
By \cref{lem:monotone} and the containment of the model in the specification, $\sem{p}_{\rul_T}(\sigma)\subseteq\sem{p}_{\rul_{S,\kappa}}(\sigma)\subseteq\sem{p}_{S,\kappa}(\sigma)$, and undefined behaviour of $p$ under $\rul_T$ implies undefined behaviour under $\rul_{S,\kappa}$. By (P0), every outcome of the artifact on a state that encodes some $\sigma\models\Phi$ is $\bot_{\mathit{rej}}$, or $p$ read with $\rul_T$ has undefined behaviour on $\sigma$, or it decodes to an element of $\sem{p}_{\rul_T}(\sigma)$; in each case \cref{def:faithful} holds for $(p,\kappa)$.
\end{proof}

\begin{proof}[Proof of \cref{cor:execution}]
An execution on a conforming implementation that satisfies $\Psi$ produces an outcome of $\sem{\loader\circ a}_{T,\kappa_T,\Psi}$, to which \cref{thm:sufficiency} applies. If the oracle is correct, an outcome that differs from every oracle-permitted result is not the decoding of an outcome in $\sem{p}_{S,\kappa}(\sigma)$, so the artifact is unfaithful, and one of the premises fails.
\end{proof}

\begin{proposition}[Converse]
\label[proposition]{prop:converse}
Suppose that (P0) holds with equality for every program of $S$, with $T$'s exact rules, and that the translator maps an operator $\mathit{op}$ of $S$ to a construct with rule $\rul_T$. If $\rul_{S,\kappa}(\mathit{op})(v)\neq\undefrule$, $S$'s rule is exact at $v$, and either $\rul_T(v)=\undefrule$ or $\rul_T(v)\not\subseteq\rul_{S,\kappa}(\mathit{op})(v)$, for a tuple $v$ that can be loaded from an input buffer, then on the kernel that loads $v$, applies $\mathit{op}$ and stores the result, the translator refuses, or its artifact rejects the state that encodes $v$, or it is unfaithful under $(\mathit{true},\mathit{true})$.
\end{proposition}

\begin{proof}[Proof of \cref{prop:converse}]
Let $K$ be the kernel that loads $v$, applies $\mathit{op}$ and stores the result, and $\sigma_v$ the state that holds $v$. $K$ is defined on $\sigma_v$, and every outcome of $K$ stores an element of $\rul_{S,\kappa}(\mathit{op})(v)$. If the translator neither refuses $K$ nor rejects the encoding of $\sigma_v$, then by (P0) with equality the artifact's outcomes on it are those of $K$ read with $\rul_T$, which are not empty; one of them goes wrong or stores an element of $\rul_T(v)\setminus\rul_{S,\kappa}(\mathit{op})(v)$, and neither decodes to an outcome of $K$.
\end{proof}

\begin{proof}[Proof of \cref{prop:provenance}]
(a) If $\rul^{\circ}$ refines every member of $D(\mathit{op})$, then at each $v$ at which some member is defined, $\rul^{\circ}(v)$ is a non-empty subset of each defined $\rul(v)$, so these share an element. Conversely, let $(\sqcap D)(v)$ be their intersection where some member is defined, and $\undefrule$ elsewhere; it refines every member, and every common refinement refines it. A least element of $D(\mathit{op})$ refines every member and so equals $\sqcap D(\mathit{op})$. The claim about one annotation follows by transitivity.
(b) Each witness kernel $K_i$ is defined on the state $\sigma_v$ and, since its source's rule is exact at $v$, stores only elements of $\rul_i(v)$, and these sets have no common element. An artifact for the shared hub program that is not refused and does not reject $\sigma_v$ has an outcome there, since outcome sets are not empty, and faithfulness to every $K_i$ would require that outcome to decode to a stored value common to all the $\rul_i(v)$.
(c) $\sqcap D(\mathit{op})\refines\rul_{S,\kappa}(\mathit{op})$, so transitivity gives the first claim. For WGSL and \code{min}, GLSL's rule (any value for a NaN operand) is refined by WGSL's own \code{min}, but no WGSL construct refines the NaN-ignoring rule.
\end{proof}

\paragraph{Variants needed.}
The number of carried variants of $\mathit{op}$ that a hub needs is the least number of groups into which $D(\mathit{op})$ splits with a common refinement in each group. It is one for every rule that \crosstlr carries; for unsigned division it would be two with a WGSL importer, and for binary32 subnormals under the legacy HLSL rules it is two.

\FloatBarrier
\section{Mechanical Check of the Rule Table}
\label{app:tablecheck}

For every native and polyfill entry, the check asks Z3 for operands $v$ and choices $c$ such that
$\mathit{def}_S(v)\wedge\neg\bigl(\mathit{def}_T(v,c)\wedge\mathit{allowed}_S(v,\mathit{res}_T(v,c))\bigr)$,
in the style of Alive2~\citep{lopes2021alive2}. Operands are 32-bit bit-vectors, and binary32 values are given by their bit patterns; $c$ are the free choices of the target's unspecified results; $\mathit{def}_S$ and $\mathit{allowed}_S$ describe the carried rule, the meet of the defining sources' rules (\cref{prop:provenance}); and $\mathit{res}_T$, $\mathit{def}_T$ evaluate the native operation, or the polyfill's \crossgl text, transcribed next to its encoding, under the target's rules for the trusted base. A satisfiable query is a counterexample; an unsatisfiable one shows the entry sound under the model.
The model reads \cref{tab:hazard-matrix} and the NaN and infinity rows of \cref{tab:hazard-matrix-env}, with amendments A12 and A13; like contraction, subnormal flushing and rounding are outside it and inside (P0). GLSL~ES is read with GLSL's rules. Signed overflow is undefined on Metal, CUDA, HIP and HLSL (DXC's \code{nsw}) and wraps elsewhere; the right shift of a negative value is arithmetic except on CUDA and HIP, where it is unspecified unless the assumption \emph{ashr} ($\Psi$) is made; on SPIR-V, an operation on a NaN or infinite operand, bit reinterpretation included, is undefined, and so it is on Metal without safe math (recorded by +C) and on GLSL without NaN support (recorded by +A); on HLSL, a comparison with a NaN or infinite operand has an unspecified result unless \code{/Gis} is given (reading it as undefined changes no verdict), whereas DXIL's \code{FMin} and \code{FMax} keep the NaN rule that DXIL defines for them; conversions of out-of-range or NaN values to integers are undefined everywhere. Unsigned remainder is left uninterpreted in the four polyfill entries of the truncated remainder (the Direct3D remainder entries use the real operation), in the polyfill and in the specification alike, where SMT-LIB defines the truncated remainder through it; each application satisfies $r<d$ and $r\le a$ for $d\neq0$. A result \emph{sound} then holds for the real operation a fortiori, and the queries take milliseconds instead of exceeding a 300-second limit.
\Cref{tab:tablecheck} gives the results. For each unsound entry the solver returns a counterexample: an infinite or NaN operand for the SPIR-V and HLSL entries, $-2^{31}\,\%\,1$ on Metal, $x\,\%\,0$ on CUDA and HIP, and $-2^{31}\gg0$ for the shift polyfill on CUDA and HIP. Re-checking the +C and +A entries without their records makes each of them unsound, so the model cannot drop any record; for five of them (\code{isnan} on Metal, GLSL and GLSL~ES, and NaN-to-zero conversion on GLSL and GLSL~ES) only because it reads the bit reinterpretation of a NaN as undefined under the target's NaN latitude, which MSL's fast math supports and GLSL's text leaves open.
Two sensitivity runs bound the model's choices. If a comparison may see a subnormal operand flushed to a zero of its sign, as GLSL and MSL permit and DXIL's \code{FMin} does, 112 entries are sound and 19 are not: \code{min} and \code{max} also fail on GLSL, GLSL~ES and Metal, and \code{/Gis} no longer rescues them on HLSL, only the saturating conversion. Under the stricter reading of DXIL in which a non-precise \code{FMin} may also be optimised unsafely, \code{h\_min\_nan} would carry no obligation under the current rules and \crosstl's three native-\code{min} matches on HLSL would hold by implementation.
The check is bounded by its model: it trusts the catalogue's reading of each target, the transcription of the polyfills, and each target's base operations at the level of its specification; it does not examine which construct each emitter chooses.

\begin{table}[h]
\caption{Mechanical check of \crosstlr's rule table, per operator and target. Entries as in \cref{tab:rule-table}; $\times$: unsound as recorded (bold: no configuration or assumption in the model makes it sound; in parentheses: the unrecorded condition under which it is sound); --: not carried for the target; Ref: refusal, not checked.}
\label{tab:tablecheck}
\footnotesize
\begin{tabular}{@{}lcccccccc@{}}
\toprule
Operator & Metal & SPIR-V & HLSL & GLSL & ESSL & WGSL & CUDA & HIP \\
\midrule
wrap $+$ & P & N & P & N & N & N & P & P \\
wrap $-$ & P & N & P & N & N & N & P & P \\
wrap $\times$ & P & N & P & N & N & N & P & P \\
\code{<<} (u) & N & P & N & P & P & N & P & P \\
\code{>>} (u) & N & P & N & P & P & N & P & P \\
\code{<<} (i) & N & P & N & P & P & N & P & P \\
\code{>>} (i) & N & P & N & P & P & N & P$^{\times}$ (ashr) & P$^{\times}$ (ashr) \\
$x/0$ & P & P & P & P & P & P & P & P \\
$x\,\%\,0$ & P & P & P & P & P & P & P & P \\
trunc.\ \code{\%} & \textbf{N}$^{\times}$ & P & P & P & P & N & \textbf{N}$^{\times}$ & \textbf{N}$^{\times}$ \\
8-bit store & N & P & P & P & P & P & N & N \\
16-bit store & N & P & P & P & P & P & N & N \\
f$\to$i sat. & P+C & \textbf{P}$^{\times}$ & P$^{\times}$ (\code{/Gis}) & P+A & P+A & Ref & P & P \\
f$\to$i NaN$\to$0 & N+C & \textbf{P}$^{\times}$ & P & P+A & P+A & Ref & P & P \\
\code{min} & P+C & \textbf{P}$^{\times}$ & P$^{\times}$ (\code{/Gis}) & P+A & P+A & Ref & P & P \\
\code{max} & P+C & \textbf{P}$^{\times}$ & P$^{\times}$ (\code{/Gis}) & P+A & P+A & Ref & P & P \\
\code{isnan} & P+C & \textbf{P}$^{\times}$ & P & P+A & P+A & Ref & P & P \\
\bottomrule
\end{tabular}

\end{table}

\FloatBarrier
\section{Architecture and an Example of Rule Carrying}
\label{app:example}

\Cref{fig:cgl} shows the kernel of \cref{fig:motivation} as the GLSL importer prints it, and \crosstlr's Metal output for it.

\begin{figure}[h]
  \centering
  \includegraphics[width=0.64\linewidth]{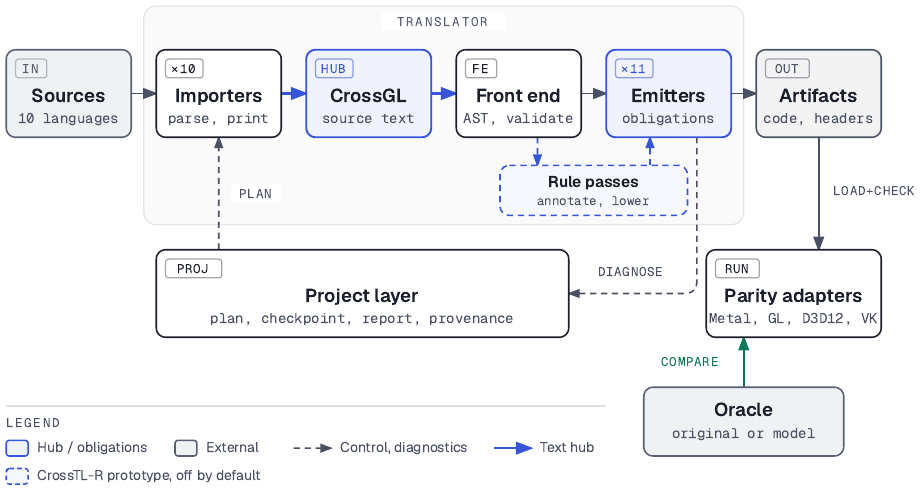}
  \caption{Architecture of \crosstl. The rule-carrying passes of \cref{sec:rulehub} (dashed) are enabled by a switch.}
  \label{fig:architecture}
  \Description{Block diagram: native sources flow through importers into CrossGL text, then through a parser into an object AST, then through per-target emitters into target artifacts; a project layer and native parity adapters surround the translator.}
\end{figure}

\begin{figure}[h]
\centering
\begin{minipage}[t]{0.42\linewidth}
\begin{lstlisting}[style=paper,language=CrossGL,numbers=none,basicstyle=\ttfamily\scriptsize]
shader main {
  layout(std430, binding = 0)
    buffer Inp { int a[]; } inp;
  layout(std430, binding = 1)
    buffer Outp { int o[]; } outp;
  compute {
    layout(local_size_x = 64) in;
    void main() {
      uint i = gl_GlobalInvocationID.x;
      int x = inp.a[i];
      outp.o[i] = (((x + 1) > x) ? 1 : 0);
    }
  }
}
\end{lstlisting}
\end{minipage}\hfill
\begin{minipage}[t]{0.55\linewidth}
\begin{lstlisting}[style=paper,language=Metal,numbers=none,basicstyle=\ttfamily\scriptsize]
int __cgl_wrap_add(int cgl_x, int cgl_y) {
  return as_type<int>(as_type<uint>(cgl_x)
                    + as_type<uint>(cgl_y));
}
kernel void kernel_main(uint3 thread_position_in_grid
                          [[thread_position_in_grid]],
    device uchar *inp [[buffer(0)]],
    device uchar *outp [[buffer(1)]]) {
  uint i = thread_position_in_grid.x;
  int x = (*reinterpret_cast<const device int *>
             (inp + (i * 4)));
  (*reinterpret_cast<device int *>(outp + (i * 4)))
      = __cgl_wrap_add(x, 1) > x ? 1 : 0;
}
\end{lstlisting}
\end{minipage}
\caption{The kernel of \cref{fig:motivation} as the GLSL importer prints it in \crossgl (left), and \crosstlr's Metal output for it (right). The annotation pass has replaced the GLSL addition by the rule-carrying \code{\_\_cgl\_wrap\_add}, and the lowering for Metal has injected its polyfill. \crosstl's own Metal output keeps \code{x + 1}, which the Metal compiler folds. Both outputs are abridged only in their headers.}
\label{fig:cgl}
\Description{Two code listings: CrossGL text with a shader block, two storage buffers and a compute entry that computes (x + 1 > x), and the Metal translation produced by CrossTL-R, in which the addition is a call to a function that adds the operands as unsigned integers.}
\end{figure}

\FloatBarrier
\section{The Hazard Catalogue and Suite}
\label{app:hazards}

The 66 cases comprise 18 in MSL, 15 in CUDA, 13 in HLSL, 9 in OpenCL~C, 7 in GLSL and 4 in SPIR-V.

\Cref{tab:hazard-matrix-env} completes the catalogue of \cref{tab:hazard-matrix} with the floating-point environment, memory and execution.

\begin{table}[h]
\caption{How GPU languages define the floating-point environment, memory and execution. Codes as in \cref{tab:hazard-matrix}.}
\label{tab:hazard-matrix-env}
\footnotesize
\setlength{\tabcolsep}{2.0pt}
\renewcommand{\arraystretch}{1.08}
\begin{tabularx}{\linewidth}{@{}>{\raggedright\arraybackslash}p{1.85cm}>{\raggedright\arraybackslash}X>{\raggedright\arraybackslash}p{1.6cm}*{5}{>{\raggedright\arraybackslash}X}@{}}
\toprule
 & GLSL & HLSL (DXBC / DXC) & MSL & WGSL & SPIR-V (Vulkan) & CUDA / HIP & OpenCL~C \\
\midrule
\multicolumn{8}{@{}l}{\emph{Floating-point environment}}\\
Default mode & contraction allowed & fast / fast & fast (API: fast)$^{a}$ & reassoc., fusion & contraction allowed; NaN/Inf assumed absent & FMA on & \code{FP\_CONTRACT} on \\
Opt-out & \code{precise} & \code{precise} & \code{\#pragma METAL fp}, \code{-fno-\allowbreak fast-math} & none & \code{No\allowbreak Contraction}, float controls & \code{\_\_fmul\_rn}, \code{--fmad=\allowbreak false} & \code{\#pragma} \\
\code{isnan}, NaN handling & may be false$^{b}$ & defined / NaN may be assumed absent & undefined in fast mode & NaN may be assumed absent & assumed absent without \code{SignedZero\allowbreak InfNan\allowbreak Preserve} & defined & defined \\
Rounding of $+,-,\times,\div$ & \textsf{n/s} (${\leq}1$ ULP; $\div$ 2.5 ULP) & RNE for $+,-,\times$ ($\div$ may be $x\cdot(1/y)$) & correctly rounded (safe); RTE or RTZ & \textsf{n/s} (up or down; $\div$ 2.5 ULP) & \textsf{impl} unless \code{Rounding\allowbreak ModeRTE} ($\div$ 2.5 ULP) & RNE & RNE ($\div$ ${\leq}2.5$ ULP) \\
Subnormals (fp32) & may flush & flush / \textsf{impl} & may flush & may flush & \textsf{impl} unless \code{Denorm\allowbreak Preserve} & preserved & \textsf{impl} \\
\multicolumn{8}{@{}l}{\emph{Memory and execution}}\\
3-vector in a buffer & array stride 16; member offset 12$^{c}$ & stride 12 (structured) & \code{float3} 16, \code{packed\_\allowbreak float3} 12 & array stride 16 & explicit & 12 & 16 \\
Out-of-bounds read & \textsf{UB}$^{d}$ & 0 / descriptor-dependent & \textbf{\textsf{UB}} & bounded & \textsf{UB}$^{d}$ & \textbf{\textsf{UB}} & \textbf{\textsf{UB}} \\
Partial work-groups & no & no & yes (\code{dispatch\allowbreak Threads}) & no & no & no & optional (2.0+) \\
Subgroup width & ext. & 4--128 & \textsf{impl} & ext. & \textsf{impl} & 32 / 32 or 64 & \textsf{impl} \\
Pointer arithmetic & none$^{e}$ & none & full & none & logical$^{e}$ & full & full \\
\bottomrule
\end{tabularx}
\par\smallskip
{\raggedright\scriptsize
$^{a}$\,a new \code{MTLCompileOptions} object on our host reports \code{mathMode}~=~fast, with fast math functions, although Apple's documentation calls relaxed mode the default on Apple silicon. $^{b}$\,GLSL~4.60 \S8.3: ``always returns false if NaNs are not implemented''. $^{c}$\,a \code{float} after a \code{vec3} in a std430 struct sits at offset 12. $^{d}$\,defined (zero or in-bounds values) when the API's robustness features are enabled. $^{e}$\,available through extensions (\code{GL\_EXT\_buffer\_reference}, \code{PhysicalStorageBuffer64}).\par}
\end{table}

The suite (\cref{tab:hazard-cases}) covers the 18 catalogue classes, plus \code{isnan}, \code{fmod} and atomics, although under the current HLSL rules no case remains for out-of-bounds reads (H11). Twenty-two cases come from an earlier, purposive suite, whose outcomes the authors already knew, and 44 are new; the registered selection rule covers both.
Every kernel processes 64 elements. The inputs place the critical values at fixed positions among ordinary values, so that an error shows as a mismatch at a known index. Each oracle is a Python function that implements the cited rule. The comparison is exact, bit-exact where sign or NaN matters, or within the stated ULP bound.
Every reference run matched its oracle except in four places, which the source's rules, its configuration or the runtime explain: DXC folds the three HLSL overflow kernels to constants; lavapipe, after dxil-spirv, preserves the subnormals that the legacy rules flush, which the current rules allow; outside safe mode, the MSL kernels lose signed zeros and \code{half} rounding; and on MoltenVK the suite's own SPIR-V overflow kernels, which are correct, do not wrap, nor, in the exploratory KosmicKrisp run, its addition kernels (\cref{app:probes}).

{\scriptsize\setlength{\tabcolsep}{3pt}
\begin{longtable}{@{}l l l >{\raggedright\arraybackslash}p{3.5cm} >{\raggedright\arraybackslash}p{4.2cm} l@{}}
\caption{The 66 cases of the hazard suite. \emph{Class}: catalogue class (\cref{tab:hazard-matrix,tab:hazard-matrix-env}). \emph{Rule}: the clause the oracle implements; for HLSL, A is the Direct3D bytecode rule and B the behaviour of DXC/DXIL. \emph{Cmp.}: exact, bit-exact (sign and NaN payload class), or within the stated ULP bound.}\label{tab:hazard-cases}\\
\toprule
Case & Class & Source & Kernel & Rule & Cmp. \\
\midrule
\endfirsthead
\toprule
Case & Class & Source & Kernel & Rule & Cmp. \\
\midrule
\endhead
\bottomrule
\endfoot
\code{g\_add\_wrap} & H1 & GLSL & x + 1 $>$ x with x near INT\_MAX & GLSL 4.60 \S{}4.1.3 WRAP & exact \\
\code{g\_mul\_wrap} & H1 & GLSL & (x * 2) / 2 == x & GLSL 4.60 \S{}4.1.3 WRAP & exact \\
\code{g\_loop\_wrap} & H1 & GLSL & for (k = s; k $<$ s + 4; ++k) & GLSL 4.60 \S{}4.1.3 WRAP & exact \\
\code{h\_add\_wrap} & H1 & HLSL & x + 1 $>$ x & A: D3D11.3 \S{}22.12.1 iadd WRAP; B: DXIL add nsw & exact \\
\code{h\_mul\_wrap} & H1 & HLSL & (x * 2) / 2 == x & A: D3D11.3 \S{}22.12.6 imul WRAP; B: DXIL mul nsw & exact \\
\code{h\_loop\_wrap} & H1 & HLSL & loop bound s + 4 & A: DXBC WRAP; B: DXIL add nsw & exact \\
\code{s\_add\_wrap} & H1 & SPIR-V & OpIAdd without NoSignedWrap & SPIR-V \S{}3.3.13 OpIAdd & exact \\
\code{s\_mul\_wrap} & H1 & SPIR-V & OpIMul without NoSignedWrap & SPIR-V \S{}3.3.13 OpIMul & exact \\
\code{h\_shift\_mask} & H2 & HLSL & x $<<$ s with s $\geq$ 32 & D3D11.3 \S{}22.11.7 ishl; DXC masks (SPIR-V.rst) & exact \\
\code{m\_shift\_mask} & H2 & MSL & (x $<<$ s) \^{} (x $>>$ s) with s $\geq$ 32 & MSL 4.1 \S{}3.1 item 11 MASK & exact \\
\code{c\_shift\_mask} & H2 & OpenCL~C & (x $<<$ s) \^{} (x $>>$ s) with s $\geq$ 32 & OpenCL C \S{}6.5.10 MASK & exact \\
\code{h\_udiv\_zero} & H3 & HLSL & a / 0u & A: D3D11.3 \S{}22.12.9 0xFFFFFFFF; B: DXIL udiv (n/f) & exact \\
\code{m\_cmp\_mixed} & H4 & MSL & int $<$ uint & C++17 usual arithmetic conversions (MSL \S{}3.1) & exact \\
\code{g\_cmp\_mixed} & H4 & GLSL & int $<$ uint & GLSL 4.60 \S{}4.1.10 & exact \\
\code{h\_cmp\_mixed} & H4 & HLSL & int $<$ uint & HLSL draft \S{}5.1 [Expr.conv] & exact \\
\code{u\_cmp\_mixed} & H4 & CUDA & int $<$ unsigned & C++ [expr.arith.conv] & exact \\
\code{c\_cmp\_mixed} & H4 & OpenCL~C & int $<$ uint & C99 \S{}6.3.1.8 & exact \\
\code{m\_uchar\_wrap} & H15 & MSL & uchar v = x; v = v + 1 & C++ integral conversion mod 256 & exact \\
\code{u\_uchar\_wrap} & H15 & CUDA & unsigned char v = x; v = v + 1 & C++ integral conversion mod 256 & exact \\
\code{c\_uchar\_wrap} & H15 & OpenCL~C & uchar v = x; v = v + 1 & C99 \S{}6.3.1.3 & exact \\
\code{m\_ulong\_mul} & H15 & MSL & high word of ulong(x) * 0xFFFFFFFF & C++ unsigned 64-bit arithmetic & exact \\
\code{u\_u64\_mul} & H15 & CUDA & high word of x * 0xFFFFFFFFull & C++ unsigned 64-bit arithmetic & exact \\
\code{u\_srem\_sign} & H3 & CUDA & a \% b with negative operands & C++ [expr.mul]/4 & exact \\
\code{m\_srem\_sign} & H3 & MSL & a \% b with negative operands & MSL 4.1 3.1 item 2: undefined for negative operands (no obligation, A12) & exact \\
\code{h\_ftoi\_sat} & H5 & HLSL & (int)x for NaN, Inf, out of range & A: D3D11.3 \S{}22.13.3 ftoi; B: DXIL fptosi & exact \\
\code{m\_ftoi\_nan} & H5 & MSL & int(NaN) == 0 & MSL 4.1 \S{}8.6 & exact \\
\code{c\_convert\_sat} & H5 & OpenCL~C & convert\_int\_sat(x) & OpenCL C \S{}6.4.3.3 & exact \\
\code{h\_min\_nan} & H6 & HLSL & min(x, NaN) & D3D11.3 \S{}3.1.3.1 minimumNumber; DXIL FMin & bits \\
\code{m\_fmin\_nan} & H6 & MSL & fmin(x, NaN) & MSL 4.1 Table 6.4 (safe math) & bits \\
\code{u\_fminf\_nan} & H6 & CUDA & fminf(x, NaN) & CUDA Math API fminf & bits \\
\code{c\_fmin\_nan} & H6 & OpenCL~C & fmin(x, NaN) & OpenCL C \S{}6.15.2 & bits \\
\code{s\_nmin\_nan} & H6 & SPIR-V & NMin(x, NaN) with SignedZeroInfNanPreserve & GLSL.std.450 NMin; no NaN result under Vulkan (no obligation, A12) & bits \\
\code{g\_precise} & H7 & GLSL & precise r = a*b - c & GLSL 4.60 \S{}4.9 & exact \\
\code{h\_precise} & H7 & HLSL & precise r = a*b - c & HLSL precise (DXIL) & exact \\
\code{m\_contract\_off} & H7 & MSL & \#pragma METAL fp contract(off): a*b - c & MSL 4.1 \S{}1.6.3 & exact \\
\code{s\_nocontraction} & H7 & SPIR-V & NoContraction a*b - c & SPIR-V NoContraction & exact \\
\code{u\_fmul\_rn} & H7 & CUDA & \_\_fadd\_rn(\_\_fmul\_rn(a, b), -c) & CUDA Math API intrinsics & exact \\
\code{c\_fp\_contract\_off} & H7 & OpenCL~C & FP\_CONTRACT OFF: a*b - c & OpenCL C \S{}6.15.2.1 & exact \\
\code{m\_precise\_sin} & H8 & MSL & precise::sin(x), x in [1e2, 1e5] & MSL precise sin (bound used: 4 ulp) & $\leq$4 \\
\code{m\_atan2} & H8/H10 & MSL & precise::atan2(y, x) with signed zeros & MSL precise atan2 (6 ulp); C99 Annex F & $\leq$6+sign \\
\code{u\_sinf} & H8 & CUDA & sinf(x), |x| $\leq$ 1e5 & CUDA Math API sinf: 2 ulp & $\leq$2 \\
\code{c\_sin} & H8 & OpenCL~C & sin(x), |x| $\leq$ 1e5 & OpenCL C \S{}7.4: 4 ulp & $\leq$4 \\
\code{u\_atan2f\_zero} & H10 & CUDA & atan2f with signed zeros & CUDA Math API atan2f special values & $\leq$2+sign \\
\code{u\_denorm} & H9 & CUDA & x * 0.5f for subnormal x & nvcc default -ftz=false & bits \\
\code{h\_denorm} & H9 & HLSL & x * 0.5 for subnormal x & A: D3D11.3 fp32 flush-to-zero; B: DXIL default 'any' & bits \\
\code{m\_isnan} & isnan & MSL & isnan(x) & MSL (safe math) & exact \\
\code{g\_isnan} & isnan & GLSL & isnan(x) & GLSL 4.60 \S{}8.3: false if NaNs are not implemented (negative control) & exact \\
\code{m\_fmod\_sign} & fmod & MSL & fmod(x, y) sign of x & C/MSL fmod & bits \\
\code{u\_fmodf\_sign} & fmod & CUDA & fmodf(x, y) sign of x & CUDA fmodf (C99) & bits \\
\code{m\_half\_mul} & H15 & MSL & half(x) * 3.0h & MSL half (binary16) & $\leq$0 \\
\code{m\_packed\_float3} & H16 & MSL & packed\_float3: 12-byte stride & MSL packed\_float3 & $\leq$0 \\
\code{h\_float3\_stride} & H16 & HLSL & StructuredBuffer$<$float3$>$: 12-byte stride & HLSL structured buffer layout & $\leq$0 \\
\code{u\_float3\_stride} & H16 & CUDA & float3 array: 12-byte elements & CUDA float3 & $\leq$0 \\
\code{g\_std430\_vec3\_float} & H16 & GLSL & struct {vec3 a; float b;}: b at offset 12 & GLSL std430 rules & $\leq$0 \\
\code{h\_oob\_read} & H11 & HLSL & a[i + 60] with 64 elements & A: D3D11.3 out-of-bounds reads return 0 & $\leq$0 \\
\code{u\_ptr\_arith} & H17 & CUDA & p = inp + 3; p[i] & C++ pointer arithmetic & exact \\
\code{m\_ptr\_arith} & H17 & MSL & p = inp + 3; p[i] & MSL device pointers & exact \\
\code{c\_ptr\_arith} & H17 & OpenCL~C & p = inp + 3; p[i] & OpenCL C global pointers & exact \\
\code{m\_partial\_tg} & H12 & MSL & dispatchThreads(100) with threadgroup 64 & MSL non-uniform threadgroups & exact \\
\code{c\_partial\_wg} & H12 & OpenCL~C & global size 100, work-group 64 & OpenCL C 2.0 non-uniform work-groups & exact \\
\code{u\_warp\_reduce} & H13 & CUDA & \_\_shfl\_down\_sync warp reduction & CUDA warpSize = 32 & exact \\
\code{m\_simd\_sum} & H13 & MSL & simd\_sum and threads\_per\_simdgroup & MSL SIMD-group functions (width impl.-defined) & exact \\
\code{h\_wave\_sum} & H13 & HLSL & WaveActiveSum and WaveGetLaneCount & HLSL SM 6.0 wave intrinsics & exact \\
\code{m\_simd\_divergent} & H14 & MSL & simd\_sum in SIMD-divergent control flow & MSL: reduces over active threads & exact \\
\code{u\_recursion} & H18 & CUDA & recursive \_\_device\_\_ fib & CUDA recursion & exact \\
\code{u\_atomic\_add} & atomics & CUDA & atomicAdd(\&out[0], 1) by 256 threads & CUDA atomicAdd (sanity) & exact \\
\end{longtable}
}

\FloatBarrier
\section{Registration, Amendments and Coding Protocol}
\label{app:prereg}

The hazard suite and the rule-carrying experiment were registered in a document written before any new case or \crosstlr ran. Its SHA-256, \code{416c68b6318bc1a4\ldots 9593dd6015}, was recorded at 18:34:02~UTC on 16~October 2026.
The registered text was not kept as a separate file at registration time; we recovered it from the log of the authoring session, and the recovered text reproduces the recorded hash. The artifact contains this file and the current one, which differs from it only in its Amendments section. The registered case table has 65 rows; its SPIR-V overflow row became two kernels (\code{s\_add\_wrap} and \code{s\_mul\_wrap}), which gives the 66 cases.
The document contains:
\begin{itemize}
\item \textbf{a chronology disclosure}: the first, purposive suite was written by authors who already knew its outcomes;
\item \textbf{the selection rule}: a source language gets one case per catalogue class if it defines the behaviour under its reference configuration, a compute kernel can observe the behaviour on our runtimes, and the language can express it as a compute kernel; two negative controls check that the classification does not invent obligations; the rule excludes memory-model litmus tests, graphics-pipeline conventions, and recursion beyond a direct recursive function;
\item \textbf{the routes, runtimes, oracles and outcome definitions};
\item \textbf{three hypotheses about \crosstlr}, H-R1 to H-R3 (\cref{sec:eval:setup});
\item \textbf{the analysis plan}.
\end{itemize}
Thirteen dated amendments followed. A1--A3 precede the runs they concern, and A4 was written during the held-out run, before its results; A5 and A6 correct the harness after the runs, A7 records a registered route that ran last, A8 and A9 correct the basis coding, A10 corrects A4's statement of its timing, A11 records where the implementation departs from the registered design, and A12 and A13 correct the catalogue after review.
\begin{description}[leftmargin=1.2em, itemsep=1pt]
\item[A1] Two implementation defects of \crosstlr were fixed after its first run: polyfill parameters captured a program buffer, and named scalar types were not recognised. The rule set did not change, and both runs are reported.
\item[A2] The emitter-fix and combined arms were added, as planned, together with an all-switches-off arm that must reproduce \crosstl.
\item[A3] H-R1 is falsified on MoltenVK. A table keyed by the runtime stack would avoid the failure, but any such variant is post hoc and labelled exploratory.
\item[A4] The held-out random kernels were generated, from seed 20261017, against a frozen \crosstlr identified by hash. The amendment was written after the kernels had been generated and translated, and before their execution results existed.
\item[A5] The Metal runner's relaxed configuration was corrected and its 436 cells re-run. No outcome changed.
\item[A6] The Metal runner's buffer bindings were corrected for 24 third-party artifacts and the 72 affected cells re-run; two outcomes changed (\cref{app:harness}). One artifact entered the basis coding and one left it, and likewise for the mechanism coding; the added artifacts were coded under the same protocol.
\item[A7] The registered route through Apple's Metal Shader Converter had been omitted from the first runs; it ran after A6, on all 13 HLSL cases. Its binary artifacts are outside the two codings.
\item[A8] The registered rule that a match changing between Metal modes or between lavapipe and MoltenVK holds by configuration or by implementation had not been applied to the basis coding. It is applied now; it moves 13 signed-wrap matches to the implementation basis, and the coders' own figures are reported beside it.
\item[A9] A codebook correction: a NaN test on the bit pattern counts as a construction only where the target's specification guarantees it. \crosstlr's 7 SPIR-V NaN tests, which read a value loaded as a float without \code{SignedZeroInfNanPreserve}, move to the implementation basis, and its 7 GLSL matches are reported as holding under the assumption that the artifact records.
\item[A10] A4 is headed ``before any run'', but it was written after the held-out kernels had been generated and translated, though before any execution result existed; the generator was also changed between a first generation of 34 kernels and the final one of 102, before any translation. A10 records this order.
\item[A11] The implementation departs from the registered intervention in four ways: it does not emit the registered SPIR-V \code{SignedZeroInfNanPreserve} mode for NaN-dependent rules, it implements none of the registered no-contraction rules, it carries an exact \code{isnan}, which the registration does not list, and its polyfills use a wider trusted base than the registered one.
\item[A12] Catalogue corrections found in review and checked against the specifications. MSL~4.1 leaves \code{\%} with a negative operand undefined (\S3.1, item~2), so MSL does not define the truncated remainder: \code{m\_srem\_sign} and the six held-out MSL remainder kernels carry no obligation, the rule table's native Metal entry for the remainder is unsound for the sources that define the rule, and the two matches of \code{u\_srem\_sign} through MSL's \code{\%} hold by implementation. Under Vulkan's rules SPIR-V's \code{NMin} defines no result for a NaN operand, so \code{s\_nmin\_nan} carries no obligation. MSL defines the arithmetic right shift; GLSL's \code{min} with a NaN gives an unspecified value; and the remedy for the SPIR-V NaN entries is \code{OpIsNan} under \code{SignedZeroInfNanPreserve}, not the bit-pattern test (\cref{sec:rulehub:impl}).
\item[A13] Found while encoding DXC's comparisons for the mechanical check of the rule table (\cref{app:tablecheck}). DXC marks every non-precise floating-point comparison with LLVM's fast-math flags unless the value is \code{precise} or \code{/Gis} is given, so under the current HLSL rules \crosstlr's HLSL polyfills for saturating conversion and for the NaN-ignoring \code{min} and \code{max} compare possibly infinite values without a defined result. Five of \crosstlr's HLSL matches, on cases whose inputs include an infinity, hold by implementation. No execution outcome changes, and the legacy reading is unaffected.
\end{description}

\paragraph{Coding protocol.}
The coders were separate instances of a large language model (Claude Opus~5.5), each given only the codebook, the specification notes and its packets (see AI Use). For the bases, the packets were split into three parts, and each part was coded by one informed and one blind instance, six in all. The registered coding of the mechanism of each non-preserved artifact used the same protocol, with four further instances: one informed and one blind for each half of the packets. For the refusal diagnostics of RQ1, a stratified sample of 706 of the 6,355 diagnostics that \crosstl gave on feasible cells of all 2,713 units, two blind instances first coded one half of the sample each, and a third then coded all of it, so that every sampled diagnostic has two blind codes. Disagreements were settled by the codebooks' rules, by the coordinating agent, and an author has yet to confirm the adjudications.
For the refusal diagnostics, the two blind codings agreed on 95\,\% of the sample (Cohen's $\kappa=0.92$, 95\,\% interval 0.89--0.94), but each agreed with the rule-based coder that labels a diagnostic from its message on only 66 and 67\,\% ($\kappa=0.54$ and $0.57$): they accepted its label ``semantic'' for only 10 and 22 of 128 such messages, and they agree on a meaning-preservation reason for 1.4\,\% of the population.
On the coders' codes alone, before A8, A9, A12 and A13, the supported shares of \cref{tab:bases} are 65\,\% for \crosstl, 82\,\% for \crosstlr and 100\,\% for glslang.
Two deviations from this protocol occurred. The instances shared a scratch directory, in which helper scripts of the same name collided; the coders' reports, and a re-check by one blind instance, found no reading of another coder's packets or codes. And one informed instance read source files of the experiment harness, which the protocol did not provide.
Four artifacts were longer than the 12,000 characters that a packet carried, and were coded from their visible part.

\FloatBarrier
\section{Ledger of Registered Items}
\label{app:ledger}

\Cref{tab:ledger} lists every registered hypothesis and prediction with its outcome. Figures are given as corrected by amendments A12 and A13, which the main text reports, and as registered; the artifact regenerates both (the registered figures with \code{HAZARD\_AS\_REGISTERED=1}).

\begin{table}[h]
\caption{Registered items, their outcomes under the current HLSL rules, and their status. \emph{Corrected}: with amendments A12 and A13; \emph{registered}: with the catalogue as registered.}
\label{tab:ledger}
\footnotesize
\setlength{\tabcolsep}{3pt}
\begin{tabularx}{\linewidth}{@{}>{\raggedright\arraybackslash}p{2.6cm}>{\raggedright\arraybackslash}X>{\raggedright\arraybackslash}p{2.5cm}>{\raggedright\arraybackslash}p{2.1cm}>{\raggedright\arraybackslash}p{1.6cm}@{}}
\toprule
Item & Registered prediction or plan & Corrected & Registered & Status \\
\midrule
Suite & 66 cases by the selection rule & 56 with an obligation & 58 & corrected (A12) \\
H-R1, outcomes & every validated \crosstlr output preserves the rule on every executed runtime & 5 divergences, all MoltenVK wrap & 5 & falsified \\
H-R1, bases & every match on the target's rule or a construction, except recorded assumptions & fails for 18 of 102 & 15 of 109 & falsified \\
H-R2 & a source-dependent (class, target) cell under \crosstl & 2 of 26 cells & 2 of 31 & holds, narrowly \\
H-R3 & no accepted share changes by more than 1 point & at most $-0.17$ points & same & holds \\
A4, \crosstlr & no held-out divergence except MoltenVK wrap & 0 of 462 executions & 0 of 492 & holds \\
A4, \crosstl & divergence on every target with a suite loss & SPIR-V only & same & failed \\
Factorial arms & four arms on the full suite & 37, 23, 5, 7 divergences & 39, 25, 5, 9 & reported \\
Configuration rule & changing matches count as configuration or implementation & applied (A8) & not applied & late (A8) \\
Blind coding & translator identity removed & \crosstlr's names visible in 75 of 225 packets & -- & deviation \\
Inventory, GPU time & table entries by kind; micro-kernels & \cref{sec:rulehub:impl,app:costs} & same & reported \\
\bottomrule
\end{tabularx}
\end{table}

\FloatBarrier
\section{Outcome Matrix, Held-Out Kernels and Drop-Freedom}
\label{app:hazard-matrix}

\Cref{tab:hazard-full} gives the outcome of every case on every runtime for \crosstl and \crosstlr. The third-party routes and the factorial arms are in the artifact.

\paragraph{Findings by class.}
\crosstl's SPIR-V loses contraction control under HLSL's \code{precise}, SPIR-V's \code{NoContraction}, CUDA's \code{\_\_fmul\_rn} and OpenCL's \code{fp contract(off)}, because the emitter omits \code{NoContraction} for these forms (it keeps it for GLSL's \code{precise}). The outputs match on lavapipe, which happens not to fuse, and diverge on MoltenVK, which fuses, whereas DXC's decorated SPIR-V is preserved on both.
Subnormals are kept by lavapipe and flushed by MoltenVK, so MoltenVK breaks CUDA's rule, which keeps them.
Signed zeros survive only in some configurations. Outside Metal's safe mode, MSL's own \code{fmod} returns $+0$ where C requires $-0$, and \crosstl's translation of CUDA's \code{atan2f}$(\pm0,\pm0)$ returns NaN; on llvmpipe and lavapipe the same translations return $3\pi/4$ where CUDA specifies $\pi$, because GLSL leaves \code{atan(0,0)} undefined.
C's \code{fmod} truncates, so its result has the sign of $x$, but \crosstl maps CUDA's \code{fmodf} to a floored remainder on every target, and MSL's \code{fmod} to a floored remainder on SPIR-V (\code{OpFMod}) and to an inexact formula on Direct3D~12.
\crosstl refuses recursion for GLSL and HLSL, and subgroup operations and pointer arithmetic for WGSL. Because its single-file translation of Metal, CUDA and OpenCL~C kernels uses a work-group size of one (S29), partial work-groups never arise on these routes, so the cases for them (H12) test nothing there. A recursive CUDA function hangs the GPU on Metal.

\paragraph{Configuration, runtime stack and bases.}
Because \crosstlr's MSL is byte-identical to \crosstl's in five of the 12 cells that change between Metal's configurations, they are 7 of 154 distinct artifacts. The 33 cells that differ between lavapipe and MoltenVK concern the wrap rule (15), contraction (10), accuracy (4), subnormals (2) and the sign of \code{atan2} at zero (2), and \crosstlr's SPIR-V is byte-identical to \crosstl's in 14 of them.
MoltenVK compiles SPIR-V with Metal's fast math by default. Forcing it off changed 4 of the 152 MoltenVK outcomes of the main routes, those of the two \code{sin} cases, which \crosstl and \crosstlr compile to the same binary, and 12 of 351 outcomes across all arms, from three binaries.
The 115 matches on the implementation basis are most often contraction that the runtime happened not to perform (35 matches), NaN handling in \code{min} and in conversions (29), accuracy (15), signed wrap that MoltenVK did not keep (13), shift counts (7) and subnormals (6).
For H-R2, the local fixes leave none of the 26 cells source-dependent. Of the 23 divergences that the registered control leaves, 5 are the MoltenVK failures that \crosstlr shares; the other 18 are narrowing to \code{uchar} (6), signed remainder (2) and saturating conversion (2) on SPIR-V, and signed wrap on Metal and Direct3D (8).

\paragraph{Defects in the suite.}
The SPIR-V emitter mistranslates \code{for (k = s; k < s + 4; ++k)}: the induction variable is initialised to 0 and only incremented, never tested, and the condition re-reads \code{s} from the buffer, so the loop does not terminate unless \code{s + 4} overflows (S3; \code{h\_loop\_wrap} hangs on both Vulkan implementations).
The SPIR-V emitter also replaces \code{fmin} and \code{isnan} by a constant, selects \code{OpFMod} where C's \code{fmod} needs \code{OpFRem}, and compiles \code{int < uint} as a signed comparison.
The HLSL and WGSL emitters reject or mistranslate several hub dialects: SPIR-V-sourced programs fail to HLSL and WGSL, and Metal- and CUDA-sourced ones fail to WGSL. The HLSL emitter, for example, turns the SPIR-V importer's struct-wrapped runtime array into a 4,096-byte structured-buffer element, which DXC rejects.

\paragraph{Held-out kernels.}
Each held-out kernel is a random expression over operations whose rule \crosstlr carries, generated, as registered, so that no operation is undefined in its source language; the six MSL remainder kernels violate this, because MSL leaves \code{\%} with a negative operand undefined, and carry no obligation (A12). Under the current HLSL rules, 84 kernels carry an obligation; the 12 HLSL kernels for wrap-around and division by zero carry one only under the legacy rules (\cref{app:readingA}), under which 96 kernels carry an obligation.
\crosstl lost mixed comparisons, \code{uchar} wrap and the truncated remainder on SPIR-V for every source language that defines them. The outcomes of neither variant depended on the source language: where \crosstl lost a rule, it lost it for every source that defines the rule. Because no kernel compares an overflowing expression, \crosstl preserved the wrap rule on every target.

{\scriptsize\setlength{\tabcolsep}{2.6pt}
\begin{longtable}{@{}l c cccccc cccccc@{}}
\caption{Outcomes of every hazard case on every runtime, for \crosstl and CrossTL-R (Metal: safe configuration). Cases marked $\dagger$ carry an obligation only under reading~A of HLSL, and do not count under reading~B, the primary reading; $\ddagger$ marks the negative control, which carries none. \emph{Ref}: the source run natively, as a check on its oracle; HLSL sources run through DXC, so their overflow and subnormal kernels show reading B; MoltenVK's wrap failure on the SPIR-V sources appears as \xmark. \cmark{} preserved, \xmark{} diverged, $\infty$ hang, i invalid, R refused, E translator error, e runtime error, c compiled only (CUDA and HIP references, which have no runtime here), -- not run, blank: no such route. lvp: lavapipe; MVK: MoltenVK; D3D: DXIL on lavapipe; GL: llvmpipe.}\label{tab:hazard-full}\\
\toprule
 & & \multicolumn{6}{c}{\crosstl} & \multicolumn{6}{c}{CrossTL-R} \\
\cmidrule(lr){3-8}\cmidrule(l){9-14}
Case & Ref & Metal & lvp & MVK & D3D & GL & WGSL & Metal & lvp & MVK & D3D & GL & WGSL \\
\midrule
\endfirsthead
\toprule
Case & Ref & Metal & lvp & MVK & D3D & GL & WGSL & Metal & lvp & MVK & D3D & GL & WGSL \\
\midrule
\endhead
\bottomrule
\endfoot
\code{g\_add\_wrap} & \cmark & \xmark & \cmark & \xmark & \xmark & \cmark & \cmark & \cmark & \cmark & \xmark & \cmark & \cmark & \cmark \\
\code{g\_mul\_wrap} & \cmark & \xmark & \cmark & \xmark & \xmark & \cmark & \cmark & \cmark & \cmark & \xmark & \cmark & \cmark & \cmark \\
\code{g\_loop\_wrap} & \cmark & \xmark & \cmark & \xmark & \xmark & \cmark & \cmark & \cmark & \cmark & \xmark & \cmark & \cmark & \cmark \\
\code{h\_add\_wrap}$^\dagger$ & \xmark & \xmark & \cmark & \xmark & \xmark & \cmark & \cmark & \cmark & \cmark & \xmark & \cmark & \cmark & \cmark \\
\code{h\_mul\_wrap}$^\dagger$ & \xmark & \xmark & \cmark & \xmark & \xmark & \cmark & \cmark & \cmark & \cmark & \xmark & \cmark & \cmark & \cmark \\
\code{h\_loop\_wrap}$^\dagger$ & \xmark & \xmark & $\infty$ & $\infty$ & \xmark & \cmark & \cmark & \cmark & $\infty$ & $\infty$ & \cmark & \cmark & \cmark \\
\code{s\_add\_wrap} & \xmark & \xmark & \cmark & \xmark &  & \cmark & i & \cmark & \cmark & \xmark &  & \cmark & i \\
\code{s\_mul\_wrap} & \xmark & \xmark & \cmark & \xmark &  & \cmark & i & \cmark & \cmark & \xmark &  & \cmark & i \\
\code{h\_shift\_mask} & \cmark & \cmark & \cmark & \cmark & \cmark & \cmark & \cmark & \cmark & \cmark & \cmark & \cmark & \cmark & \cmark \\
\code{m\_shift\_mask} & \cmark & \cmark & \cmark & \cmark & \cmark & \cmark & i & \cmark & \cmark & \cmark & \cmark & \cmark & i \\
\code{c\_shift\_mask} & \cmark & \cmark & \cmark & \cmark & \cmark & \cmark & \cmark & \cmark & \cmark & \cmark & \cmark & \cmark & \cmark \\
\code{h\_udiv\_zero}$^\dagger$ & \cmark & \xmark & \cmark & \xmark & \cmark & \cmark & \xmark & \cmark & \cmark & \cmark & \cmark & \cmark & \cmark \\
\code{m\_cmp\_mixed} & \cmark & \cmark & \xmark & \xmark & \cmark & \cmark & i & \cmark & \cmark & \cmark & \cmark & \cmark & i \\
\code{g\_cmp\_mixed} & \cmark & \cmark & \xmark & \xmark & \cmark & \cmark & i & \cmark & \cmark & \cmark & \cmark & \cmark & \cmark \\
\code{h\_cmp\_mixed} & \cmark & \cmark & \xmark & \xmark & \cmark & \cmark & i & \cmark & \cmark & \cmark & \cmark & \cmark & \cmark \\
\code{u\_cmp\_mixed} & c & \cmark & \xmark & \xmark & \cmark & \cmark & i & \cmark & \cmark & \cmark & \cmark & \cmark & i \\
\code{c\_cmp\_mixed} & \cmark & \cmark & \xmark & \xmark & \cmark & \cmark & i & \cmark & \cmark & \cmark & \cmark & \cmark & \cmark \\
\code{m\_uchar\_wrap} & \cmark & \cmark & \xmark & \xmark & \cmark & \cmark & i & \cmark & \cmark & \cmark & \cmark & \cmark & i \\
\code{u\_uchar\_wrap} & c & \cmark & \xmark & \xmark & \cmark & \cmark & i & \cmark & \cmark & \cmark & \cmark & \cmark & i \\
\code{c\_uchar\_wrap} & \cmark & \cmark & \xmark & \xmark & \cmark & \cmark & i & \cmark & \cmark & \cmark & \cmark & \cmark & i \\
\code{m\_ulong\_mul} & \cmark & \cmark & \cmark & \cmark & \cmark & \cmark & i & \cmark & \cmark & \cmark & \cmark & \cmark & i \\
\code{u\_u64\_mul} & c & i & \cmark & \cmark &  & \cmark & i & i & \cmark & \cmark &  & \cmark & i \\
\code{u\_srem\_sign} & c & \cmark & \xmark & \xmark & \cmark & \cmark & i & \cmark & \cmark & \cmark & \cmark & \cmark & i \\
\code{m\_srem\_sign}$^\ddagger$ & \cmark & \cmark & \xmark & \xmark & \cmark & \cmark & i & \cmark & \cmark & \cmark & \cmark & \cmark & i \\
\code{h\_ftoi\_sat}$^\dagger$ & \cmark & \cmark & \cmark & \cmark & \cmark & \cmark & \xmark & \cmark & \cmark & \cmark & \cmark & \cmark & R \\
\code{m\_ftoi\_nan} & \cmark & \cmark & \cmark & \cmark & \cmark & \cmark & i & \cmark & \cmark & \cmark & \cmark & \cmark & R \\
\code{c\_convert\_sat} & \cmark & i & \xmark & \xmark &  & i & i & \cmark & \cmark & \cmark & \cmark & \cmark & R \\
\code{h\_min\_nan} & \cmark & \cmark & \cmark & \cmark & \cmark & \cmark & \cmark & \cmark & \cmark & \cmark & \cmark & \cmark & R \\
\code{m\_fmin\_nan} & \cmark & \cmark & \xmark & \xmark & \cmark & \cmark & i & \cmark & \cmark & \cmark & \cmark & \cmark & R \\
\code{u\_fminf\_nan} & c & \cmark & \cmark & \cmark & \cmark & \cmark & i & \cmark & \cmark & \cmark & \cmark & \cmark & R \\
\code{c\_fmin\_nan} & \cmark & \cmark & \cmark & \cmark & \cmark & \cmark & \cmark & \cmark & \cmark & \cmark & \cmark & \cmark & R \\
\code{s\_nmin\_nan}$^\ddagger$ & \cmark & \cmark & \cmark & \cmark &  & \cmark & i & \cmark & \cmark & \cmark &  & \cmark & i \\
\code{g\_precise} & \cmark & \cmark & \cmark & \cmark & \cmark & \cmark & \xmark & \cmark & \cmark & \cmark & \cmark & \cmark & \xmark \\
\code{h\_precise} & \cmark & \cmark & \cmark & \xmark & \cmark & \cmark & \xmark & \cmark & \cmark & \xmark & \cmark & \cmark & \xmark \\
\code{m\_contract\_off} & \cmark & \cmark & \cmark & \xmark & \cmark & \cmark & i & \cmark & \cmark & \xmark & \cmark & \cmark & i \\
\code{s\_nocontraction} & \cmark & \xmark & \cmark & \xmark &  & \cmark & i & \xmark & \cmark & \xmark &  & \cmark & i \\
\code{u\_fmul\_rn} & c & \xmark & \cmark & \xmark & \cmark & \cmark & i & \xmark & \cmark & \xmark & \cmark & \cmark & i \\
\code{c\_fp\_contract\_off} & \cmark & \xmark & \cmark & \xmark & \cmark & \cmark & \xmark & \xmark & \cmark & \xmark & \cmark & \cmark & \xmark \\
\code{m\_precise\_sin} & \cmark & \cmark & \cmark & \cmark & \cmark & \cmark & i & \cmark & \cmark & \cmark & \cmark & \cmark & i \\
\code{m\_atan2} & \cmark & \cmark & \cmark & \cmark & \cmark & \cmark & i & \cmark & \cmark & \cmark & \cmark & \cmark & i \\
\code{u\_sinf} & c & \cmark & \cmark & \xmark & \cmark & \cmark & i & \cmark & \cmark & \xmark & \cmark & \cmark & i \\
\code{c\_sin} & \cmark & \cmark & \cmark & \xmark & \cmark & \cmark & \xmark & \cmark & \cmark & \xmark & \cmark & \cmark & \xmark \\
\code{u\_atan2f\_zero} & c & \cmark & \xmark & \cmark & \cmark & \xmark & i & \cmark & \xmark & \cmark & \cmark & \xmark & i \\
\code{u\_denorm} & c & \xmark & \cmark & \xmark & \cmark & \cmark & i & \xmark & \cmark & \xmark & \cmark & \cmark & i \\
\code{h\_denorm}$^\dagger$ & \xmark & \cmark & \xmark & \cmark & \xmark & \xmark & \cmark & \cmark & \xmark & \cmark & \xmark & \xmark & \cmark \\
\code{m\_isnan} & \cmark & \cmark & \xmark & \xmark & \cmark & \cmark & i & \cmark & \cmark & \cmark & \cmark & \cmark & R \\
\code{g\_isnan}$^\ddagger$ & \cmark & \cmark & \xmark & \xmark & \cmark & \cmark & i & \cmark & \xmark & \xmark & \cmark & \cmark & i \\
\code{m\_fmod\_sign} & \cmark & \cmark & \xmark & \xmark & \xmark & \cmark & i & \cmark & \xmark & \xmark & \xmark & \cmark & i \\
\code{u\_fmodf\_sign} & c & \xmark & \xmark & \xmark & \xmark & \xmark & i & \xmark & \xmark & \xmark & \xmark & \xmark & i \\
\code{m\_half\_mul} & \cmark & \cmark & \xmark & \xmark & \cmark & \cmark & i & \cmark & \xmark & \xmark & \cmark & \cmark & i \\
\code{m\_packed\_float3} & \cmark & \xmark & \xmark & \xmark & \cmark & \xmark & i & \xmark & \xmark & \xmark & \cmark & \xmark & i \\
\code{h\_float3\_stride} & \cmark & \xmark & \xmark & \xmark & \cmark & \xmark & \xmark & \xmark & \xmark & \xmark & \cmark & \xmark & \xmark \\
\code{u\_float3\_stride} & c & \xmark & \xmark & \xmark & \cmark & \xmark & i & \xmark & \xmark & \xmark & \cmark & \xmark & i \\
\code{g\_std430\_vec3\_float} & \cmark & \cmark & \cmark & \cmark & \cmark & \cmark & \cmark & \cmark & \cmark & \cmark & \cmark & \cmark & \cmark \\
\code{h\_oob\_read}$^\dagger$ & \cmark & \cmark & \cmark & \cmark & \cmark & \cmark & \xmark & \cmark & \cmark & \cmark & \cmark & \cmark & \xmark \\
\code{u\_ptr\_arith} & c & i & \cmark & \cmark &  & i & R & i & \cmark & \cmark &  & i & R \\
\code{m\_ptr\_arith} & \cmark & \cmark & \cmark & \cmark & \cmark & \cmark & R & \cmark & \cmark & \cmark & \cmark & \cmark & R \\
\code{c\_ptr\_arith} & \cmark & i & \cmark & \cmark &  & i & R & i & \cmark & \cmark &  & i & R \\
\code{m\_partial\_tg} & \cmark & \cmark & \cmark & \cmark & \cmark & \cmark & i & \cmark & \cmark & \cmark & \cmark & \cmark & i \\
\code{c\_partial\_wg} & \cmark & \cmark & \cmark & \cmark & \cmark & \cmark & \cmark & \cmark & \cmark & \cmark & \cmark & \cmark & \cmark \\
\code{u\_warp\_reduce} & c & \cmark & \xmark & \xmark & \xmark & -- & R & \cmark & \xmark & \xmark & \xmark & -- & R \\
\code{m\_simd\_sum} & \cmark & \cmark & \xmark & \xmark & \xmark & -- & R & \cmark & \xmark & \xmark & \xmark & -- & R \\
\code{h\_wave\_sum} & \cmark & \cmark & \cmark & \cmark & \cmark & -- & R & \cmark & \cmark & \cmark & \cmark & -- & R \\
\code{m\_simd\_divergent} & \cmark & \cmark & \xmark & \xmark & \xmark & -- & R & \cmark & \xmark & \xmark & \xmark & -- & R \\
\code{u\_recursion} & c & $\infty$ & i & i & R & R & i & $\infty$ & i & i & R & R & i \\
\code{u\_atomic\_add} & c & \cmark & \cmark & \cmark & \cmark & \cmark & i & \cmark & \cmark & \cmark & \cmark & \cmark & i \\
\end{longtable}
}

\begin{table}[h]
\caption{Held-out random kernels with an obligation under the current HLSL rules (6 per family and source, generated from a fixed seed, 20261017): executions and divergences on every runtime of the primary configuration. \emph{Executed} excludes outputs that a target compiler rejected, and CUDA and HIP outputs, which are only compiled.}
\label{tab:heldout}
\small
\begin{tabular}{@{}llrrrr@{}}
\toprule
 & & \multicolumn{2}{c}{\crosstl} & \multicolumn{2}{c}{CrossTL-R} \\
\cmidrule(lr){3-4}\cmidrule(l){5-6}
Family & Source & Executed & Diverged & Executed & Diverged \\
\midrule
wrap $+,-,\times$ & GLSL & 36 & 0 & 36 & 0 \\
mixed compare & GLSL & 30 & 12 & 36 & 0 \\
 & HLSL & 30 & 12 & 36 & 0 \\
 & MSL & 30 & 12 & 30 & 0 \\
 & CUDA & 30 & 12 & 30 & 0 \\
 & OpenCL & 30 & 12 & 36 & 0 \\
shift mask & HLSL & 36 & 0 & 36 & 0 \\
 & MSL & 30 & 0 & 30 & 0 \\
 & OpenCL & 36 & 0 & 36 & 0 \\
truncated \code{\%} & CUDA & 30 & 12 & 30 & 0 \\
 & OpenCL & 36 & 12 & 36 & 0 \\
\code{uchar} wrap & MSL & 30 & 12 & 30 & 0 \\
 & CUDA & 30 & 12 & 30 & 0 \\
 & OpenCL & 30 & 12 & 30 & 0 \\
\midrule
Total & & 444 & 120 & 462 & 0 \\
\bottomrule
\end{tabular}

\end{table}

\begin{table}[h]
\caption{Drop-freedom of \crosstl and \crosstlr on the hazard suite, per class and target (primary configuration, current HLSL rules; both Vulkan implementations count for SPIR-V). Verdicts combine the outcomes with the two codings of \cref{sec:eval:hazards}. \cmark: no case dropped its obligation, and at least one was preserved on the target's rule or by an explicit construct. \cmark$^{s}$: as \cmark, but a case also diverged or hung through the runtime stack. \xmark: some case diverged or hung through a fallback, a wrong construct or a dropped obligation. $\ell$: every drop is latent, a match on an unrecorded configuration or on the particular implementation. \textsf{R}: refused, with any other case of the class invalid. \textsf{inv}: every output was rejected by the target compiler. --: the runtime could not run the output. \textsf{vac}: the class cannot arise under single-file translation (partial work-groups, H12), so the cell tests nothing. Classes as in \cref{tab:hazard-matrix,tab:hazard-matrix-env}; H8/10 is a software \code{atan2}. A \cmark{} is evidence from one to eight kernels, not a proof of drop-freedom.}
\label{tab:dropfree}
\scriptsize
\setlength{\tabcolsep}{3.2pt}
\begin{tabular}{@{}lcccccccccc@{}}
\toprule
 & \multicolumn{5}{c}{\crosstl} & \multicolumn{5}{c}{CrossTL-R} \\
\cmidrule(lr){2-6}\cmidrule(l){7-11}
Class & Metal & SPIR-V & HLSL & GLSL & WGSL & Metal & SPIR-V & HLSL & GLSL & WGSL \\
\midrule
H1 & \xmark & \cmark$^{s}$ & \xmark & \cmark & \cmark & \cmark & \cmark$^{s}$ & \cmark & \cmark & \cmark \\
H2 & \cmark & $\ell$ & \cmark & $\ell$ & \cmark & \cmark & \cmark & \cmark & \cmark & \cmark \\
H3 & $\ell$ & \xmark & $\ell$ & \cmark & \textsf{inv} & $\ell$ & \cmark & \cmark & \cmark & \textsf{inv} \\
H4 & \cmark & \xmark & \cmark & \cmark & \textsf{inv} & \cmark & \cmark & \cmark & \cmark & \cmark \\
H5 & $\ell$ & \xmark & $\ell$ & $\ell$ & \textsf{inv} & \cmark & $\ell$ & $\ell$ & \cmark & \textsf{R} \\
H6 & $\ell$ & \xmark & \cmark & $\ell$ & $\ell$ & \cmark & $\ell$ & $\ell$ & \cmark & \textsf{R} \\
H7 & \xmark & \xmark & $\ell$ & $\ell$ & \xmark & \xmark & \xmark & $\ell$ & $\ell$ & \xmark \\
H8 & $\ell$ & \xmark & $\ell$ & $\ell$ & \xmark & $\ell$ & \xmark & $\ell$ & $\ell$ & \xmark \\
H9 & \xmark & \xmark & $\ell$ & $\ell$ & \textsf{inv} & \xmark & \xmark & $\ell$ & $\ell$ & \textsf{inv} \\
H10 & $\ell$ & \xmark & \cmark & \xmark & \textsf{inv} & $\ell$ & \xmark & \cmark & \xmark & \textsf{inv} \\
H12 & \textsf{vac} & \textsf{vac} & \textsf{vac} & \textsf{vac} & \textsf{vac} & \textsf{vac} & \textsf{vac} & \textsf{vac} & \textsf{vac} & \textsf{vac} \\
H13 & $\ell$ & \xmark & \xmark & -- & \textsf{R} & $\ell$ & \xmark & \xmark & -- & \textsf{R} \\
H14 & \cmark & \xmark & \xmark & -- & \textsf{R} & \cmark & \xmark & \xmark & -- & \textsf{R} \\
H15 & $\ell$ & \xmark & \cmark & \cmark & \textsf{inv} & $\ell$ & \xmark & \cmark & \cmark & \textsf{inv} \\
H16 & \xmark & \xmark & \cmark & \xmark & \xmark & \xmark & \xmark & \cmark & \xmark & \xmark \\
H17 & \cmark & \cmark & \cmark & \cmark & \textsf{R} & \cmark & \cmark & \cmark & \cmark & \textsf{R} \\
H18 & \cmark$^{s}$ & \textsf{inv} & \textsf{R} & \textsf{R} & \textsf{inv} & \cmark$^{s}$ & \textsf{inv} & \textsf{R} & \textsf{R} & \textsf{inv} \\
fmod & \xmark & \xmark & \xmark & \xmark & \textsf{inv} & \xmark & \xmark & \xmark & \xmark & \textsf{inv} \\
isnan & $\ell$ & \xmark & \cmark & $\ell$ & \textsf{inv} & \cmark & $\ell$ & \cmark & \cmark & \textsf{R} \\
H8/10 & \cmark & \cmark & \cmark & \cmark & \textsf{inv} & \cmark & $\ell$ & \cmark & \cmark & \textsf{inv} \\
atomics & \cmark & \cmark & \cmark & \cmark & \textsf{inv} & \cmark & \cmark & \cmark & \cmark & \textsf{inv} \\
\bottomrule
\end{tabular}

\end{table}

In two of each variant's drop-free cells the only divergences are the runtime stack's: signed wrap on SPIR-V, which MoltenVK did not keep, and recursion on Metal. The registered rule A8, which moves the MoltenVK wrap matches to the implementation basis, is not applied in \cref{tab:dropfree}: it rests on the runtime stack's divergence, and with it signed wrap on SPIR-V would be a latent drop for both variants. Read from the outcomes alone, without the codings, the table would credit \crosstl with 55 drop-free cells and \crosstlr with 62.
\crosstl is drop-free or refuses on every target in only two classes: pointers (H17), and partial work-groups (H12), which single-file translation avoids by using work-groups of one, so that this cell says nothing about how \crosstl handles them. \crosstlr adds signed wrap (H1), shift ranges (H2) and mixed comparisons (H4). Its NaN tests on SPIR-V and its comparisons of infinite values on HLSL keep float-to-integer conversion (H5), the NaN-ignoring \code{min} (H6) and \code{isnan} out of the set.

\FloatBarrier
\section{Runtime Probes}
\label{app:probes}

\paragraph{KosmicKrisp.}
After the registered runs, we executed on KosmicKrisp every SPIR-V artifact that had run on MoltenVK: 356 of the suite and 339 of the held-out test, with the suite's harness and oracles (LunarG Vulkan SDK 1.4.363.0, driver 26.2.99, reporting conformance version 1.4.3.2; Khronos lists KosmicKrisp as conformant to Vulkan~1.4 on Apple silicon, submission 997~\citep{khronos2026conformant}). The binaries are byte-identical to those of the suite, and a MoltenVK control in the same session reproduced all 695 recorded outcomes. Over the 55 signed-wrap cells of all SPIR-V routes, MoltenVK preserves 7, diverges in 43 and hangs in 5 (\crosstl's non-terminating loop); KosmicKrisp preserves 21, diverges in 29 and hangs in 5; lavapipe preserves 50 and hangs in 5. KosmicKrisp diverges on every SPIR-V route of the four addition cases, the suite's reference SPIR-V and glslang's and DXC's output included, and on DXC's loop case; none of these binaries carries \code{NoSignedWrap}. The MSL that it generates computes the addition in a signed \code{int} temporary (\code{t28 = t24 + int(1);}), so Apple's compiler may fold \code{x + 1 > x}; the stack-keyed polyfill's unsigned addition is printed the same way. Its 21 multiplication cells match because NIR rewrites \code{x * 2} and the division by 2 into shifts before printing, not because the MSL guarantees the rule.
Outside the wrap rule, KosmicKrisp differs from MoltenVK only in \code{u\_atan2f\_zero}, where it follows Mesa's lowering of SPIR-V's \code{Atan2}, which is undefined at $(0,0)$, as lavapipe does; it differs from lavapipe only where SPIR-V leaves the result open: contraction without \code{NoContraction}, \code{sin} accuracy, unsigned division by zero and subnormals. On the held-out kernels, KosmicKrisp matches MoltenVK in all 339 executions.
Khronos lists lavapipe as conformant to Vulkan~1.3 for an earlier Mesa release, and the Mesa release we ran reports Vulkan~1.4 without a conformance entry; MoltenVK is not listed~\citep{khronos2026conformant}.

\paragraph{Metal's math-mode pragma.}
Nine one-statement kernels ran under the API default, under explicit fast math and under safe math, each without \code{\#pragma METAL fp math\_mode(safe)}, with it at file scope, where \crosstlr emits it, and with it in a block. Under both fast configurations without the pragma, four kernels give results that IEEE~754 forbids, because the compiler folds them: \code{x != x} to false, \code{x * 0.0f} and \code{x - x} to zero, and \code{x + 0.0f} to \code{x}. With the pragma at either scope, all four give the IEEE results; \code{isnan}, \code{isinf}, an unordered comparison, \code{fmin} with a NaN and $1/x$ showed no fast-math effect. The compiler rejects an unknown mode, so it parses the pragma rather than ignoring it.

\FloatBarrier
\section{The Corpus}
\label{app:corpora}

The 22 slices come from 12 projects, each pinned to an upstream revision and permissively licensed (\cref{tab:corpora}). The 2,713 units comprise 654 GLSL, 429 HLSL, 331 Slang, 346 SPIR-V, 240 Metal, 190 CUDA, 108 HIP, 234 OpenCL~C, 150 Rust-GPU and 31 Mojo files. Of the 17,567 feasible cells for validated targets, 8 could not be validated, so the shares are over 17,559. \crosstl imports GLSL, Slang, SPIR-V and Rust-GPU almost completely (98 to 99\,\%), but Metal only at 35\,\% and none of the Mojo puzzles, and nvcc accepts 14 of 355 CUDA and 13 of 349 HIP outputs, mostly because hub type names and source built-ins pass through unchanged. On the hazard cases that each route accepts, SPIRV-Cross diverges in 10 of 27 executions and Slang in 4 of 16, against 18 of 47 and 9 of 33 for \crosstl on the same cases. Of the 6,986 cells that ask a target for a stage it cannot express, \crosstl refused 97\,\% of those for GLSL~ES but 14\,\% for CUDA and HIP, and validators accepted 87 of the artifacts it emitted for the others, such as 38 Metal outputs for ray-tracing stages. The SPIR-V slice is the GLSL slice compiled by glslang, because public SPIR-V is distributed as binaries.
Of the 2,713 units, 128 did not compile with their project's own build. Most of the rejections are of two kinds: some of ggml's OpenCL kernels (68) expect compile-time constants that ggml's host code supplies per device, such as the SIMD width, or use vendor extensions; HeCBench's HIP benchmarks (39) need ROCm libraries or intrinsics that HIP's NVIDIA platform lacks.
The 2,585 translated units form 719 program families (794 among all 2,713 units). In 8 of the 22 slices we found no grouping above the file, so each unit there is its own family and the intervals treat those files as independent.
The harness never attempted 764 cells: the Apple, Candle and MLX Metal programs to Slang, Mojo, GLSL~ES and Rust, the MLX kernels also to CUDA and HIP, and 50 such cells of the PyTorch programs. The harness began to skip these targets for Metal sources during the run, so the ggml programs and the other PyTorch cells were translated.
56 feasible cells and 7 imports timed out, all from Metal programs; we did not record the host's load during the run, so time-outs near the limit may partly reflect contention.
For WGSL, naga and Tint disagree on 40 outputs: naga accepts 10 that Tint rejects, and rejects 30 that Tint accepts. The outcomes use naga's verdict.
Slang's HLSL output names its entry points after the Slang source (\code{vertexMain}, \ldots) without stage attributes. Our first analysis recognised entry points by name, compiled these outputs as libraries, and accepted 290 of 328 (88\,\%) before its entry-point check failed them. Compiled per entry point with the stage of the source's \code{[shader]} attribute, as every other tool's HLSL is, 41 pass (12.5\,\%); most of the others lack the semantics that DXC requires on every input and output of a stage's interface. \Cref{tab:baselines} uses the per-entry verdicts, which change no comparison's significance. For the Vulkan-Samples Slang units, the Slang chain passed the samples' own profile (\code{spirv\_1\_4}) for every target.

\begin{table}[h]
\caption{Success rates (\%) of \crosstl and of each third-party chain on the same units, per (source, target) pair. $\dagger$: the difference from \crosstl is significant (sign-flip test over program families, Holm's correction over 39 comparisons, $\alpha=0.05$). --: the chain does not offer the pair.}
\label{tab:baselines}
\footnotesize
\setlength{\tabcolsep}{4pt}
\begin{tabular}{@{}ll r rrrrrr@{}}
\toprule
Source & Target & Units & \crosstl & SPIRV-Cross & naga (SPIR-V) & naga (GLSL) & Tint & Slang \\
\midrule
GLSL & Metal & 654 & 58 & 87$^\dagger$ & 46$^\dagger$ & 30$^\dagger$ & 71$^\dagger$ & -- \\
GLSL & HLSL & 654 & 65 & 82$^\dagger$ & 40$^\dagger$ & 27$^\dagger$ & 71$^\dagger$ & -- \\
GLSL & GLSL & 654 & 69 & 81$^\dagger$ & -- & -- & 70 & -- \\
GLSL & WGSL & 654 & 28 & -- & 52$^\dagger$ & 32 & 71$^\dagger$ & -- \\
HLSL & Metal & 429 & 24 & 89$^\dagger$ & 55$^\dagger$ & -- & 78$^\dagger$ & 53$^\dagger$ \\
HLSL & GLSL & 429 & 37 & 88$^\dagger$ & -- & -- & 76$^\dagger$ & 93$^\dagger$ \\
HLSL & WGSL & 429 & 20 & -- & 58$^\dagger$ & -- & 78$^\dagger$ & 29$^\dagger$ \\
HLSL & SPIR-V & 429 & 31 & -- & -- & -- & -- & 95$^\dagger$ \\
Slang & Metal & 328 & 20 & -- & -- & -- & -- & 64$^\dagger$ \\
Slang & HLSL & 328 & 66 & -- & -- & -- & -- & 12$^\dagger$ \\
Slang & GLSL & 328 & 35 & -- & -- & -- & -- & 95$^\dagger$ \\
Slang & WGSL & 328 & 12 & -- & -- & -- & -- & 63$^\dagger$ \\
Slang & SPIR-V & 328 & 26 & -- & -- & -- & -- & 100$^\dagger$ \\
SPIR-V & Metal & 344 & 44 & 88$^\dagger$ & 50$^\dagger$ & -- & 75$^\dagger$ & -- \\
SPIR-V & HLSL & 344 & 16 & 83$^\dagger$ & 44$^\dagger$ & -- & 75$^\dagger$ & -- \\
SPIR-V & GLSL & 344 & 67 & 87$^\dagger$ & -- & -- & 74$^\dagger$ & -- \\
SPIR-V & WGSL & 344 & 63 & -- & 56$^\dagger$ & -- & 75$^\dagger$ & -- \\
\bottomrule
\end{tabular}

\end{table}

\begin{figure}[h]
\centering
\includegraphics[width=0.86\linewidth]{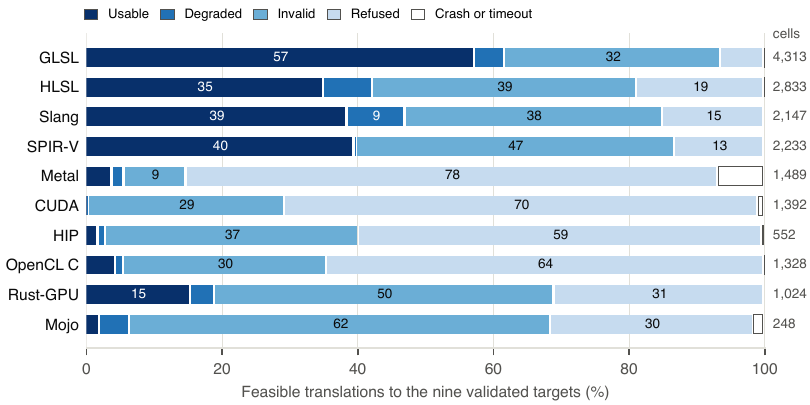}
\caption{Outcomes of \crosstl's translations of third-party programs by source language, over feasible (unit, target) cells for the nine targets with a validator (Metal, HLSL, GLSL, SPIR-V, WGSL, GLSL~ES, Slang, CUDA, HIP). \emph{Crash or timeout} counts as refused in the tables. Numbers are percentages; the right column gives the number of cells.}
\label{fig:reach}
\Description{Horizontal stacked bars per source language. GLSL has 57 percent usable outputs, HLSL 35, Slang 39, SPIR-V 40 and Rust-GPU 15; Metal, CUDA, HIP, OpenCL C and Mojo have under 5 percent usable, with most cells refused or invalid.}
\end{figure}
Whole-file translation of MLX's own kernel files yields 4 usable cells out of 245: \crosstl refuses 163 and times out on 54, because the MLX integration works per kernel family, with configuration that the files themselves do not state.

\begin{table}[h]
\caption{Validity by target, over feasible cells. \emph{Output}: \crosstl produced an artifact. \emph{Accepted}: the validator accepted it. \emph{Fallback}: accepted, but the raw emitter output contains a fallback marker. \emph{Usable}: accepted, without a fallback and with an entry point. The interval is a 95\,\% cluster bootstrap over program families. Rust and Mojo outputs could not be validated on our host.}
\label{tab:reach-target}
\small
\setlength{\tabcolsep}{4pt}
\begin{tabular}{@{}llrrrrrrl@{}}
\toprule
Target & Validator & Cells & Output & Accepted & Fallback & Usable & \% & 95\,\% CI (families) \\
\midrule
Metal & Metal compiler & 2,358 & 1,760 & 772 & 42 & 727 & 30.8 & 28.2--33.3 \\
HLSL & DXC & 2,585 & 1,846 & 956 & 14 & 942 & 36.4 & 33.4--39.4 \\
GLSL & glslang & 2,419 & 1,581 & 986 & 0 & 984 & 40.7 & 37.5--43.5 \\
SPIR-V & spirv-val & 2,585 & 1,901 & 1,489 & 603 & 886 & 34.3 & 31.3--37.1 \\
WGSL & naga, Tint & 2,343 & 1,274 & 530 & 0 & 526 & 22.4 & 20.0--24.7 \\
GLSL~ES & glslang (ES) & 1,579 & 1,324 & 796 & 0 & 796 & 50.4 & 47.1--53.8 \\
Slang & slangc & 2,418 & 1,916 & 594 & 12 & 582 & 24.1 & 21.7--26.3 \\
CUDA & nvcc & 640 & 355 & 14 & 0 & 12 & 1.9 & 0.7--3.1 \\
HIP & nvcc & 640 & 349 & 13 & 0 & 11 & 1.7 & 0.6--2.9 \\
Rust & none & 2,419 & 1,878 & -- & -- & -- & -- & \\
Mojo & none & 597 & 286 & -- & -- & -- & -- & \\
\bottomrule
\end{tabular}

\end{table}

\begin{table}[h]
\caption{Evaluation corpora. Units are translation units with an entry point in their include closure. \emph{Native}: units that the project's own build command accepted (Rust-GPU and Mojo: no native check). \emph{Seen}: used during \crosstl's development. HeCBench: the same 100 benchmarks, drawn with a fixed seed from the 459 that exist in both CUDA and HIP. The artifact rebuilds each slice from its pinned upstream revision and checks every file against a recorded hash.}
\label{tab:corpora}
\footnotesize
\setlength{\tabcolsep}{4pt}
\begin{tabular}{@{}llrrlll@{}}
\toprule
Language & Project & Units & Native & Commit & Licence & Note \\
\midrule
GLSL & Willems Vulkan examples & 347 & 347 & \code{4ee2501c} & MIT &  \\
GLSL & Khronos Vulkan-Samples & 307 & 307 & \code{177edebf} & Apache-2.0 &  \\
\addlinespace[2pt]
HLSL & Willems Vulkan examples & 308 & 308 & \code{4ee2501c} & MIT &  \\
HLSL & Khronos Vulkan-Samples & 121 & 121 & \code{177edebf} & Apache-2.0 &  \\
\addlinespace[2pt]
Slang & Willems Vulkan examples & 173 & 170 & \code{4ee2501c} & MIT &  \\
Slang & Khronos Vulkan-Samples & 158 & 158 & \code{177edebf} & Apache-2.0 &  \\
\addlinespace[2pt]
SPIR-V asm. & Willems Vulkan examples & 346 & 344 & \code{4ee2501c} & MIT & compiled from GLSL \\
\addlinespace[2pt]
Metal & llama.cpp / ggml & 36 & 36 & \code{f3363593} & MIT &  \\
Metal & PyTorch (MPS back end) & 50 & 50 & \code{1e233a2e} & BSD-3-Clause &  \\
Metal & Hugging Face Candle & 16 & 16 & \code{12bb5d1f} & MIT/Apache-2.0 &  \\
Metal & Apple Metal sample code & 89 & 89 & \code{0bc50e5b} & MIT &  \\
Metal & Apple MLX & 49 & 49 & \code{9c3d3557} & MIT & seen \\
\addlinespace[2pt]
CUDA & Hugging Face Candle & 27 & 27 & \code{12bb5d1f} & MIT/Apache-2.0 &  \\
CUDA & PolyBench/ACC & 21 & 21 & \code{70ea4ca9} & BSD-style &  \\
CUDA & Rodinia & 35 & 24 & \code{9c10d3ea} & BSD-style &  \\
CUDA & HeCBench & 107 & 102 & \code{f0c817aa} & BSD-3-Clause & sample of 459 \\
\addlinespace[2pt]
HIP & HeCBench & 108 & 69 & \code{f0c817aa} & BSD-3-Clause & sample of 459 \\
\addlinespace[2pt]
OpenCL~C & llama.cpp / ggml & 182 & 114 & \code{f3363593} & MIT &  \\
OpenCL~C & PolyBench/ACC & 21 & 21 & \code{70ea4ca9} & BSD-style &  \\
OpenCL~C & Rodinia & 31 & 31 & \code{9c10d3ea} & BSD-style &  \\
\addlinespace[2pt]
Rust-GPU & Rust-GPU Vulkan examples & 150 & -- & \code{b29a37eb} & MIT &  \\
\addlinespace[2pt]
Mojo & Mojo GPU puzzles & 31 & -- & \code{d360bc4c} & Apache-2.0 & seen \\
\midrule
Total & 22 slices & 2,713 & 2,404 & & & \\
\bottomrule
\end{tabular}

\end{table}

\begin{table}[h]
\caption{Reach by source language. \emph{Strict import}: the importer's \crossgl parses with the strict front end and declares an entry point; the interval is a 95\,\% cluster bootstrap over program families. The outcome shares are over feasible cells for the nine validated targets. \emph{Refused} includes crashes and time-outs (1.9 and 0.9\,\% of all stops). $^{w}$: where the bootstrap interval reaches 0 or 100\,\%, the Wilson interval is given instead.}
\label{tab:reach-source}
\small
\begin{tabular}{@{}lrrrrrrrr@{}}
\toprule
 & & \multicolumn{2}{c}{Strict import} & & \multicolumn{4}{c}{Feasible cells, validated targets (\%)} \\
\cmidrule(lr){3-4}\cmidrule(l){6-9}
Source & Units & \% & 95\,\% CI & Cells & Usable & Degr. & Invalid & Refused \\
\midrule
GLSL & 654 & 99.4 & 98.4--99.8$^{w}$ & 4,313 & 57.3 & 4.4 & 31.9 & 6.4 \\
HLSL & 429 & 65.5 & 62.5--68.5 & 2,833 & 35.1 & 7.2 & 38.9 & 18.8 \\
Slang & 328 & 99.4 & 97.8--99.8$^{w}$ & 2,147 & 38.5 & 8.5 & 38.0 & 15.0 \\
SPIR-V & 344 & 98.8 & 97.0--99.5$^{w}$ & 2,233 & 39.5 & 0.3 & 46.9 & 13.3 \\
Metal & 240 & 35.4 & 29.6--41.2 & 1,489 & 3.9 & 1.7 & 9.1 & 85.2 \\
CUDA & 174 & 60.3 & 51.6--69.2 & 1,392 & 0.1 & 0.2 & 29.0 & 70.7 \\
HIP & 69 & 58.0 & 45.5--70.8 & 552 & 1.8 & 1.1 & 37.3 & 59.8 \\
OpenCL~C & 166 & 86.7 & 80.5--92.6 & 1,328 & 4.4 & 1.2 & 29.9 & 64.5 \\
Rust-GPU & 150 & 98.0 & 94.3--99.3$^{w}$ & 1,024 & 15.4 & 3.5 & 50.1 & 31.0 \\
Mojo & 31 & 0.0 & 0.0--11.0$^{w}$ & 248 & 2.0 & 4.4 & 62.1 & 31.5 \\
\midrule
All & 2,585 & & & 17,559 & 31.1 & 3.9 & 35.0 & 30.0 \\
\bottomrule
\end{tabular}

\end{table}

\begin{table}[h]
\caption{Outcomes by corpus slice: shares (\%) of the feasible cells for the nine validated targets, as in \cref{tab:reach-target}. \emph{Families} are the program families that the bootstrap intervals resample. \emph{Seen}: used during \crosstl's development.}
\label{tab:reach-slices}
\footnotesize
\setlength{\tabcolsep}{4pt}
\begin{tabular}{@{}lrrrrrrr@{}}
\toprule
Slice & Units & Families & Cells & Usable & Degraded & Invalid & Stopped \\
\midrule
Sascha Willems, GLSL & 347 & 95 & 2,254 & 61 & 4 & 29 & 6 \\
Vulkan Samples, GLSL & 307 & 78 & 2,059 & 54 & 5 & 35 & 7 \\
Vulkan Samples, Slang & 158 & 48 & 1,033 & 49 & 6 & 33 & 12 \\
Sascha Willems, SPIR-V & 344 & 94 & 2,233 & 40 & 0 & 47 & 13 \\
Sascha Willems, HLSL & 308 & 80 & 2,032 & 36 & 7 & 40 & 18 \\
Vulkan Samples, HLSL & 121 & 41 & 801 & 33 & 9 & 37 & 21 \\
Sascha Willems, Slang & 170 & 94 & 1,114 & 29 & 11 & 43 & 18 \\
PolyBench, OpenCL C & 21 & 21 & 168 & 20 & 0 & 8 & 71 \\
Rust-GPU shaders & 150 & 79 & 1,024 & 15 & 4 & 50 & 31 \\
Candle, Metal & 16 & 16 & 108 & 10 & 4 & 24 & 62 \\
Rodinia, OpenCL C & 31 & 21 & 248 & 6 & 4 & 25 & 65 \\
Apple Metal samples & 89 & 89 & 464 & 6 & 1 & 17 & 76 \\
PyTorch MPS & 50 & 50 & 387 & 4 & 0 & 7 & 88 \\
Mojo GPU puzzles (seen) & 31 & 31 & 248 & 2 & 4 & 62 & 31 \\
HeCBench, HIP & 69 & 64 & 552 & 2 & 1 & 37 & 60 \\
MLX kernels (seen) & 49 & 49 & 242 & 2 & 7 & 0 & 91 \\
ggml, OpenCL C & 114 & 114 & 912 & 1 & 1 & 35 & 63 \\
ggml, Metal & 36 & 36 & 288 & 0 & 0 & 0 & 99 \\
HeCBench, CUDA & 102 & 88 & 816 & 0 & 0 & 34 & 66 \\
Candle, CUDA & 27 & 27 & 216 & 0 & 0 & 34 & 66 \\
PolyBench, CUDA & 21 & 21 & 168 & 0 & 0 & 0 & 100 \\
Rodinia, CUDA & 24 & 17 & 192 & 0 & 0 & 28 & 72 \\
\bottomrule
\end{tabular}

\end{table}

\begin{table}[h]
\caption{H-R3 on the corpus. Registered run: accepted shares (\%) of \crosstl and \crosstlr on the same feasible cells, per validated target, with the change in percentage points, its exact one-sided 95\,\% lower bound (\emph{Bound}, from a Clopper--Pearson bound on the share of cells that lost acceptance; an unregistered non-inferiority reading), and the cells in which \crosstlr crashed on an enumeration declaration. No change is significant after Holm's correction of the exact McNemar tests. Crash fixed: the change in the accepted share and in the output share (percentage points) after an exploratory re-run of the crashed jobs with a one-line fix. The median ratio of output bytes, \crosstlr to \crosstl, is 1.00 on every target. Slang has 73 fewer cells than in \cref{tab:reach-target}, because the \crosstlr run did not translate the ggml and PyTorch Metal programs to Slang.}
\label{tab:hr3}
\footnotesize
\setlength{\tabcolsep}{4pt}
\begin{tabular}{@{}lrrrrrrrr@{}}
\toprule
 & & \multicolumn{5}{c}{Registered run} & \multicolumn{2}{c}{Crash fixed} \\
\cmidrule(lr){3-7}\cmidrule(l){8-9}
Target & Cells & \crosstl & CrossTL-R & $\Delta$ & Bound & Crashes & $\Delta$ acc. & $\Delta$ out. \\
\midrule
Metal & 2,358 & 32.7 & 32.6 & -0.17 & -0.39 & 57 & +0.00 & +0.00 \\
HLSL & 2,585 & 37.0 & 37.0 & +0.00 & -0.12 & 68 & +0.00 & -0.27 \\
GLSL & 2,419 & 40.8 & 40.7 & -0.04 & -0.20 & 68 & +0.00 & +0.00 \\
SPIR-V & 2,585 & 57.6 & 57.5 & -0.08 & -0.24 & 68 & +0.00 & +0.00 \\
WGSL & 2,343 & 22.6 & 22.5 & -0.13 & -0.39 & 68 & -0.13 & -0.47 \\
GLSL~ES & 1,579 & 50.4 & 50.3 & -0.13 & -0.40 & 0 & -0.13 & +0.00 \\
Slang & 2,345 & 25.3 & 25.3 & +0.00 & -0.13 & 14 & +0.00 & -0.04 \\
CUDA & 640 & 2.2 & 2.2 & +0.00 & -0.47 & 28 & +0.00 & +0.00 \\
HIP & 640 & 2.0 & 2.0 & +0.00 & -0.47 & 28 & +0.00 & +0.00 \\
\bottomrule
\end{tabular}

\end{table}

\paragraph{H-R3 in detail.}
The \crosstlr run lacks the 294 cells of ggml and PyTorch Metal programs to Slang, Rust, Mojo and GLSL~ES that the \crosstl run translated before the harness began to skip them.
The largest change in the accepted share, on Metal, is $-0.17$ points (cluster interval $-0.36$ to $-0.04$), and four targets have no discordant cell at all.
Thirteen cells lost acceptance, and one gained it. Seven of the losses are crashes of \crosstlr on enumeration declarations, a one-line defect of its annotation pass. Four are refusals that the design intends: HLSL fragment shaders whose NaN-ignoring \code{min} and \code{max} WGSL cannot express. The last two are GLSL~ES outputs in which \crosstlr writes its record of an assumption above the \code{\#version} line, which the ES profile forbids. The one gain is a WGSL output that became valid because the explicit conversion that \crosstlr inserts for a mixed comparison is what WGSL requires.
In 133 cells an output disappeared, which lowers the output share by 1.1 to 2.5 points on Metal, HLSL, CUDA and HIP; 114 of these are the enumeration crash (91 of them in cells whose output the validator had rejected, 16 in Rust and Mojo outputs, which have no validator, and the 7 accepted ones above), 11 are refusals for WGSL, 7 are a new refusal of the HLSL emitter, which cannot lower the explicit integer narrowing that the annotation pass inserts, and the last is a time-out: \crosstlr exceeded the limit translating a Candle CUDA kernel to Slang, which \crosstl had translated in 8 seconds into an output that slangc rejected. In an exploratory re-run of the 521 crashed jobs with the one-line fix, every crash of the annotation pass disappeared. The accepted shares then change by at most 0.13 points and the output share by at most 0.47 points, through the refusals for WGSL and HLSL, the GLSL~ES defect and one time-out.
Carrying the rules leaves outputs almost unchanged: only 4 to 12\,\% of the graphics outputs and 37\,\% of the CUDA and HIP outputs change at all.

\FloatBarrier
\section{Monomorphisation of Metal's C++ Subset}
\label{app:mono}

\Cref{alg:mono} sketches how the Metal importer rewrites the reachable templates of a program into named, concrete copies before parsing it.
The procedure runs as ordered source-to-source steps: C preprocessing, extended for Metal's includes and features; MLX-style and explicit instantiation; function- and struct-template materialisation, with partial specialisation; free operator overloads; elision of provably stateless compile-time globals; lowering of struct methods to free functions with an explicit \code{self} reference; constrained (\code{enable\_if}) overload resolution; and integral constant folding.
Template argument deduction is conservative. Argument types are inferred from buffer element types, locals, casts and struct fields, and an argument that cannot be inferred is reported, not guessed.

\begin{figure}[h]
\begin{algorithmic}[1]
\Require preprocessed Metal text $T$; kernel entry roots $R$; specialisation budget $B$
\State $I \gets$ instantiation requests in $T$: \code{instantiate\_kernel} macros, \code{template [[host\_name]]} declarations, explicit \code{template f<A>;}
\ForAll{$(f, A, h) \in I$} \Comment{deduplicated by $(f, \mathrm{norm}(A), h)$}
  \State bind $A$, defaults and variadics to $f$'s parameters; substitute identifiers in $f$'s source span
  \State rename the copy to the host name $h$ and mark it \code{[[host\_name(h)]]}
\EndFor
\Repeat \Comment{explicit calls \code{f<A>(...)} reachable from $R$}
  \For{each unique key $(f, \mathrm{norm}(A))$, at most $B$}
     \State append a copy of $f$ with $A$ substituted as $\mathrm{mangle}(f, A)$; rewrite its call sites
  \EndFor
\Until{$T$ no longer changes}
\State materialise struct templates $S\langle A\rangle$ (selecting partial specialisations); lower methods to free functions
\State resolve calls to constrained (\code{enable\_if}) templates: the unique enabled candidate, else the native overload, else \textbf{refuse}
\State prune unreferenced templates; fold integral constants
\end{algorithmic}
\caption{Monomorphisation of Metal's C++ subset (simplified from 14 ordered steps). Every step rewrites source spans. For targets without templates, a request that cannot be resolved within the budget raises a typed diagnostic instead of emitting partially specialised code; for \crossgl and Metal output, the unresolved templates are kept.}
\label{alg:mono}
\end{figure}

\FloatBarrier
\section{Costs}
\label{app:costs}

\paragraph{Translation time.}
We timed every unit's import to \crossgl, and its translation to GLSL, with the options of the corpus run and without output formatting, in four warm worker processes, and recorded the CPU time of each call. Calls that exceeded the 300-second limit, which the worker stops after a further 10~s, are censored and left out of the slope fits; only Metal has such calls, 5 imports and 13 translations, and the 90th percentile of Metal imports is 23~s. Bytes are those of the unit's own file, not of its include closure.
The host was shared with other work during the measurement, with a median load average of 64 to 94 on its 12 cores, so the absolute times are upper estimates; their growth with input size is less affected.
The slopes of log import time on log bytes lie between 0.66 for GLSL and 1.05 for Slang, and the 95\,\% intervals of the slopes for Slang, Metal, HIP, OpenCL~C and Mojo extend above 1, up to 1.19 for Metal. Translation to GLSL grows faster than linearly from Slang (slope 1.22, interval 1.11 to 1.33) and SPIR-V (1.08, 1.00 to 1.16).

\paragraph{Carrying rules.}
We wrote one micro-kernel per carried rule, in a source language that defines the rule. Each kernel applies its operation 256 times per thread to $2^{20}$ threads, and we timed both translations on the GPU over 30 rounds in a shuffled order. On Metal, the NaN-aware \code{fmin} costs 1.39 [1.39, 1.40]; the \code{isnan} and conversion kernels request the same safe mode and cost nothing measurable, so the mode alone does not explain it. Through MoltenVK, the truncated remainder costs 1.21 [1.18, 1.23], \code{isnan} 5.85 [5.28, 6.52] and the NaN-aware \code{fmin} 7.33 [6.75, 7.89]; for the last two, \crosstl's own SPIR-V computes a constant (S2, S9 in \cref{tab:defects}), so \crosstlr is compared with a program that does no work. Wrapping arithmetic, shift masking, division by zero, narrow stores and float-to-integer conversion cost nothing measurable on either runtime, because their polyfills compile to the operation the hardware performs anyway, or add a select next to an operation that dominates.
The annotation pass rewrote 0.16 of every 100 arithmetic, shift, division and comparison operations on 1,840 corpus units, from 3.4 per 100 in GLSL to 0.03 in CUDA; on the 2,477 feasible validated cells of the 337 units in which it rewrote at least one operation, the accepted share fell from 13.2 to 13.0\,\%. It failed on 75 corpus units: on 62 because it does not handle enumerations (44 of them Metal programs), and on 13 because the saved hub program did not parse again. It failed on 47 of the 90 Metal programs.

\paragraph{Recording the source.}
After lowering, 2.7 to 22.8 times as many polyfill call sites remain per target under the strongest rules as under the sources' own (\cref{tab:provenance}). Of the 2,345 units whose import succeeded, 2,326 were measured: the saved hub programs of 15 do not parse again, and 4 timed out. The 58 units that need Metal's safe mode, a recorded NaN assumption and a refusal for WGSL only under the strongest rules are GLSL, Slang, Rust-GPU and SPIR-V programs whose \code{min} and \code{max} (47 programs) or float-to-integer conversions (11) promise nothing for NaN.

\paragraph{Source-owned functions.}
On all $2^{32}$ binary32 inputs, \crosstl's own \code{sin}, \code{cos}, \code{exp} and \code{log} meet MSL's bound for precise functions on Metal (at most 3.19~ULP), whereas GLSL's built-ins, which a namesake mapping would use, miss it by up to $10^{14}$~ULP (\cref{tab:exhaustive}).
We evaluated the six functions on four routes: Apple's precise functions, \crosstl's Metal output, \crosstl's SPIR-V on MoltenVK, and, for comparison, the GLSL built-in function on MoltenVK. Because MSL and Vulkan allow binary32 subnormals to be flushed, we accept a flushed input or result. \code{log2} and \code{tan}, which \crosstl maps to the target's functions, match Apple's accuracy on Metal; through MoltenVK, which compiles with fast math by default, they miss the bound on 3.2 million and 3.2 billion inputs (S34). The GLSL built-ins miss it by up to 64~ULP for \code{exp}, $10^{6}$ for \code{log}, and $10^{14}$ for \code{sin}, \code{cos} and \code{tan}, whose large arguments receive no accurate range reduction.
On Metal, \code{sin} takes 0.68 against 0.37~ms per $2^{24}$ inputs, and \code{exp} 0.32 against 0.33~ms. On MoltenVK, \crosstl's SPIR-V takes 25 to 65~ms per $2^{24}$ inputs, against under 2~ms for glslang's. Most of that gap is not the algorithms: \crosstl's SPIR-V declares a work-group size of one (S29), so MoltenVK runs one thread per SIMD group, and even the natively mapped \code{log2} takes 25~ms. The algorithms add 1.5 to 2.6 times on top of that. The importer replaces the functions before it knows the target, so Metal-to-Metal translation pays this cost too, although Apple's functions already meet the bound there.

\begin{table}[h]
\caption{What recording the source's rule saves (exploratory), on the 2,326 corpus units whose saved hub program parses again and whose annotation did not time out, with the one-line enumeration fix applied. The annotation pass runs once with each unit's own source rules (\emph{Source}) and once with the strongest rule for every operation of every source (\emph{Strongest}), and the rule table lowers both versions for each target. The pass rewrites 5,390 operations in 355 units under the first and 50,027 in 612 under the second. \emph{Polyfill call sites}: rule-carrying calls left after lowering. \emph{Units marked}: units whose artifact requests Metal's safe mode (C), records an assumption that the implementation supports NaN (A), or is refused (R). The SPIR-V lowering of NaN rules records neither a configuration nor an assumption (\cref{sec:rulehub}).}
\label{tab:provenance}
\small
\begin{tabular}{@{}lrrrrr@{}}
\toprule
 & \multicolumn{3}{c}{Polyfill call sites} & \multicolumn{2}{c}{Units marked} \\
\cmidrule(lr){2-4}\cmidrule(l){5-6}
Target & Source & Strongest & Ratio & Source & Strongest \\
\midrule
Metal & 1{,}550 & 30{,}337 & 19.6 & 34 C & 92 C \\
HLSL & 4{,}070 & 32{,}229 & 7.9 & 0 & 0 \\
GLSL & 3{,}015 & 18{,}472 & 6.1 & 34 A & 92 A \\
GLSL ES & 3{,}015 & 18{,}472 & 6.1 & 34 A & 92 A \\
SPIR-V & 3{,}015 & 18{,}472 & 6.1 & 0 & 0 \\
WGSL & 1{,}216 & 3{,}314 & 2.7 & 34 R & 92 R \\
CUDA & 1{,}949 & 44{,}522 & 22.8 & 0 & 0 \\
HIP & 1{,}949 & 44{,}522 & 22.8 & 0 & 0 \\
\bottomrule
\end{tabular}

\end{table}

\begin{table}[h]
\caption{Exhaustive binary32 accuracy of \crosstl's source-owned functions (\code{sin}, \code{cos}, \code{exp}, \code{log}) and of two functions it maps to the target ($^{*}$). \emph{max}: maximum error in ULP over all $2^{32}$ inputs, tolerating flushed subnormals. \emph{over}: inputs whose result exceeds the bound, or whose NaN, infinity or zero result does not match (M: millions, G: billions). Apple: \code{precise::f} compiled by Apple's compiler in safe mode. \crosstl: its Metal output. \crosstl/MVK: its SPIR-V on MoltenVK. GLSL/MVK: GLSL's built-in \code{f} compiled by glslang, on MoltenVK. Bounds are MSL's for precise functions.}
\label{tab:exhaustive}
\small
\setlength{\tabcolsep}{3.5pt}
\begin{tabular}{@{}lrrrrrrrrr@{}}
\toprule
 & \multicolumn{2}{c}{Apple} & \multicolumn{2}{c}{\crosstl} & \multicolumn{2}{c}{\crosstl/MVK} & \multicolumn{2}{c}{GLSL/MVK} & Bound \\
\cmidrule(lr){2-3}\cmidrule(lr){4-5}\cmidrule(lr){6-7}\cmidrule(lr){8-9}
Function & max & over & max & over & max & over & max & over & (ULP) \\
\midrule
\code{sin} & 3.86 & 0 & 3.19 & 0 & 3.19 & 0 & $10^{14}$ & 3.2\,G & 4 \\
\code{cos} & 3.95 & 0 & 3.15 & 0 & 3.15 & 0 & $10^{14}$ & 2.1\,G & 4 \\
\code{exp} & 1.38 & 0 & 0.91 & 0 & 0.93 & 0 & 64.46 & 37\,M & 4 \\
\code{log} & 0.86 & 0 & 1.97 & 0 & 2.81 & 0 & $10^{6}$ & 3.1\,M & 4 \\
\code{log2}$^{*}$ & 1.80 & 0 & 1.80 & 0 & $10^{6}$ & 3.2\,M & $10^{6}$ & 3.2\,M & 4 \\
\code{tan}$^{*}$ & 5.27 & 0 & 5.27 & 0 & $10^{14}$ & 3.2\,G & $10^{14}$ & 3.2\,G & 6 \\
\bottomrule
\end{tabular}

\end{table}

\begin{table}[h]
\caption{Translation time by source language: median CPU time of the import to \crossgl and of the translation to GLSL, 90th percentile of the import, calls censored at 300~s (both kinds), and the slope of $\log_{10}$ import time on $\log_{10}$ source bytes with a 95\,\% bootstrap interval. Measured on a shared host (median load average 64 to 94 on 12 cores).}
\label{tab:latency}
\footnotesize
\setlength{\tabcolsep}{4pt}
\begin{tabular}{@{}lrrrrrl@{}}
\toprule
 & & \multicolumn{2}{c}{Median CPU time (ms)} & & & \\
\cmidrule(lr){3-4}
Source & Units & Import & + GLSL & 90th pct. & Cens. & Slope (95\,\% CI) \\
\midrule
GLSL & 654 & 2 & 41 & 6 & 0 & 0.66 (0.60--0.72) \\
HLSL & 429 & 6 & 41 & 14 & 0 & 0.81 (0.76--0.85) \\
Slang & 328 & 3 & 54 & 8 & 0 & 1.05 (0.98--1.13) \\
SPIR-V & 344 & 2 & 62 & 3 & 0 & 0.89 (0.86--0.92) \\
Metal & 240 & 705 & 1436 & 22,798 & 18 & 0.92 (0.68--1.19) \\
CUDA & 174 & 82 & 416 & 844 & 0 & 0.81 (0.66--0.97) \\
HIP & 69 & 81 & 283 & 383 & 0 & 0.85 (0.61--1.12) \\
OpenCL~C & 166 & 83 & 181 & 358 & 0 & 0.95 (0.87--1.01) \\
Rust-GPU & 150 & 7 & 100 & 15 & 0 & 0.97 (0.94--1.00) \\
Mojo & 31 & 14 & 44 & 44 & 0 & 1.00 (0.95--1.05) \\
\bottomrule
\end{tabular}

\end{table}

\FloatBarrier
\section{Fault Injection and the MLX Case Study}
\label{app:faults}\label{app:mlx}

\paragraph{Fault injection.}
Around a kernel whose scalar input \code{count} carries the checked precondition $1\le\code{count}\le4$, we built runtime packages for Metal, OpenGL and Direct3D, and mutated every contract line of each package (delete, duplicate, move, truncate, add an unknown field, remove each field, alter each value), each time updating the artifact's hash and size in the package record so that the integrity check could not catch the change. Package inspection rejected all 315 mutations. The checks of the artifact's own contract lines rejected 263; the other 52 are well-formed contracts with another meaning, such as a precondition bound moved by one, which only the comparison with the package record caught. Dispatch requests with \code{count} outside the precondition were rejected and those inside it accepted, on all three targets; these checks run in each package's host-side request builder, and no request reached a GPU or a Direct3D runtime.

\paragraph{The MLX case study.}
The integration translates kernels from MLX's 49 \code{.metal} files, organised as 46 kernel families, and supplies per family the configuration that the files do not state. With native execution mandatory, 42 family modules executed translated kernels on Metal in 1,493 execution tests, and 38 on OpenGL in 1,495, with no failures apart from 8 setup errors in two families that have no OpenGL path; 21 families run under caller-trusted index ranges on OpenGL; on Metal, 37 families also compare with upstream MLX code run on the same GPU, unchanged except for one four-line instantiation. The integration consists mostly of per-family knowledge that the source does not state, so the hub does not make a second target cheap in code. Whole-file translation of MLX's own kernel files, without that knowledge, yields 4 usable cells out of 245 (\cref{app:corpora}).

\FloatBarrier
\section{Implementation Inventory and Defects}
\label{app:inventory}
\label{app:defects}

Importers print one of two dialects of \crossgl: C-style stage blocks (Metal, HLSL, GLSL, SPIR-V, Rust-GPU, Slang and Mojo) or WGSL-like \code{fn} declarations with bindings (CUDA, HIP and OpenCL~C). The GLSL and HLSL emitters, and the GLSL~ES emitter that reuses them, run shared lowering passes and the C-family conversion model; the others map constructs directly. Refusals are exception classes that carry a stable diagnostic code, and fallbacks emit a neutral value with a comment.
The project's contract tests cover metadata preconditions, resource-storage encodings, dispatch regions, frozen specialisations and index-range contracts; we rest no claim on them, and test the loader independently (\cref{sec:eval:check}). The fault injection of \cref{sec:eval:check} made 315 mutations (228 of preconditions, 69 of specialisations and 18 of storage encodings); the checks of the artifact's own contract lines rejected all 20 that add an unknown field or duplicate a line verbatim. The kernel has no dispatch-region contract, so that kind was not tested.

\Cref{tab:defects} lists the defects of the evaluated implementation that we found while preparing this paper, grouped by effect: \emph{silent} (S), where the target compiler accepts the output or an interface loads but the meaning changes; \emph{invalid} (V), where the target compiler rejects the output; and \emph{crash} (X), an unstructured exception.
By component, 12 of the 34 silent defects lie in the SPIR-V emitter and 11 in the importers. By mechanism, they map a construct to a namesake with a different rule (for example S1, S8, S27, S32), replace it by a neutral value (S2, S9, S23), lose a type or a decoration (S4, S6, S15, S21), or lose a binding, a stage or control flow (S3, S16, S18, S19).
The artifact records, for each row, the experiment, probe or code location that exhibits it, and we checked every silent defect that a hazard case exhibits against its raw artifact.
We have not yet reported these defects, or the third-party findings of \cref{sec:eval:hazards,sec:eval:rulehub} (the wrap failures of MoltenVK and KosmicKrisp, SPIRV-Cross and Slang copying overflow into MSL, naga and Tint dropping \code{NoContraction}, DXC's \code{float3} stride), to their maintainers.

{\scriptsize\setlength{\tabcolsep}{3pt}
\begin{longtable}{@{}l >{\raggedright\arraybackslash}p{2.9cm} >{\raggedright\arraybackslash}p{\dimexpr\linewidth-4.2cm\relax}@{}}
\caption{Defects of the evaluated implementation. S: silent; V: invalid output; X: crash. $^{\circ}$: found by reading the code; every other row was exhibited by an executed artifact, probe or run.}\label{tab:defects}\\
\toprule
ID & Component & Defect \\
\midrule
\endfirsthead
\toprule
ID & Component & Defect \\
\midrule
\endhead
\bottomrule
\endfoot
S1 & Metal emitter & Signed $+,-,\times$ from GLSL and HLSL emitted as C++ operators, which Metal treats as undefined on overflow (\code{g\_add\_wrap}). \\
S2 & SPIR-V emitter & \code{isnan} becomes an ordered comparison of the constant 0.0 with itself, which is always false. Unknown functions (\code{isinf}, \code{transpose}, \code{inverse}, \code{\_\_shfl\_down\_sync}) become a neutral value with a warning comment, which formatting can strip. The modules pass \code{spirv-val} (\code{m\_isnan}, \code{g\_isnan}). \\
S3 & SPIR-V emitter & In \code{for (k = s; k < s + 4; ++k)} from HLSL, \code{k} is initialised to 0 and only incremented, never tested, and the condition re-tests \code{s < s + 4}, so the loop does not terminate unless \code{s + 4} overflows (\code{h\_loop\_wrap}). \\
S4 & SPIR-V emitter & \code{precise} from HLSL is not lowered to \code{NoContraction}: the emitter recognises only the attribute form (\code{h\_precise}, MoltenVK). \\
S5 & WGSL emitter & \code{precise} is dropped without a diagnostic, although WGSL cannot forbid contraction. \\
S6 & SPIR-V emitter & 8- and 16-bit integers are widened to 32 bits without re-narrowing (\code{m\_uchar\_wrap}). \\
S7 & SPIR-V, WGSL emitters & \code{half} arithmetic is computed in binary32 without binary16 rounding (\code{m\_half\_mul}). \\
S8 & SPIR-V emitter & \code{fmod} becomes \code{OpFMod} (sign of the divisor) instead of \code{OpFRem} (\code{m\_fmod\_sign}). \\
S9 & SPIR-V emitter & \code{fmin} cannot be lowered: the output stores the constant 0.0, with a warning comment (\code{m\_fmin\_nan}). \\
S10 & SPIR-V emitter & A mixed \code{int < uint} comparison is emitted as a signed comparison (\code{m\_cmp\_mixed}). \\
S11 & HLSL, SPIR-V emitters & HLSL's matrix construction and \code{mul} orientation are transposed relative to GLSL, Metal and SPIR-V; the SPIR-V emitter zero-pads \code{mat4(1.0)}, which is then not the identity. \\
S12 & GLSL, WGSL emitters & \code{long}, \code{ulong} and \code{size\_t} are narrowed to 32 bits by the GLSL emitter; the WGSL emitter narrows only \code{long} and passes the others through. \\
S13 & SPIR-V emitter & bfloat16 products are not re-rounded to bfloat16. \\
S14$^{\circ}$ & CUDA, HIP emitters & The CUDA \code{WaveActiveSum} helper is a shuffle-down tree that is correct only in lane 0 and under full masks; the HIP helper returns the lane's own value. \\
S15 & Metal importer & \code{packed\_float3} becomes a 16-byte \code{float3}, changing strides and member offsets (\code{m\_packed\_float3}). \\
S16 & GLSL emitter & In whole-file translation of a multi-kernel Metal file the second kernel loses its entry status and bindings collide. \\
S17 & SPIR-V importer & An \code{OpLoad} is forwarded as a variable reference and observes a later store. \\
S18 & SPIR-V importer & Sequential selection constructs are flattened into unconditional stores. \\
S19 & HLSL importer & A helper function whose name begins with a stage prefix (\code{as}, \code{ms}, \code{ps}, \code{vs}, \ldots) becomes an entry of that stage. \\
S20 & GLSL importer & A loose \code{std140} uniform block loses its binding and block name. \\
S21 & CUDA importer & \code{\_\_nv\_bfloat16} becomes binary16, and \code{size\_t} 32 bits. \\
S22$^{\circ}$ & HIP importer & \code{warpSize} is the constant 32 (64 on wave64 GPUs). \\
S23$^{\circ}$ & \crossgl front end & In the lenient default mode, a function body that fails to parse is replaced by an empty block. \\
S24$^{\circ}$ & WGSL emitter & \code{if (c) return;} with a non-block body is emitted as \code{if (c) \{\}}. \\
S25$^{\circ}$ & SPIR-V emitter & Loose uniforms become \code{Private} variables that the host cannot set. \\
S26 & Metal importer & The source-owned \code{precise::pow} handles special cases itself but calls the target's \code{pow} for the rest. Under Apple's default options its error (23.9~ULP) exceeds MSL's 16-ULP bound. \\
S27 & SPIR-V emitter & Signed integer \code{\%} becomes \code{OpSMod} (sign of the divisor) instead of \code{OpSRem}. For a CUDA kernel, \code{-7 \% 2} yields 1 instead of $-1$. \\
S28 & CUDA, HIP, OpenCL importers & \code{fmodf(x, y)} and \code{fmod(x, y)} become the hub's floored \code{mod(x, y)}, so every target computes $x - y\lfloor x/y\rfloor$ instead of C's truncated remainder (\code{u\_fmodf\_sign}). \\
S29 & Single-file translation & The work-group size of Metal, CUDA and OpenCL~C kernels is a host parameter, and single-file translation emits a size of one. Subgroup collectives (\code{simd\_sum}, \code{\_\_shfl\_down\_sync} reductions) then see one lane (\code{u\_warp\_reduce}, \code{m\_simd\_sum}). \\
S30 & Metal emitter & A recursive device function is emitted unchanged, which MSL permits in kernels from Metal~2.4. The Metal compiler accepts it and the GPU hangs, so the cause may lie in the runtime; the GLSL and HLSL emitters refuse recursion that they cannot rewrite as a loop (\code{u\_recursion}). \\
S31 & SPIR-V importer & GLSL.std.450's \code{NMin} and \code{NMax} are imported as \code{min} and \code{max}, the spelling used for \code{FMin} and \code{FMax}, so the NaN rule of the SPIR-V instruction set is lost; the SPIR-V round trip emits \code{FMin} (\code{s\_nmin\_nan}, which carries no obligation under Vulkan's rules). \\
S32 & HLSL emitter & Signed $+,-,\times$ from GLSL are emitted as HLSL operators, which DXC marks \code{nsw} and folds, so the wrap rule is lost on Direct3D~12 (\code{g\_add\_wrap}, \code{g\_mul\_wrap}, \code{g\_loop\_wrap}). \\
S33$^{\circ}$ & GLSL emitter & binary16 buffers are widened to 32-bit elements without a contract, so the host interface changes silently. \\
S34 & Metal importer & \code{precise::log2} and \code{precise::tan} map to the target's functions, which miss MSL's bound for precise functions on 3.2~million and 3.2~billion inputs through MoltenVK (\cref{tab:exhaustive}). \\
\midrule
V1--V3 & WGSL emitter & Illegal \code{var<storage, write>}; \code{global\_invocation\_id} typed \code{u32} for Metal sources; atomics on non-atomic types, \code{mat4x4(1.0)}, undefined \code{isnan}, \code{fmod}, \code{uint8}. \\
V4--V5 & GLSL emitter & Name collisions in a vertex shader; guard macros and renamed inputs in combined multi-stage files. \\
V6 & HLSL, GLSL, Metal, WGSL, CUDA, HIP & Unknown identifiers are passed through to the target compiler. \\
V7 & CUDA, HIP emitters & Module-scope buffers emitted as host globals used in device code; \code{uchar} and \code{bfloat} are not CUDA types. \\
V8$^{\circ}$ & Rust, Mojo emitters & \crossgl \code{char} is mapped to Rust's Unicode \code{char} and to Mojo's \code{String}. \\
V9 & SPIR-V, WGSL emitters & Recursion is emitted instead of refused, and the validators reject the cyclic call graph (\code{u\_recursion}). \\
V10 & CUDA, HIP emitters & A graphics entry point is emitted as a \code{\_\_device\_\_} function, named \code{main} when the source entry is \code{main}, so the output has no kernel, instead of being refused; nvcc rejects \code{main} in 2,119 graphics-stage outputs of the corpus run. \\
V11 & Emitters, Mojo sources & Generic type arguments of Mojo programs are printed as Python representations of syntax-tree nodes (\code{LiteralNode(...)}), in the outputs of 22 of the 31 Mojo puzzles. \\
\midrule
X1 & CUDA, HIP emitters & \code{KeyError} in the lowering of vector arithmetic, the second most frequent crash (54) in the corpus run. \\
X2 & WGSL emitter & \code{AttributeError} on enumeration declarations, the most frequent crash in the corpus run (55). \\
X3 & Rust emitter & \code{KeyError} on \code{u8} and \code{i8} (38 crashes in the corpus run). \\
\end{longtable}}

\FloatBarrier
\section{Harness Corrections}
\label{app:harness}

We found and corrected seven errors in our own harness, none of which changed a conclusion.
\begin{itemize}
\item The Metal runner treated a request for relaxed math as fast math. We re-ran all 910 affected cells of the suite and the held-out kernels, and no outcome changed.
\item The Metal runner bound the buffers of 24 third-party artifacts incorrectly: it swapped the input and output of one Tint kernel, left unbound the buffer from which Tint's kernels read their buffer sizes, and filled naga's size buffer in the wrong order. We re-ran the 72 affected cells of the suite and 108 of the held-out kernels. Two outcomes changed: Tint keeps the \code{std430} layout rule on Metal, and, under the legacy HLSL rules, its bounds clamp now loses Direct3D's out-of-bounds rule.
\item The suite never recorded the native compile of its CUDA sources. All 15 compile.
\item Artifact maps of early runs were overwritten, and paths were rebuilt from the suite's deterministic naming.
\item The corpus records no stage for Slang units and some Metal units, so 309 GLSL outputs could not be validated. Re-validated with the stages their sources declare, 200 were accepted and 102 rejected; 7 remain unvalidated.
\item The corpus recorded every Rust-GPU unit as a compute program, although 121 of the 150 are vertex and fragment programs. Feasibility was recomputed from the source.
\item The MLX runs did not set three shared switches that gate some of the families' execution tests, and 68 of them were skipped. We re-ran the four affected families with the switches set (\cref{sec:eval:mlx}).
\end{itemize}

\FloatBarrier
\section{Toolchains and Reproduction}
\label{app:repro}

\paragraph{Toolchains.}
All experiments ran on one Apple M2~Max (12 CPU cores, 32\,GB, macOS~27.2). Metal programs were compiled by the Xcode Metal compiler (32023.918); Vulkan programs ran on MoltenVK~1.4.2 and, exploratorily, on KosmicKrisp~26.2.99 from LunarG's Vulkan SDK~1.4.363.0, and WGSL programs through wgpu-native~29.0.1 (via wgpu-py~0.32) on Metal. Mesa~25.2.8 provided lavapipe and llvmpipe (OpenGL~4.5) in a Linux container. The Direct3D~12 route used DXC~1.9.2609 and dxil-spirv, and Apple's Metal Shader Converter~2.0.5 converted DXIL into Metal libraries, which we bound through its documented top-level argument buffer. OpenCL~C ran on Apple's OpenCL~1.2. CUDA and HIP outputs were compiled by nvcc~13.0 for \code{sm\_80}.
The mechanical check of the rule table used Z3~5.3.1. The validators were the Metal compiler; DXC with each entry point's stage profile; glslangValidator~16.6.0; \code{spirv-val} from SPIRV-Tools v2026.4; naga~30.0.1 and Tint (Dawn \code{4a4e5d06}) for WGSL; and slangc~2026.19.
The catalogue's sources are GLSL~4.60, Direct3D~11.3 (DXBC) and DXIL as compiled by DXC~1.9, MSL~4.1, WGSL, SPIR-V~1.6 with the Vulkan~1.4 environment, CUDA~13 and HIP (ISO~C++ where MSL and CUDA are silent), and OpenCL~C~3.0 (C99 where it is silent).
The baselines ran with these options: glslang and DXC compiled each unit with its project's own flags and profiles; SPIRV-Cross emitted MSL~3.0, HLSL for Shader Model~6.6 and GLSL~450; the other tools used their default options for each output language. SPIRV-Cross's and Tint's GLSL was validated as OpenGL GLSL, and Slang's as Vulkan GLSL, the dialect it emits. Without the requirement that its raw output carry no fallback marker, \crosstl's success rates in \cref{tab:baselines} rise by at most 7 points, except from HLSL and Slang to SPIR-V (43 and 53 points).

\paragraph{Reproduction.}
The artifact contains a manifest of the corpora (repository, upstream revision, licence and the SHA-256 of every unit file), with scripts that fetch the pinned revisions and rebuild the units and the native filter; the hazard suite, its registered and current registration documents, the held-out generator, the harness and the container files for the Linux runtimes; the \crosstlr passes as a patch against the evaluated implementation of \crosstl; every raw result as JSON lines; the coding packets, codebooks, codings and adjudications; and the analysis scripts that generate every table, figure and number in this paper.
The analysis scripts regenerate their outputs from the raw results, and a further script checks every number quoted in the text against those outputs and fails if any differs. An ordered run list gives every step with its inputs, outputs and host, and a requirements file pins the Python environment; local paths are set through environment variables.
Every table, figure and number can be regenerated from the shipped results on any host with Python. Re-running the experiments needs the deposited raw outputs, the corpus rebuilt from its pinned revisions, a container runtime, and, for the GPU runs, an Apple-silicon Mac.
\FloatBarrier

\end{document}